\documentclass[iop]{emulateapj}

\def\M1{M_1^\prime}
\def\Msun{M_\odot}

\def\hMsun{$h^{-1}M_\odot$}

\def\hMpc{$h^{-1}{\rm Mpc}$}

\def\wp{$w_p(r_p)$}

\gdef\Vdisp{$\sigma$}
\gdef\Mdyn{M$_{\rm dyn}$}
\gdef\Mstar{M$_{\rm star}$}
\gdef\Mden{$\Sigma$}
\gdef\Re{$R_{e}$}
\gdef\Pel{P$_{\rm el}$}
\gdef\Mhalo{M$_{\rm halo}$}

\slugcomment{To appear in ApJ.}

\shorttitle{}
\shortauthors{Wake et al.}

\begin{document}

%% LaTeX will automatically break titles if they run longer than
%% one line. However, you may use \\ to force a line break if
%% you desire.

%%\title{The relationship between galaxy velocity dispersion and dark matter halo properties}
\title{Which galaxy property is the best indicator of its host dark matter halo properties?}

%% Use \author, \affil, and the \and command to format
%% author and affiliation information.
%% Note that \email has replaced the old \authoremail command
%% from AASTeX v4.0. You can use \email to mark an email address
%% anywhere in the paper, not just in the front matter.
%% As in the title, use \\ to force line breaks.

\author{David A. Wake\altaffilmark{1},
Marijn Franx\altaffilmark{2},
Pieter G.\ van Dokkum\altaffilmark{1}
}

%% Notice that each of these authors has alternate affiliations, which
%% are identified by the \altaffilmark after each name.  Specify alternate
%% affiliation information with \altaffiltext, with one command per each
%% affiliation.

\altaffiltext{1}{Department of Astronomy, Yale University, New Haven, CT 06520}
\altaffiltext{2}{Sterrewacht Leiden, Leiden University, NL-2300 RA Leiden,
The Netherlands.}

%% Mark off your abstract in the ``abstract'' environment. In the manuscript
%% style, abstract will output a Received/Accepted line after the
%% title and affiliation information. No date will appear since the author
%% does not have this information. The dates will be filled in by the
%% editorial office after submission.

\begin{abstract}
In this work we investigate the link between galaxy velocity dispersion, mass and other properties (color, morphology) with the properties of dark matter halos by comparing the clustering of galaxies at both fixed mass and velocity dispersion (\Vdisp). We use the Sloan Digital Sky Survey to define a volume limited sample of massive galaxies complete in both stellar mass (\Mstar)~and \Vdisp~with \Mstar~$> 6 \times 10^{10}$ \hMsun~and \Vdisp~$>$ 75 km/s. Using this sample we show that at fixed \Vdisp~there is no dependence of the clustering amplitude on stellar or dynamical mass (\Mdyn). Conversely when \Mstar~or \Mdyn~are fixed there is a clear dependence of the clustering amplitude on \Vdisp~with higher \Vdisp~galaxies showing a higher clustering amplitude. We also show that whilst when \Mstar~or \Mdyn~are fixed there remains a dependence of clustering amplitude on morphology, there is no such dependency when \Vdisp~is fixed. However, we do see a dependence of the clustering amplitude on $g-r$ color when both mass and \Vdisp~are fixed. Despite this, even when we restrict our samples to only ellipticals or red galaxies the relationship between \Vdisp~and clustering amplitude at fixed mass remains. It seems likely that the residual correlation with color is driven by satellite galaxies in massive halos being redder at fixed \Vdisp. The lack of a similar residual morphology dependence implies that the mechanism turning satellites red is not changing their morphology. Our central result is that \Vdisp~is more closely related to the clustering amplitude of galaxies than either \Mstar~or \Mdyn. This implies that \Vdisp~is more tightly correlated than \Mstar~or \Mdyn~with the halo properties that determine clustering, either halo mass or age, and supports the notion that the star formation history of a galaxy is more closely related to its halo properties than its overall mass.  

\end{abstract}

%% Keywords should appear after the \end{abstract} command. The uncommented
%% example has been keyed in ApJ style. See the instructions to authors
%% for the journal to which you are submitting your paper to determine
%% what keyword punctuation is appropriate.


%% From the front matter, we move on to the body of the paper.
%% In the first two sections, notice the use of the natbib \citep
%% and \citet commands to identify citations.  The citations are
%% tied to the reference list via symbolic KEYs. The KEY corresponds
%% to the KEY in the \bibitem in the reference list below. We have
%% chosen the first three characters of the first author's name plus
%% the last two numeral of the year of publication as our KEY for
%% each reference.

%% Authors who wish to have the most important objects in their paper
%% linked in the electronic edition to a data center may do so by tagging
%% their objects with \objectname{} or \object{}.  Each macro takes the
%% object name as its required argument. The optional, square-bracket 
%% argument should be used in cases where the data center identification
%% differs from what is to be printed in the paper.  The text appearing 
%% in curly braces is what will appear in print in the published paper. 
%% If the object name is recognized by the data centers, it will be linked
%% in the electronic edition to the object data available at the data centers  
%%
%% Note that for sources with brackets in their names, e.g. [WEG2004] 14h-090,
%% the brackets must be escaped with backslashes when used in the first
%% square-bracket argument, for instance, \object[\[WEG2004\] 14h-090]{90}).
%%  Otherwise, LaTeX will issue an error. 

\section{Introduction}

In cosmological models of galaxy formation the formation history of galaxies, and hence their present day properties, are intimately linked with the formation history and properties of the dark matter (DM) haloes in which they reside. Therefore, if we wish to understand the process of galaxy formation it is important to observationally establish a link between the properties of galaxies and those of the halos in which they reside.

In recent years many studies have linked galaxy properties to halo mass using clustering, gravitational lensing, satellite kinematics, group catalogues or subhalo abundance matching, with the main focus being on linking the stellar properties (stellar luminosity, stellar mass, color, SFR) or those of the central black hole (luminosity, black hole mass) to the galaxy host dark matter halo properties. This work has shown strong correlations between halo mass and stellar mass, color, star formation rate, or morphology, such that more massive (luminous), redder (less star-forming), or more spheroidal galaxies live in more massive halos \citep[e.g.][]{Li06a,Yang08,Yang09,Zehavi11,More11,Mandelbaum06,Leauthaud11}. For AGN the relationships are less clear although correlations do exist between narrow line AGN luminosity or radio luminosity \citep{Wake04,Li06b,Wake08b,Mandelbaum09} and halo mass.
Perhaps unsurprisingly the strongest relationship occurs with stellar mass, which for central galaxies in halos show a tight correlation with a relatively small scatter of ~0.17 dex \citep{Yang11}. At fixed stellar mass properties that indicate the star formation history of a galaxy then show little dependence on halo mass \citep{More11}, when only considering central galaxies.   

In this work we focus on investigating how the dynamical properties of galaxies, such as velocity dispersion (\Vdisp) or surface mass density (\Mden), in addition to stellar mass are related to the clustering amplitude of galaxies and hence their host dark matter halo properties. This is a particularly relevant question as it is becoming increasingly clear that many galaxy properties such as their current star-formation rates, star-formation histories, metalicities, and black hole masses are more closely related to these dynamical properties than their total stellar mass \citep[e.g.][]{Bender93,Ferrarese00,Gebhardt00,Kauffmann03b,Franx08}. If the formation history and properties of the host dark matter halo are the major underlying cause of these galaxy properties, we might expect that \Vdisp~or \Mden~are better indicators than stellar mass of these halo properties.

In this paper we make use of the very large sample of galaxies with velocity dispersion, stellar mass, size, color and morphology measurements from the seventh data release of the Sloan Digital Sky Survey \citep[SDSS DR7;][]{Abazajian09}. We split these galaxies into narrow bins of one parameter, in particular mass or \Vdisp, and then see if there is any dependence of the clustering amplitude on another galaxy parameter within that narrow bin. In this way we can investigate if there are any residual dependencies on halo properties manifested by higher or lower clustering amplitudes as one parameter is varied whilst another remains fixed.  

Throughout this paper, we assume a flat $\Lambda$--dominated CDM cosmology with $\Omega_m=0.27$, $H_0=73 $km s$^{-1} $Mpc$^{-1}$, and $\sigma_8=0.8$ unless otherwise stated.

\section{Data}
\label{sec:data}

The galaxy data used in this analysis are gathered from the seventh data release of the SDSS \citep{Abazajian09}. We begin with the Large Scale Structure samples of the NYU Value Added Galaxy catalog \citep[VAGC;][]{Blanton05}. These samples have been carefully constructed for measurements of large scale structure and include corrections for fiber collisions, the tracking of spatially dependent completeness and appropriate random catalogs all of which are required to accurately measure clustering. The sample we use has an $r$ band magnitude range of $14.5 < r < 17.6$. In addition the NYU VAGC gives k-corrected (to z=0.1) absolute magnitudes \citep{Blanton03}, velocity dispersion measurements from the Princeton Spectroscopic pipeline, and circularised sersic fits for each galaxy, all of which we make use of in this analysis.

For estimates of the stellar mass we make use of the MPA-JHU value added catalog which provides stellar mass estimates based on stellar population fits to the SDSS photometry \citep{Kauffmann03, Salim07}. The overlap between the MPA-JHU and NYU VAGCs is close to but to quite 100\% and so we remove the regions where they do not match from the analysis and adjust the completeness corrections appropriately (see section \ref{sec:clus}). We also remove any region of the survey that has a spectroscopic completeness less than 70\%. This leaves an area of 7640 deg$^2$ and a total sample of 521,313 galaxies.

The SDSS velocity dispersions are measured within the 3" diameter SDSS fiber. We correct to a common aperture of one eighth of an effective radius ($r_e$), the central velocity dispersion, using the relation $\sigma_0 = \sigma_{ap}(8r_{ap}/r_e)^{0.066}$ where $r_{ap}$ = 1."5 \citep{Cappellari06}. $r_e$ is taken from the best fitting circularised sersic profile fit.

In addition to investigating how clustering depends on stellar mass and velocity dispersion, we also include the dynamical mass (\Mdyn) as an addition galaxy mass indicator. Estimates of galaxy stellar masses are uncertain due to the complex nature of the star formation history, stellar population synthesis modeling, extinction law and initial mass function \citep[e.g. see][]{Marchesini09, Muzzin09, Conroy09}. The dynamical mass, which is estimated purely from the velocity dispersion and the size of the galaxy, will not be affected by these same systematic uncertainties and will likely have errors more similar to those of the velocity dispersion. In addition there is a strong and fairly tight correlation between stellar mass and dynamical mass in the SDSS \citep[][see also Figure \ref{fig:MstarMdyn}]{Taylor10} and so both masses are largely tracing the same physical property of a galaxy.

To estimate dynamical mass we follow \citet{Taylor10} where
\begin{equation}
	 M_{dyn} = K_V \sigma_0^2 r_e/G 
\end{equation}
with
\begin{equation}
	K_V = \frac{73.32}{10.465 + (n-0.95)^2} + 0.954	
\end{equation}
where $n$ is the sersic $n$ parameter and the gravitational constant $G = 4.3 \times 10^{-3}$ pc $\Msun^{-1}$ km$^2$ s$^{-2}$.

Finally we will make use of both galaxy color and morphology to further parameterize our galaxy samples. Throughout we use $g-r$ colors from the K-corrected NYU VAGC absolute magnitudes. As already stated these are corrected to $z=0.1$; although we will refer to them as $g-r$ colors they are not quite $g-r$ at rest. The morphological classifications we use come from Galaxy Zoo \citep{Lintott11} which provides multiple visual classifications for each galaxy in the SDSS spectroscopic sample. The parameter we use is the probability that a galaxy is an elliptical ($P_{el}$) corrected for redshift bias whereby higher redshift galaxies are more likely to be classified as ellipticals due to their smaller apparent sizes \citep[see ][ for details]{Lintott11}.
 
\begin{figure}

\vspace{11cm}
\includegraphics{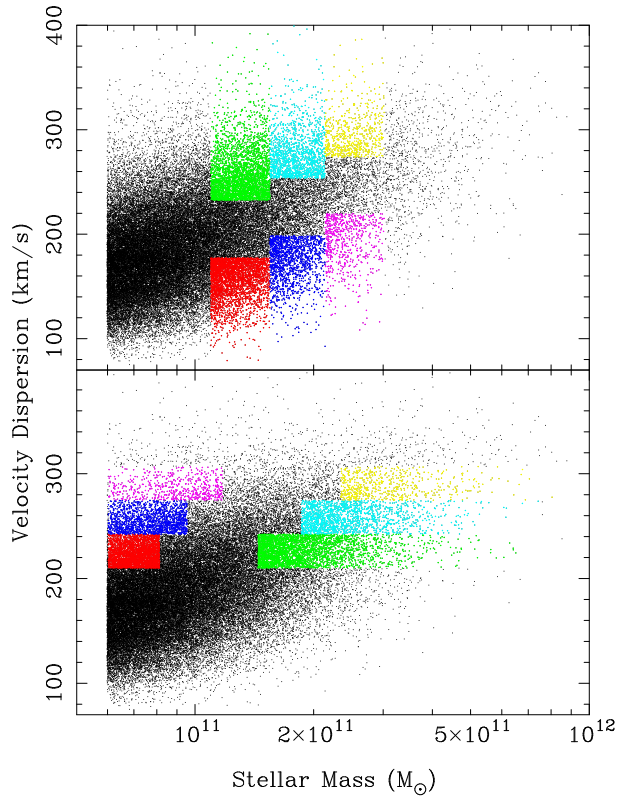}
%%\special{psfile=2ptcross_wp_ratio_SMatvd.ps hoffset=250 voffset=0  hscale=40 vscale=40 angle=0}
\caption{\small The distribution of galaxies from the parent sample with \Mstar~$> 6\times10^{10}\Msun$ in the \Mstar~-- \Vdisp~plane. The colored regions in the top panel show the high and low \Vdisp~samples at fixed \Mstar~and in the bottom panel they show the high and low \Mstar~samples at fixed \Vdisp.
\label{fig:VdSM}}
\end{figure}

\begin{figure*}

\vspace{11cm}
\includegraphics{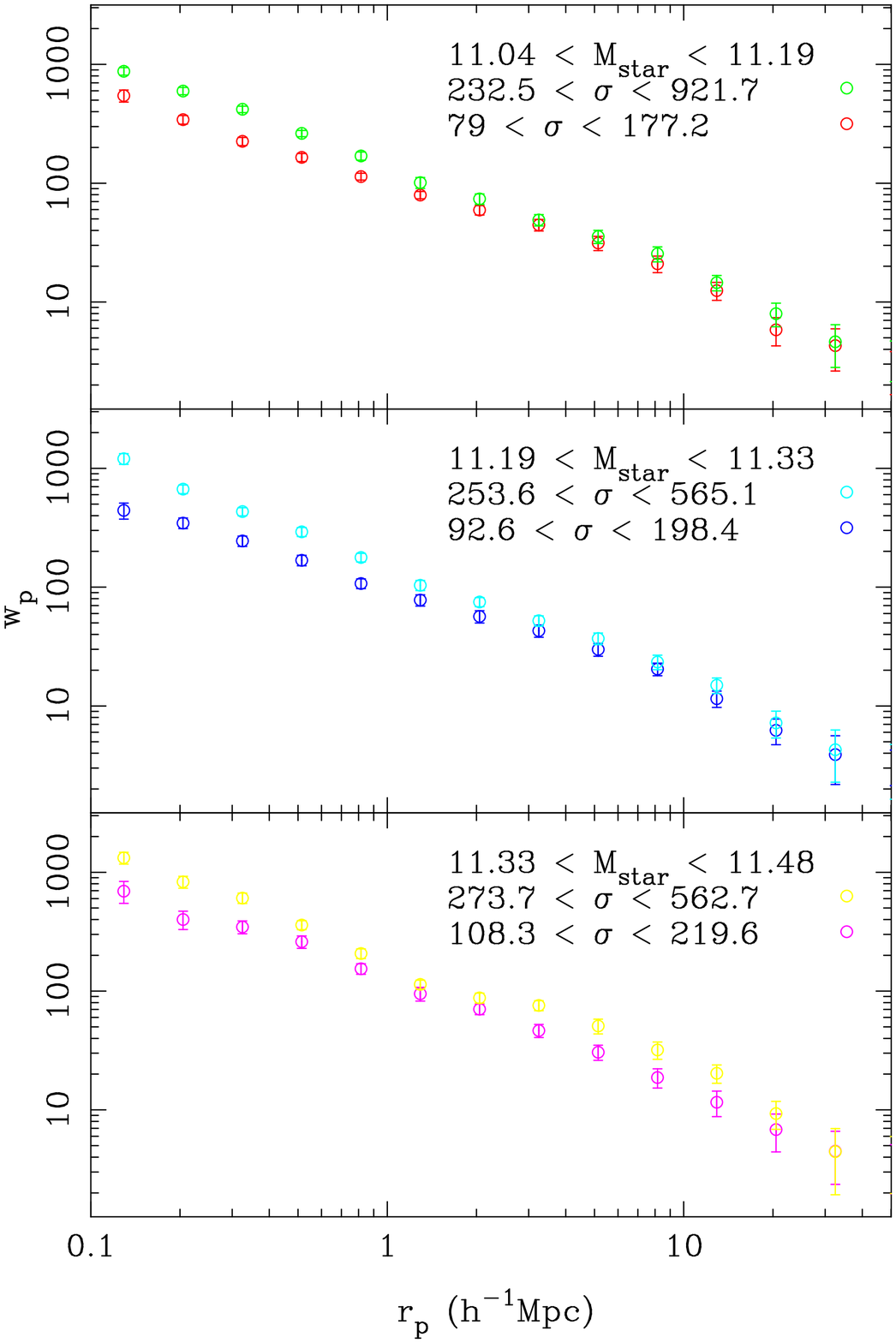}
\includegraphics{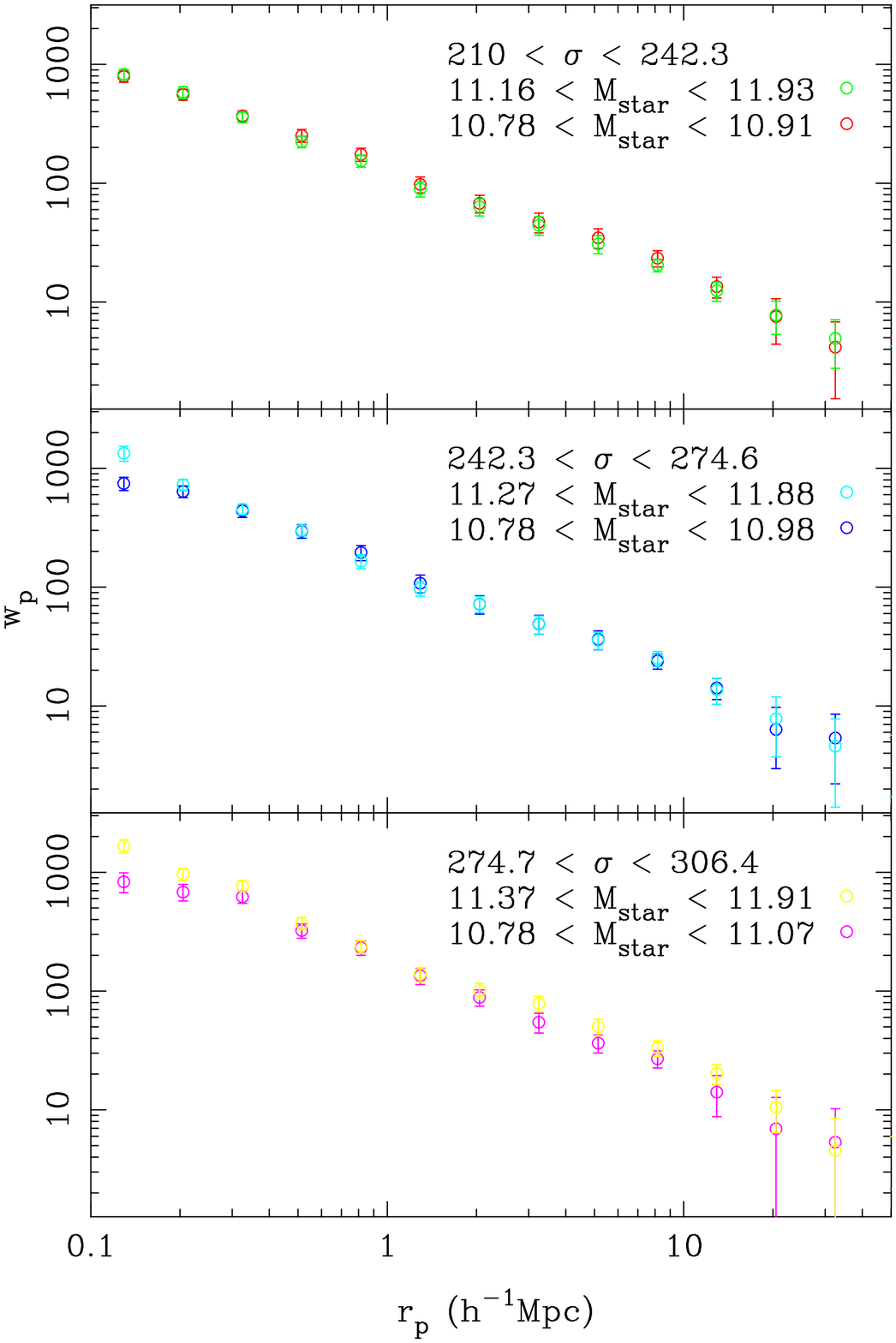}
\caption{\small The projected correlation functions of high and low \Vdisp~samples at fixed \Mstar~(left).  The projected correlation functions of high and low \Mstar~at fixed \Vdisp~(right). The correlation function measurements are colored to match the selection regions shown in Figure \ref{fig:VdSM}.
\label{fig:2ptVdSM1}}
\end{figure*}

\begin{figure*}

\vspace{11cm}
\includegraphics{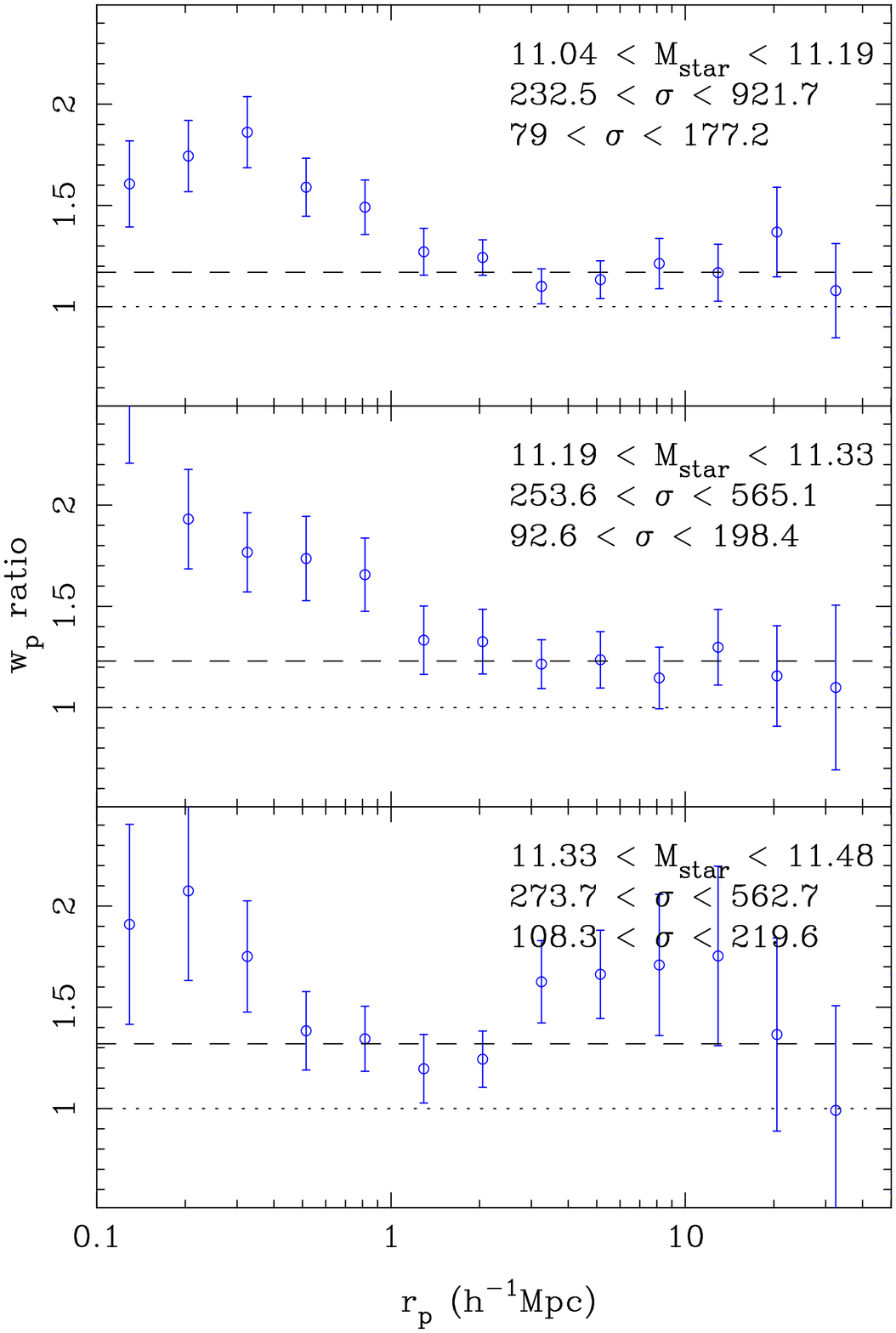}
\includegraphics{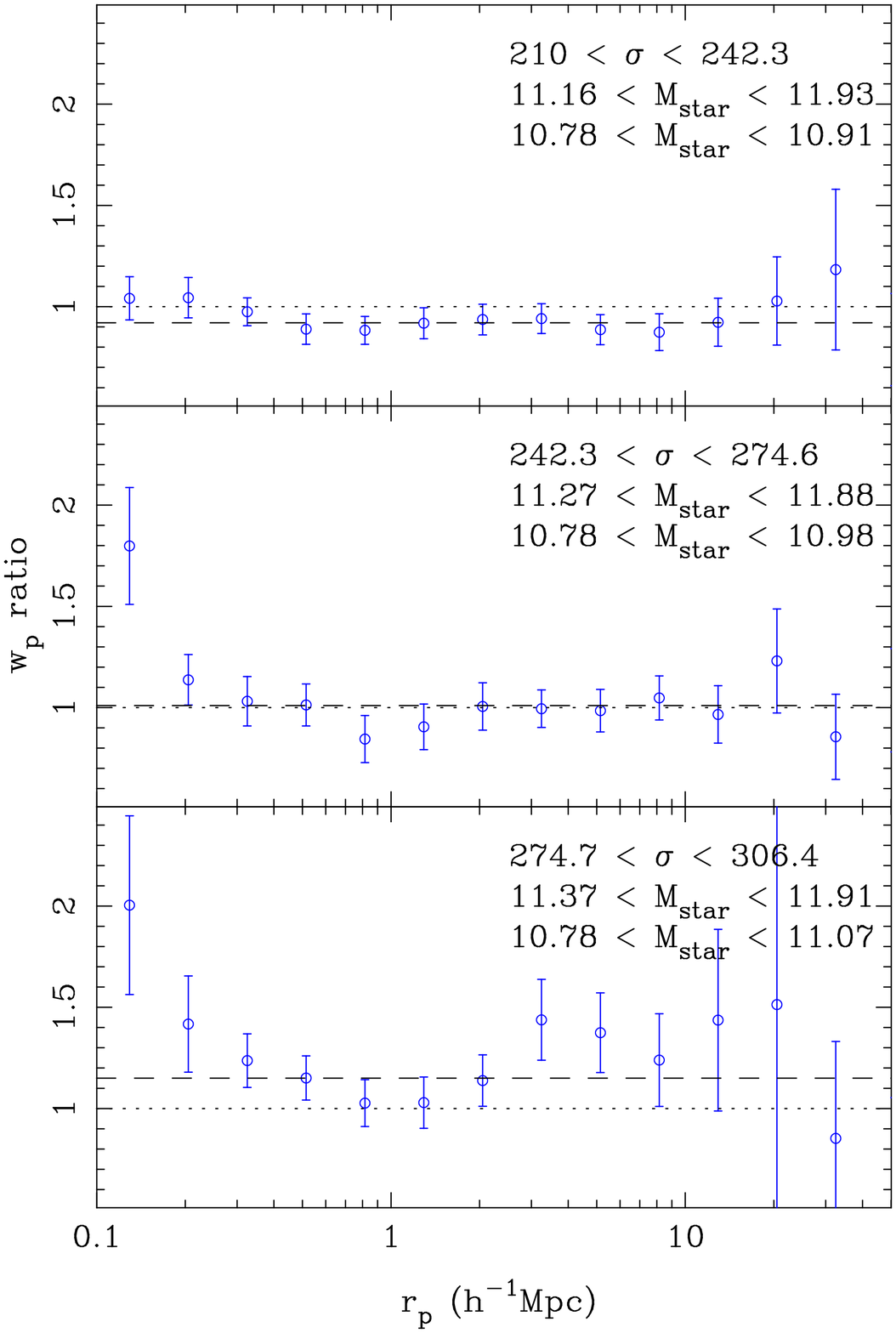}
\caption{\small The ratio of the projected correlation functions between high and low velocity dispersion samples at fixed stellar mass (left).  The ratio of the projected correlation functions between high and low stellar mass samples at fixed velocity dispersion (right). The best fit ratio on scales $1.6 < r_p < 25.1$ h$^{-1}$Mpc is shown as the dashed line.
\label{fig:2ptVdSM}}
\end{figure*}

\begin{figure*}

\vspace{11cm}
\includegraphics{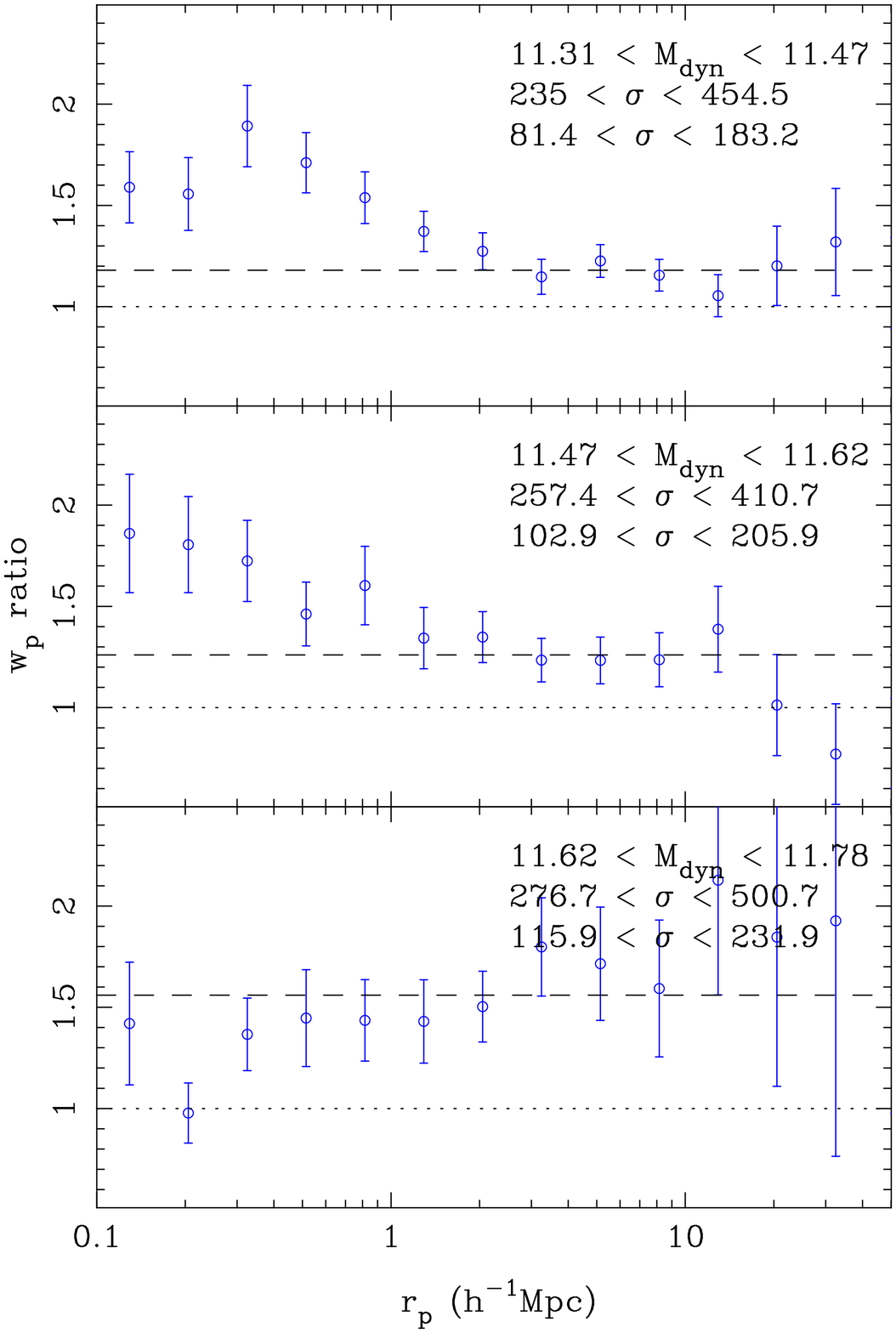}
\includegraphics{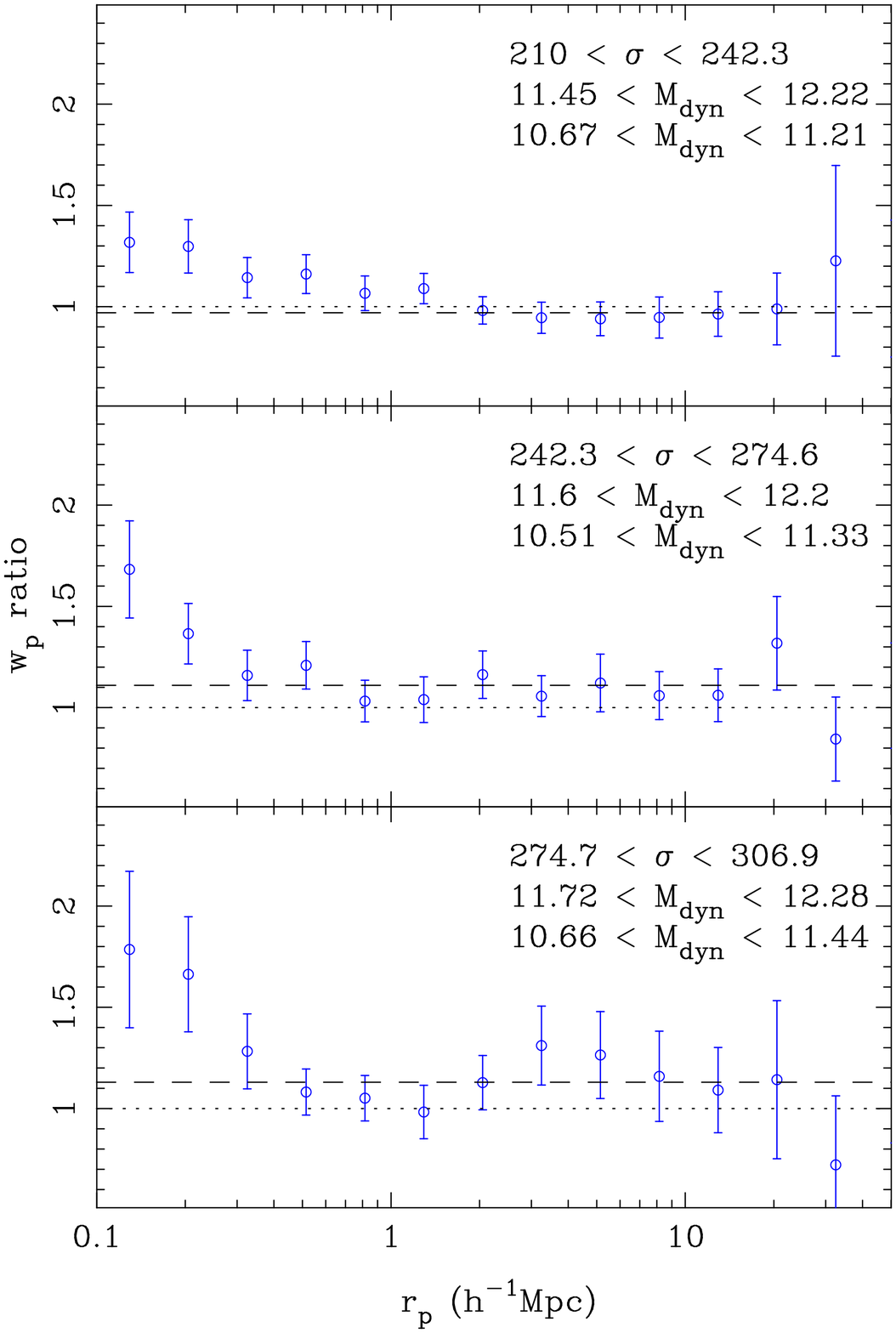}
\caption{\small The ratio of the projected correlation functions between high and low velocity dispersion samples at fixed dynamical mass (left).  The ratio of the projected correlation functions between high and low dynamical mass samples at fixed velocity dispersion (right). The best fit ratio on scales $1.6 < r_p < 25.1$ h$^{-1}$Mpc is shown as the dashed line.
\label{fig:2ptVddM}}
\end{figure*}

\begin{figure*}

\vspace{14.7cm}
\includegraphics{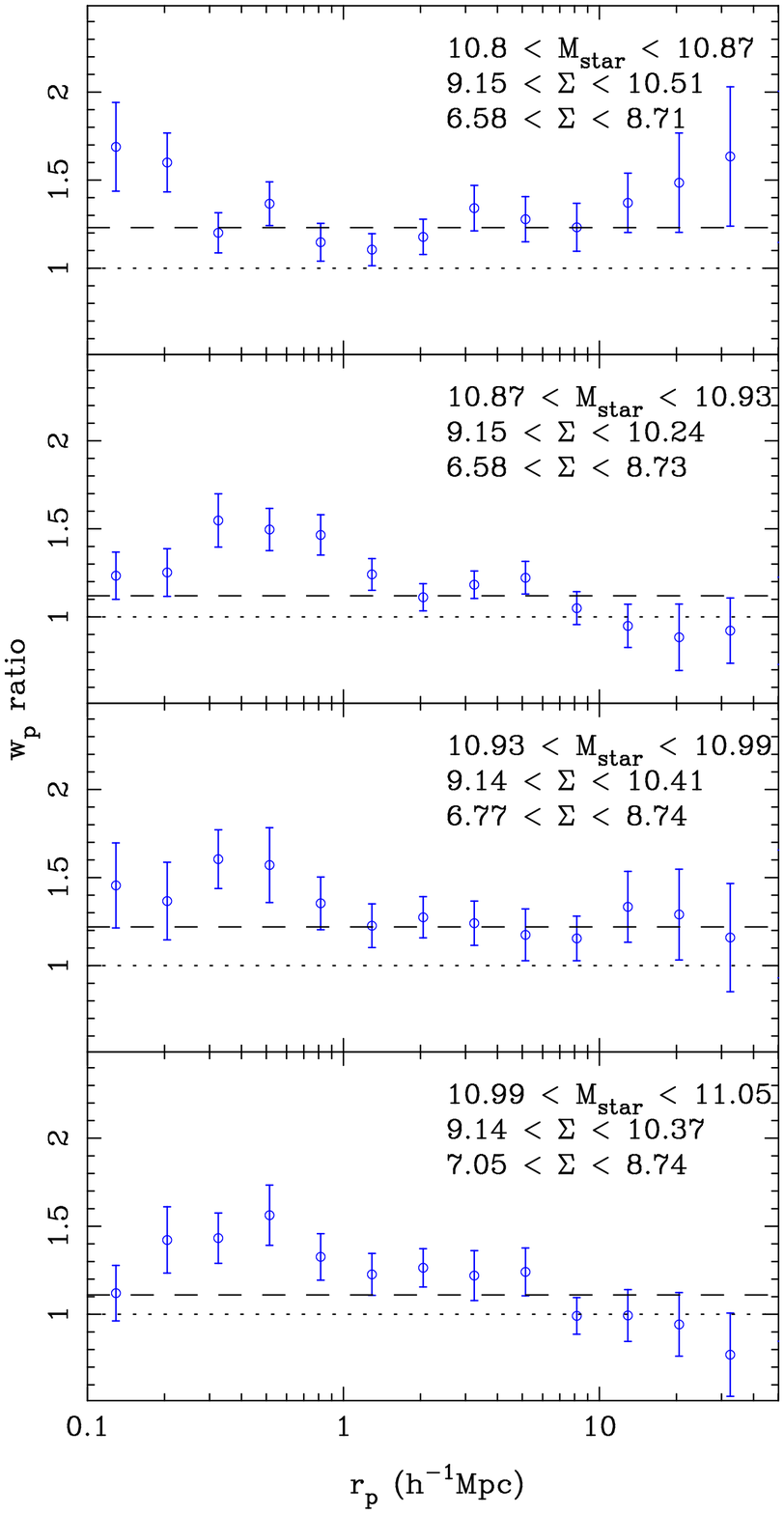}
\includegraphics{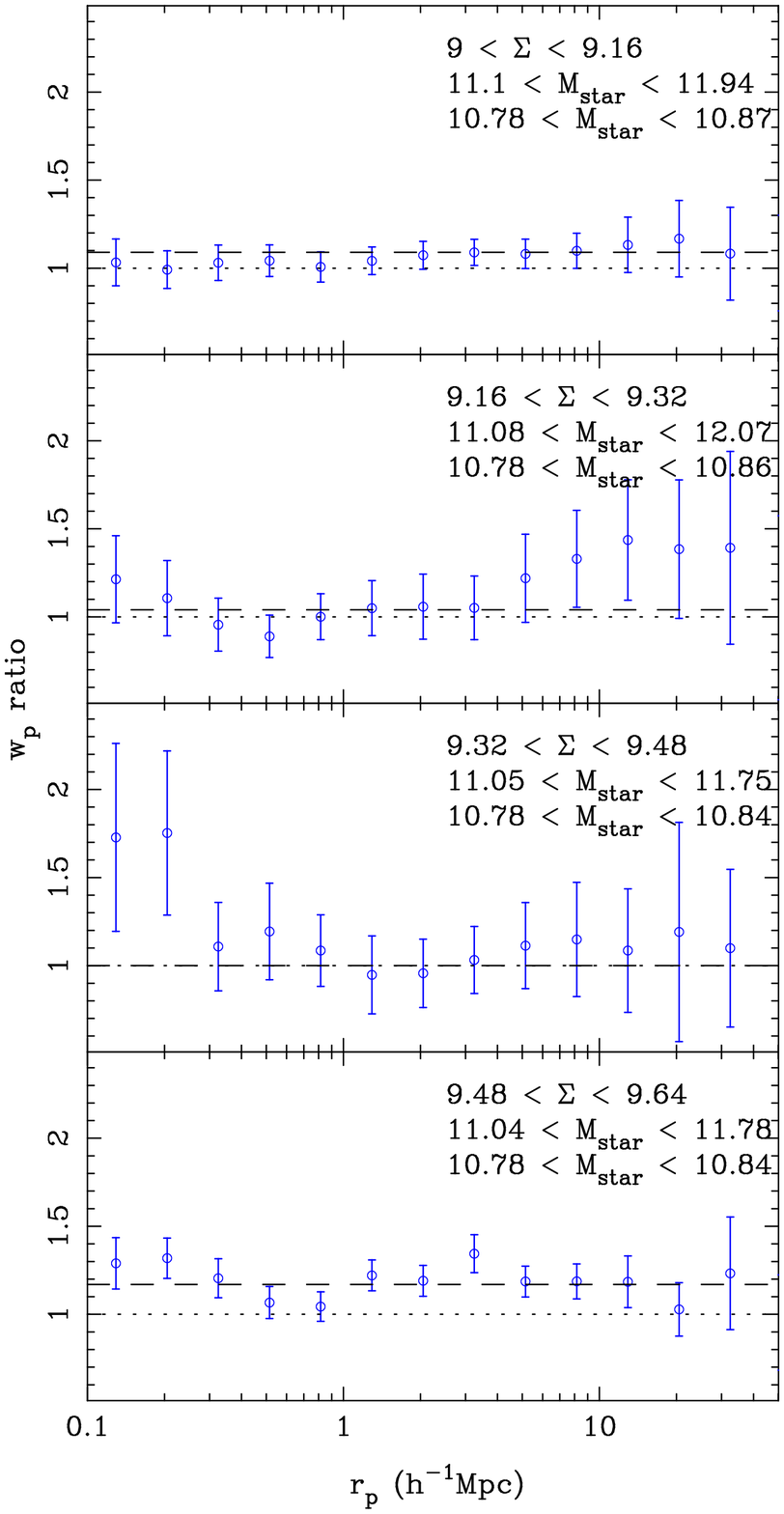}
\caption{\small The ratio of the projected correlation functions between high and low surface mass density samples at fixed stellar mass (left) and high and low stellar mass samples at fixed surface mass density (right). The best fit ratio on scales $1.6 < r_p < 25.1$ h$^{-1}$Mpc is shown as the dashed line.
\label{fig:2ptMdenSM}}
\end{figure*}

\begin{figure*}

\vspace{18cm}
\includegraphics{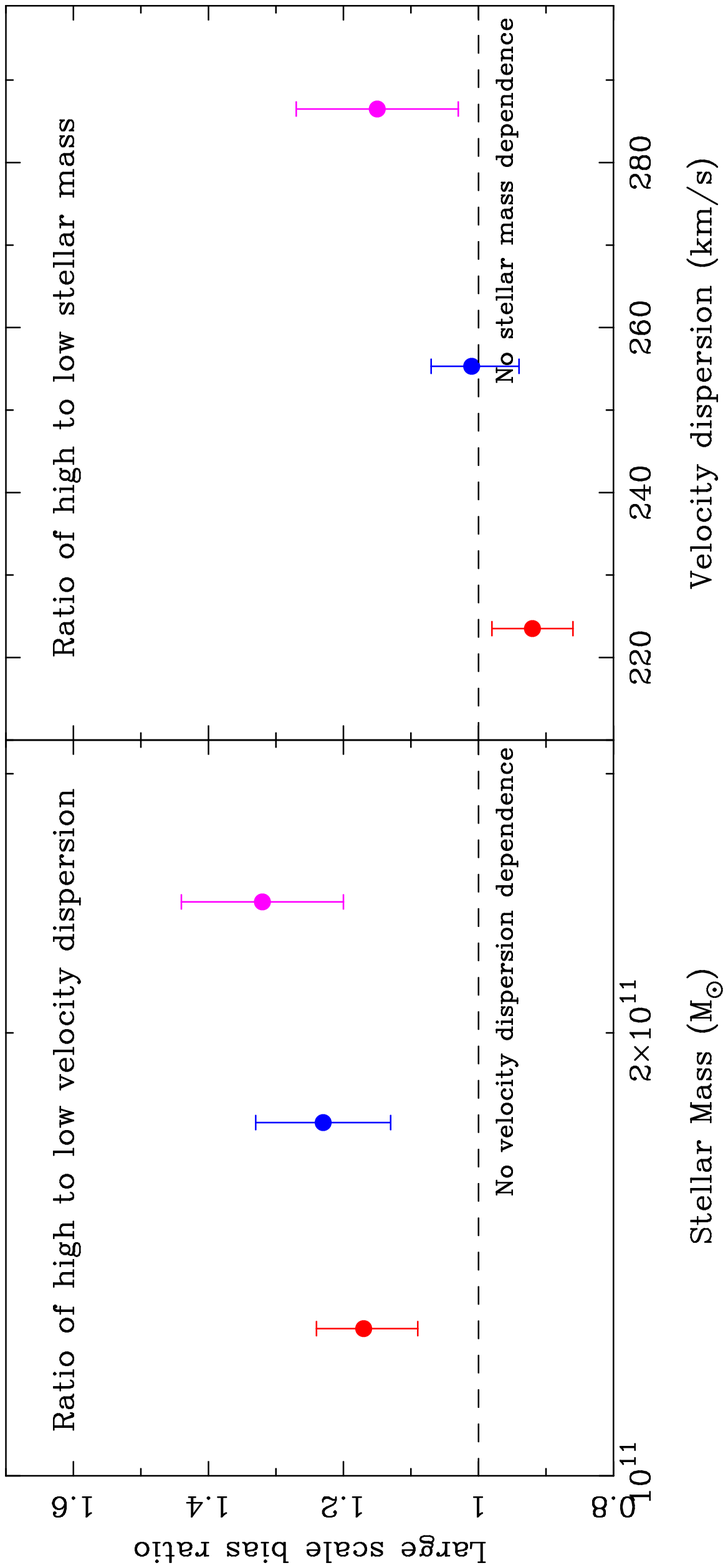}
\includegraphics{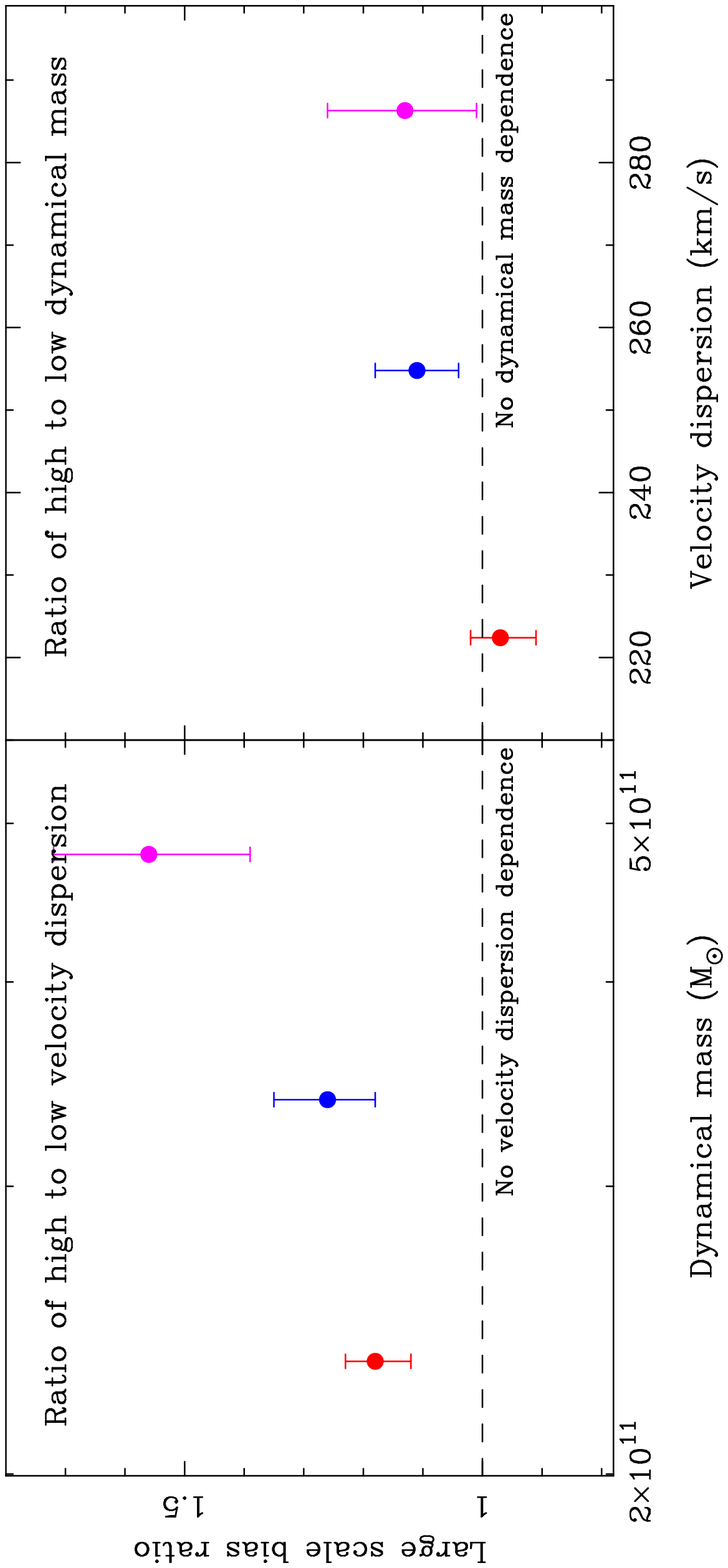}
\includegraphics{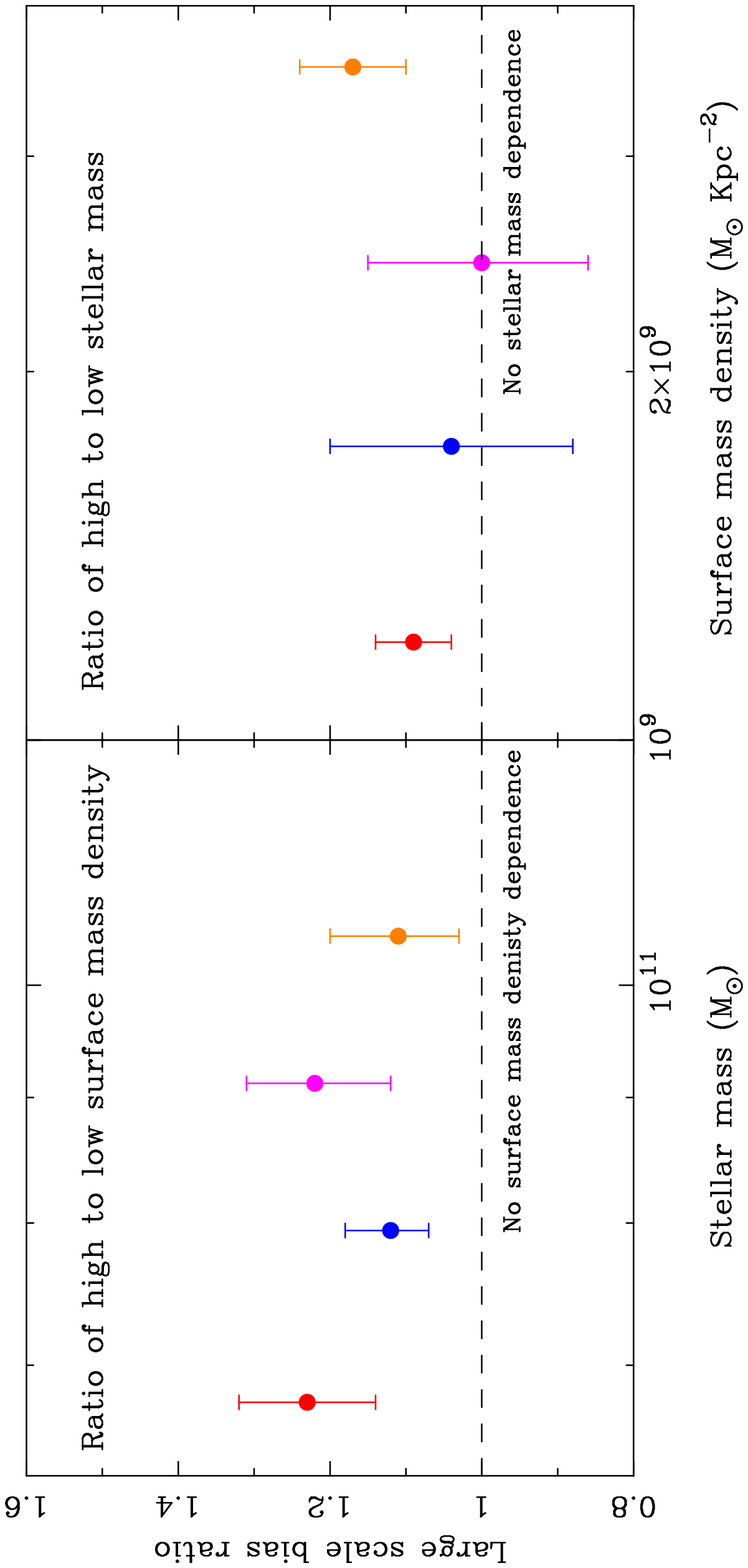}
\caption{\small The best fit large scale bias ratios for the samples plotted in Figures \ref{fig:2ptVdSM}, \ref{fig:2ptVddM} and  \ref{fig:2ptMdenSM}. The top panels show the bias ratio for high and low \Mstar~at fixed \Vdisp~and high and low \Vdisp~at fixed \Mstar. The middle panels show the bias ratio for high and low \Mdyn~at fixed \Vdisp~and high and low \Vdisp~at fixed \Mdyn. The bottom panels show the bias ratio for high and low \Mden~at fixed \Vdisp~and high and low \Vdisp~at fixed \Mden. There is a significant dependence of the bias on \Vdisp~or \Mden~at fixed mass, whereas there is little or no dependence on mass at fixed \Vdisp~or \Mden.
\label{fig:biasrat}}
\end{figure*}

\begin{table*}
  \begin{center}
    \caption{\label{tab:ratfit} Fits to correlation funtion ratios}
	\begin{tabular}{lcccc}
	\tableline
      	\multicolumn{1}{l}{Sample} &
      	\multicolumn{1}{c}{Best fit ratio} &
      	\multicolumn{1}{c}{Best fit $\chi^2$} &
      	\multicolumn{1}{c}{Ratio = 1 $\chi^2$} &
      	\multicolumn{1}{c}{Ratio = 1 prob} \\
	\tableline
\\
\Mstar~at \Vdisp~1 & 0.92 $\pm$ 0.06 & 1.69 & 3.26 & 0.661\\
\Mstar~at \Vdisp~2 & 1.01 $\pm$ 0.07 & 2.07 & 2.07 & 0.839\\
\Mstar~at \Vdisp~3 & 1.15 $\pm$ 0.12 & 5.52 & 6.97 & 0.223\\
\Mstar~at \Vdisp~All & - & - & 12.30 & 0.66\\\\
\Vdisp~at \Mstar~1 & 1.17 $\pm$ 0.07 & 7.51 & 12.10 & 0.033\\
\Vdisp~at \Mstar~2 & 1.23 $\pm$ 0.10 & 2.24 & 7.31 & 0.198\\
\Vdisp~at \Mstar~3 & 1.32 $\pm$ 0.12 & 4.88 & 11.54 & 0.042\\
\Vdisp~at \Mstar~All & - & - & 30.95 & 0.0089\\\\
\Mdyn~at \Vdisp~1 & 0.97 $\pm$ 0.05 & 0.40 & 0.74 & 0.981\\
\Mdyn~at \Vdisp~2 & 1.11 $\pm$ 0.07 & 3.15 & 5.27 & 0.384\\
\Mdyn~at \Vdisp~3 & 1.13 $\pm$ 0.12 & 2.13 & 3.18 & 0.673\\
\Mdyn~at \Vdisp~All & - & - & 9.18 & 0.87\\\\
\Vdisp~at \Mdyn~1 & 1.18 $\pm$ 0.05 & 6.81 & 14.87 & 0.011\\
\Vdisp~at \Mdyn~2 & 1.26 $\pm$ 0.09 & 4.62 & 12.80 & 0.025\\
\Vdisp~at \Mdyn~3 & 1.56 $\pm$ 0.16 & 3.06 & 13.89 & 0.016\\
\Vdisp~at \Mdyn~All & - & - & 41.57 & 0.00026\\\\
\Mstar~at \Mden~1 & 1.09 $\pm$ 0.05 & 0.22 & 2.60 & 0.762\\
\Mstar~at \Mden~2 & 1.04 $\pm$ 0.16 & 2.98 & 3.05 & 0.692\\
\Mstar~at \Mden~3 & 1.00 $\pm$ 0.14 & 0.46 & 0.46 & 0.994\\
\Mstar~at \Mden~4 & 1.17 $\pm$ 0.07 & 5.88 & 10.73 & 0.057\\
\Mstar~at \Mden~All & - & - & 16.84 & 0.66\\\\
\Mden~at \Mstar~1 & 1.23 $\pm$ 0.09 & 4.24 & 10.31 & 0.067\\
\Mden~at \Mstar~2 & 1.12 $\pm$ 0.05 & 6.55 & 10.69 & 0.058\\
\Mden~at \Mstar~3 & 1.22 $\pm$ 0.09 & 1.96 & 6.54 & 0.257\\
\Mden~at \Mstar~4 & 1.11 $\pm$ 0.08 & 9.47 & 11.23 & 0.047\\
\Mden~at \Mstar~All & - & - & 38.77 & 0.0071\\\\

	\tableline
    \end{tabular}
%%\tablecomments{}
\end{center}
\end{table*}

\section{Samples}
\label{sec:samples}

We define a parent sample with stellar mass $> 6 \times 10^{10}\Msun$ and 0.04 $< z <$ 0.113. The stellar mass limit and redshift cuts are chosen to yield the largest stellar mass limited sample of galaxies that can be defined from the NYU SDSS VAGC galaxy catalog, thus maximizing the number of galaxies available for clustering measurements. This relatively high stellar mass cut is also helpful as the SDSS velocity dispersions are only reliable at $>$ 75 km/s and thus become incomplete in \Vdisp~at low stellar masses. By inspecting the distribution of velocity dispersion in narrow stellar mass bins we estimate that 98\% of galaxies with stellar mass $> 6 \times 10^{10}\Msun$ have velocity dispersion in excess of 75 km/s. Finally we remove galaxies with unreliable \Vdisp~measurements by requiring that the error in \Vdisp~be less than 10\%.

Since there is a significant scatter in the relation between stellar mass and velocity dispersion we must make further cuts if we wish to define samples that are complete in both. To define the velocity dispersion completeness limit  
we investigate the distribution of stellar mass in narrow velocity dispersion bins for the full DR7 sample. Using these measurements we find the velocity dispersion that defines a complete sample of galaxies at the stellar mass of $6 \times 10^{10} \Msun$. We find that for galaxies with a velocity dispersion of 210 km/s 85\% have stellar masses greater than $6 \times 10^{10}\Msun$ and thus we restrict our sample to have \Vdisp~grater than this limit when defining \Vdisp~limited samples. 

The aim of this work is to investigate which is more fundamental in determining the clustering amplitude, a galaxy's mass, either stellar or dynamical, or its central velocity dispersion. We approach this question by measuring the clustering where we have fixed the mass and allowed \Vdisp~to vary and visa-versa.
We define a series of samples with a narrow range in one parameter and then take the upper and lower quartile of the second parameter. We begin by defining the first velocity dispersion range, starting at the velocity dispersion completeness limit (210 km/s) and cutting at a dispersion of 242.3 km/s, the limit that contains 20\% of the galaxies with \Vdisp~larger than 210 km/s. We then split that sample by either stellar mass or dynamical mass into upper and lower quartiles. In a similar manner we construct two more samples with the same interval in \Vdisp~with successively higher \Vdisp~limits again splitting into highest and lowest quartiles in either stellar mass or dynamical mass. Details of these samples are given in Tables \ref{tab:SMatVd} and \ref{tab:dMatVd} in Appendix \ref{sec:appsamp}. To make the reciprocal samples, i.e. highest and lowest quartiles in \Vdisp~at fixed stellar mass or dynamical mass, we find the appropriate starting minimum mass and interval that produces three samples with the same number of galaxies as in the \Vdisp~ range samples. We then spit these into the highest and lowest quartiles of \Vdisp. Details of these samples are given in Tables \ref{tab:VdatSM} and \ref{tab:VdatdM}. Figure 1 shows the distribution of galaxies in the \Vdisp~-\Mstar~plane for the parent sample as well as the samples at fixed \Vdisp~and \Mstar. 

We also define a series of samples in narrow stellar mass ranges where we take the highest and lowest quartiles in stellar surface mass density (\Mden) and their reciprocals (Tables \ref{tab:MdenatSM} and \ref{tab:SMatMden}) again ensuring that they are complete in both stellar mass and surface mass density. This is an important consistency check for our measurements. It is possible that the scatter introduced by larger errors on \Mstar~could reduce any differences in relative clustering amplitude compared to \Vdisp, something which should be reduced by using \Mdyn. Whilst any errors on \Mdyn~will be different and likely to be smaller than those on \Mstar~they will still be larger than the error on \Vdisp~due to the additional uncertainties in fitting the profile to determine \Re~and the sersic-n parameter. However, \Mden, which should show similar trends to \Vdisp, contains both the errors on \Mstar~and on \Re. Therefore, any reduction in clustering trends due to scatter in the measurement of the physical parameters should be largest for this sample.  

For each pair of samples (e.g. high and low stellar mass at fixed dispersion) we find that the mean of the fixed parameter changes very little between the samples split into high and low quartiles. For a fair comparison we compare the mass ratio to the square of the \Vdisp~ratio since mass is proportional to $\sigma^2$. For all samples the ratio of the mean of the fixed parameter varies by less than 5\% and so should have a negligible effect on the clustering amplitude. Conversely the ratio of the means of the varying parameter covers a factor of 2 to 3.5.

In the Appendix \ref{sec:appsamp} we plot redshift, velocity, mass, mass density and morphology distributions for each sample where appropriate. As expected, since we have defined a volume limited parent sample and have only made cuts where we are complete in \Vdisp, $M_{den}$ and mass, the redshift distributions for all samples are the same.   
\\

\section{Clustering measurement techniques}
\label{sec:clus}

The two-point correlation function, $\xi(r)$, is defined as the excess probability above Poisson of finding an object at a physical separation $r$ from another object. This is calculated by comparing the number of pairs as a function of $r$ in our galaxy catalogs with the number in a random catalog that covers the same volume as our data.

Since the space density of galaxies in our subsamples is very low shot noise would be a significant source of error for measurements of their auto-correlation functions. However, we can overcome this problem by using a much denser sample of galaxies to trace the underlying density field and measure the clustering of our sub-samples relative to this larger sample using the cross-correlation function.
The obvious choice for the larger sample is the parent sample we defined above containing all galaxies passing our basic selection criteria. This sample contains almost 65,000 galaxies and so has a space density almost 20 times higher than our largest sub-sample. 
Whilst the individual cross-correlation functions will reflect the intrinsic clustering amplitude of both the sub- and parent samples the ratio between any two of these cross correlations will just reflect the ratio between the clustering of the two sub-samples in question, with the clustering of the parent sample in effect being 'canceled out'.  

Measuring the line-of-sight distance using redshifts introduces distortions into the correlation function. The effect of these distortions can be overcome by separating the clustering signal into contributions perpendicular ($r_p$) and parallel ($\pi$) to the line-of-sight ($\xi(r_p,\pi)$). One can then integrate over the line of sight direction to estimate the projected correlation function 
\begin{equation}
	\label{eq:wprp}
	w_p(r_p) = 2\int^{\infty}_0 d\pi\,\xi(r_p,\pi)
	        = 2\int^{\infty}_{r_p} \frac{r\,dr\,\xi(r)}{(r^2-r_p^2)^{1/2}}.
\end{equation}
The final expression only involves the real-space correlation function $\xi(r)$ showing that $w_p(r_p)$ is not compromised by redshift space distortions \citep{Davis83}. In practice it is only possible to integrate out to some maximum $\pi$ because $\xi(r_p,\pi)$ is poorly known on very large scales resulting in additional noise being introduced to the measurement. We integrate to 60$h^{-1}$Mpc which is sufficiently large to include most correlated pairs and gives stable results. 

We make this measurement using the \citet{Landy93} estimator in the cross-correlation form, with
\begin{displaymath}
  \xi(r_p,\pi)  = \hspace{0.77\columnwidth}
\end{displaymath}
\begin{equation}
  \frac{DD1}{RR}\left(\frac{n_R^2}{n_D n_{D1}}\right) - \frac{DR}{RR}\left(\frac{n_R}{n_D}\right)  - \frac{D1R}{RR}\left(\frac{n_R}{n_{D1}}\right) + 1
\end{equation}
  where DD1 are the pair counts between the parent and sub-sample, DR are the pair counts between parent and random sample, D1R are the pair counts between sub-sample and random sample and RR are the pair counts between the random sample, all split into the $r_p$ and $\pi$ directions. We are able to use the same random sample in each term as all samples are complete throughout the same volume, i.e. they cover the same spatial extent and have the same redshift distributions. We calculate the pair counts in a grid which is logarithmically spaced in $r_p$ of width 0.2 log(\hMpc) and linearly spaced in the $\pi$ direction of width 2\hMpc~with all separations in co-moving coordinates.
The random catalogue is based on the one provided by the NYU VAGC with additional cuts to match the exact spatial coverage of the MPA/JHU stellar masses and redshifts matching the redshift distribution of our samples. The resulting random catalogue contains almost 1.9 million randoms, close to 29 times as many as in our parent galaxy catalog, which is a sufficient number such that our errors are never dominated by shot noise in the random pair counts.

Since the individual bins in a correlation function measurement are highly correlated we need to generate accurate covariance matrices if we wish to compare our clustering measurements in a meaningful way. The simplest approach and the one we chose to follow is to use jackknife resampling \citep{Scranton02}. There is some debate about the accuracy of covariance matrices generated in this way; for instance \citet{Norberg09} suggest that the structure of jackknife covariance matrices may not be entirely accurate for \wp~ measurements, although the variance is reliable. Conversely \citet{Zehavi02, Zehavi04, Zehavi05} find that they can closely reproduce the structure and amplitude of covariance matrices generated by mock catalogs using jackknife resampling. Because of this uncertainty, it might be preferable to generate covariance matrices from large numbers of mock galaxy catalogues generated by populating dark matter halo catalogs so as to reproduce the clustering and density of each sample. However, since we have so many sub-samples, each with a complex selection, it would be enormously challenging to undertake such a scheme. We have therefore chosen to use jackknife resampling which should be sufficient for our purposes. We split the SDSS area into 146 equal area regions and then repeatedly calculate \wp~removing one area at a time. These 146 \wp~measurements are then used to generate a full covariance matrix.

\section{Results}

\begin{figure}

\vspace{8.1cm}
\includegraphics{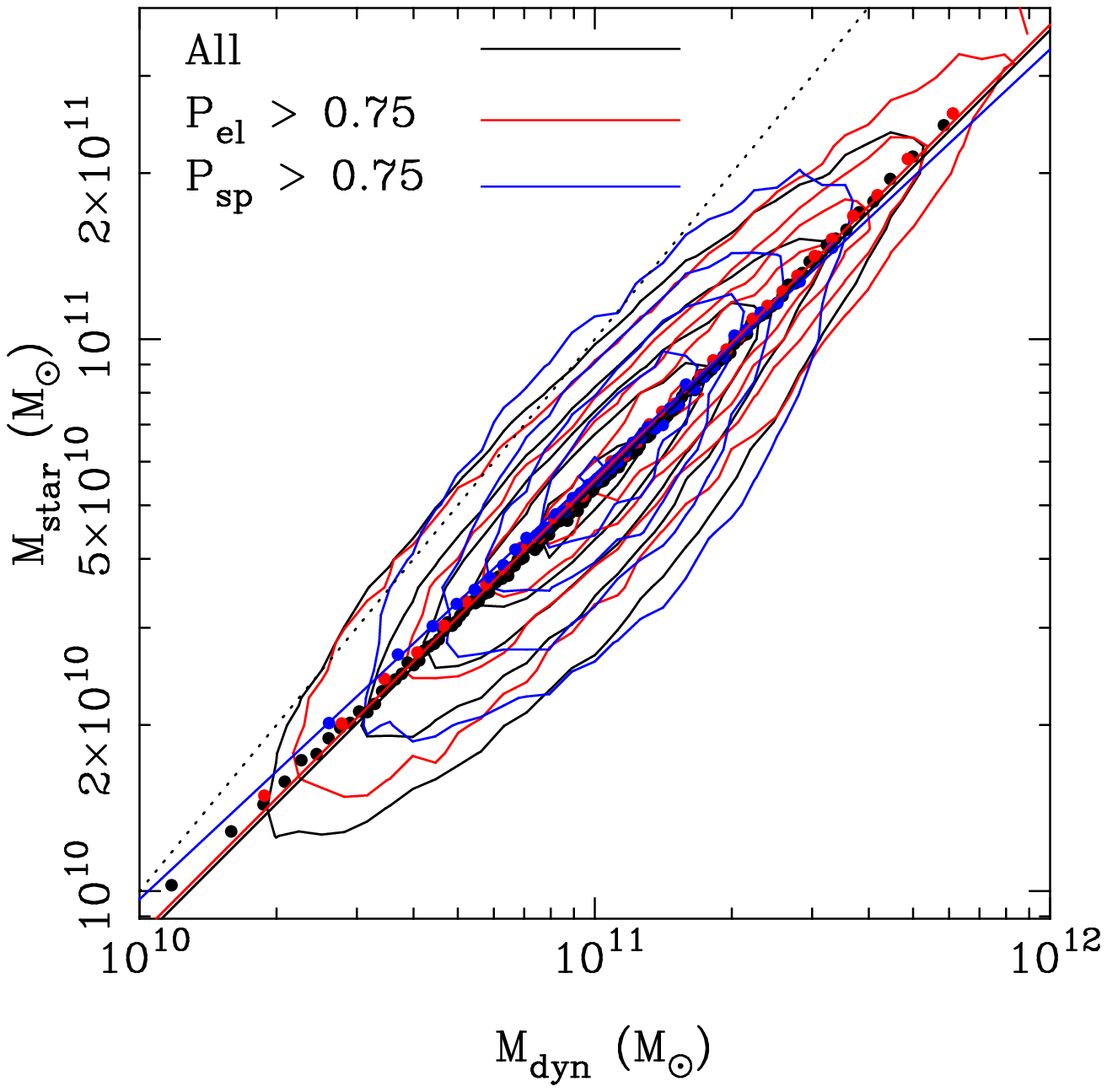}
\caption{\small The relationship between dynamical and stellar mass for all DR7 galaxies with 0.04 $< z <$ 0.113 and \Vdisp~$>$ 75 km/s (black contours) and those galaxies classified as either ellipticals (red contours) or spirals (blue contours). The points show the median \Mstar~in bins of \Mdyn~and the solid lines the biweight best fit. The dotted line shows the one-to-one relation. This relationship is almost identical independent of morphology over the whole mass range covered by this plot, and becomes indistinguishable for the mass range of interested in this paper. 
\label{fig:MstarMdyn}}
\end{figure}

\subsection{Is galaxy mass or \Vdisp~more closely related to clustering amplitude?}

Figure \ref{fig:2ptVdSM1} shows the \wp~measurements for the high and low \Vdisp~samples at fixed \Mstar~(left) and the high and low \Mstar~samples at fixed \Vdisp~(right). We find that when the mass is held fixed the high \Vdisp~samples have a higher clustering amplitude than the low \Vdisp~samples. This is true on all scales and for all mass ranges. However, when \Vdisp~is held fixed there appears to be little or no dependence of the clustering amplitude on \Mstar. This is the central result of this paper: clustering amplitude depends on \Vdisp~at fixed mass, but does not depend on mass at fixed \Vdisp. 

This result is illustrated more clearly in Figure \ref{fig:2ptVdSM} which shows the ratio of \wp~between the high and low \Vdisp~samples at fixed \Mstar~(left) and the high and low \Mstar~samples at fixed \Vdisp~(right). Figure \ref{fig:2ptVddM} is similar to Figure \ref{fig:2ptVdSM}, except \Mstar~is replaced by \Mdyn~and shows exactly the same trends as with \Mstar~. We also show in Figure \ref{fig:2ptMdenSM} the ratio of \wp~ between high and low \Mden~samples at fixed \Mstar~and high and low \Mstar~samples at fixed \Mden. This is, in effect, the same as splitting by size at fixed mass or mass at fixed size. Since \Mdyn~and \Mstar~are highly correlated one would expect galaxies with smaller \Re~at a fixed \Mstar, thus higher \Mden, to have a higher \Vdisp. These samples should then be analogous to those with varying \Vdisp~at fixed \Mstar~and indeed they show the same trend of higher clustering amplitude at higher \Mden~when \Mstar~is fixed and very little dependence with \Mstar~when \Mden~is fixed. As discussed in Section \ref{sec:samples} this also implies that these observed trends are not the result of larger measurement errors on \Mstar~or \Mdyn~compared to \Vdisp, which could have reduced the clustering dependence on mass compared to \Vdisp.

To quantify these ratios in \wp, which constitutes a ratio in galaxy bias, we find the best fitting scale independent amplitude on scales $1.6 < r_p < 25.1$ h$^{-1}$Mpc. These scales are sufficiently large as to be dominated by pair counts between DM halos, rather than galaxies within halos, and so should show a scale independence with respect to the dark matter clustering and give an indication of the relative linear bias of the two samples. For the \wp~ratios shown in Figures \ref{fig:2ptVdSM}, \ref{fig:2ptVddM} and \ref{fig:2ptMdenSM} this appears to be the case and indeed a scale independent ratio is an acceptable fit for all but one of the samples. These fits are made using the full covariance matrices for the \wp~ratio, and are shown as the dashed lines in Figures \ref{fig:2ptVdSM}, \ref{fig:2ptVddM} and \ref{fig:2ptMdenSM}. Details of the fits are given in Table \ref{tab:ratfit} and the best fit bias ratios and errors are plotted for all these samples in Figure \ref{fig:biasrat}. In addition to the best fit we also calculate the $\chi^2$ for a \wp~ratio of one over the same scales, and give the $\chi^2$ value as well as the probability of an acceptable fit in Table \ref{tab:ratfit}. In all cases the best fit ratio is higher when \Vdisp~or \Mden~is varied at fixed mass than when \Vdisp~or \Mden~are fixed and the mass allowed to vary. Similarly when \Vdisp~or \Mden~are fixed the \wp~ratios between the high and low mass samples are consistent with one in all but one case whereas the varying \Vdisp~or \Mden~samples at fixed mass are nearly all inconsistent with a \wp~ratio of one. For each parameter pair we can combine the samples and determine the probability that all are consistent with a \wp~ratio of 1. Again we find that at fixed \Vdisp~or \Mden~the samples are indeed consistent with showing no dependence of the clustering amplitude on mass at the 66\%, 87\% and 66\% levels, whereas there is only a 0.9\%, 0.03\% and 0.7\% chance that all three of the varying \Vdisp~or \Mden~samples at fixed mass are consistent with no variation in clustering amplitude on large scales. 

On small scales the differences are just as clear with the varying \Vdisp~or \Mden~samples typically showing even larger \wp~ratios than at large scales. The samples with varying mass at fixed \Vdisp~or \Mden~ occasionally show moderately larger \wp~ratios on small scales than on large scales, but often there is no difference with the \wp~ratio remaining consistent with one on all scales. 

Overall these measurements make it clear that velocity dispersion or stellar mass surface density are more closely related to galaxy bias than either stellar mass or dynamical mass.
\\
\\

\subsection{The Importance of Morphology and Color}

\begin{table*}
  \begin{center}
    \caption{\label{tab:ratfitcolel} Fits to correlation funtion ratios for sample split by morphology and color}
	\begin{tabular}{lcccc}
	\tableline
      	\multicolumn{1}{l}{Sample} &
      	\multicolumn{1}{c}{Best fit ratio} &
      	\multicolumn{1}{c}{Best fit $\chi^2$} &
      	\multicolumn{1}{c}{Ratio = 1 $\chi^2$} &
      	\multicolumn{1}{c}{Ratio = 1 prob} \\
	\tableline
\\
\Pel~at \Vdisp~1 & 0.92 $\pm$ 0.05 & 2.44 & 4.55 & 0.473\\
\Pel~at \Vdisp~2 & 1.06 $\pm$ 0.09 & 4.29 & 4.71 & 0.452\\
\Pel~at \Vdisp~3 & 1.07 $\pm$ 0.10 & 2.95 & 3.41 & 0.638\\
\Pel~at \Vdisp~All & - & - & 12.67 & 0.63\\
\\
\Pel~at \Mdyn~1 & 1.12 $\pm$ 0.05 & 2.60 & 6.16 & 0.291\\
\Pel~at \Mdyn~2 & 1.20 $\pm$ 0.08 & 1.05 & 5.77 & 0.329\\
\Pel~at \Mdyn~3 & 1.10 $\pm$ 0.12 & 1.43 & 2.02 & 0.847\\
\Pel~at \Mdyn~All & - & - & 13.95 & 0.53\\
\\
$g-r$ at \Vdisp~1 & 1.35 $\pm$ 0.11 & 3.35 & 11.81 & 0.037\\
$g-r$ at \Vdisp~2 & 1.36 $\pm$ 0.14 & 4.55 & 11.34 & 0.045\\
$g-r$ at \Vdisp~3 & 1.17 $\pm$ 0.17 & 5.69 & 6.64 & 0.248\\
$g-r$ at \Vdisp~All & - & - & 29.80 & 0.013\\
\\
$g-r$ at \Mdyn~1 & 1.34 $\pm$ 0.10 & 6.66 & 16.34 & 0.006\\
$g-r$ at \Mdyn~2 & 1.50 $\pm$ 0.16 & 2.60 & 12.76 & 0.026\\
$g-r$ at \Mdyn~3 & 1.13 $\pm$ 0.11 & 3.83 & 5.19 & 0.394\\
$g-r$ at \Mdyn~All & - & - & 34.29 & 0.0031\\
\\
\Vdisp~at \Mdyn~El 1 & 1.29 $\pm$ 0.08 & 1.51 & 12.27 & 0.031\\
\Vdisp~at \Mdyn~El 2 & 1.32 $\pm$ 0.10 & 4.73 & 13.46 & 0.019\\
\Vdisp~at \Mdyn~El 3 & 1.29 $\pm$ 0.12 & 2.82 & 7.57 & 0.181\\
\Vdisp~at \Mdyn~El All & - & - & 33.30 & 0.0043\\
\\
\Vdisp~at \Mdyn~red 1 & 1.21 $\pm$ 0.06 & 2.32 & 12.08 & 0.034\\
\Vdisp~at \Mdyn~red 2 & 1.15 $\pm$ 0.08 & 1.08 & 4.49 & 0.482\\
\Vdisp~at \Mdyn~red 3 & 1.48 $\pm$ 0.15 & 1.34 & 10.71 & 0.058\\
\Vdisp~at \Mdyn~red All & - & - & 27.28 & 0.027\\
	\tableline
    \end{tabular}
%%\tablecomments{}
\end{center}
\end{table*}

\begin{figure*}

\vspace{12cm}
\includegraphics{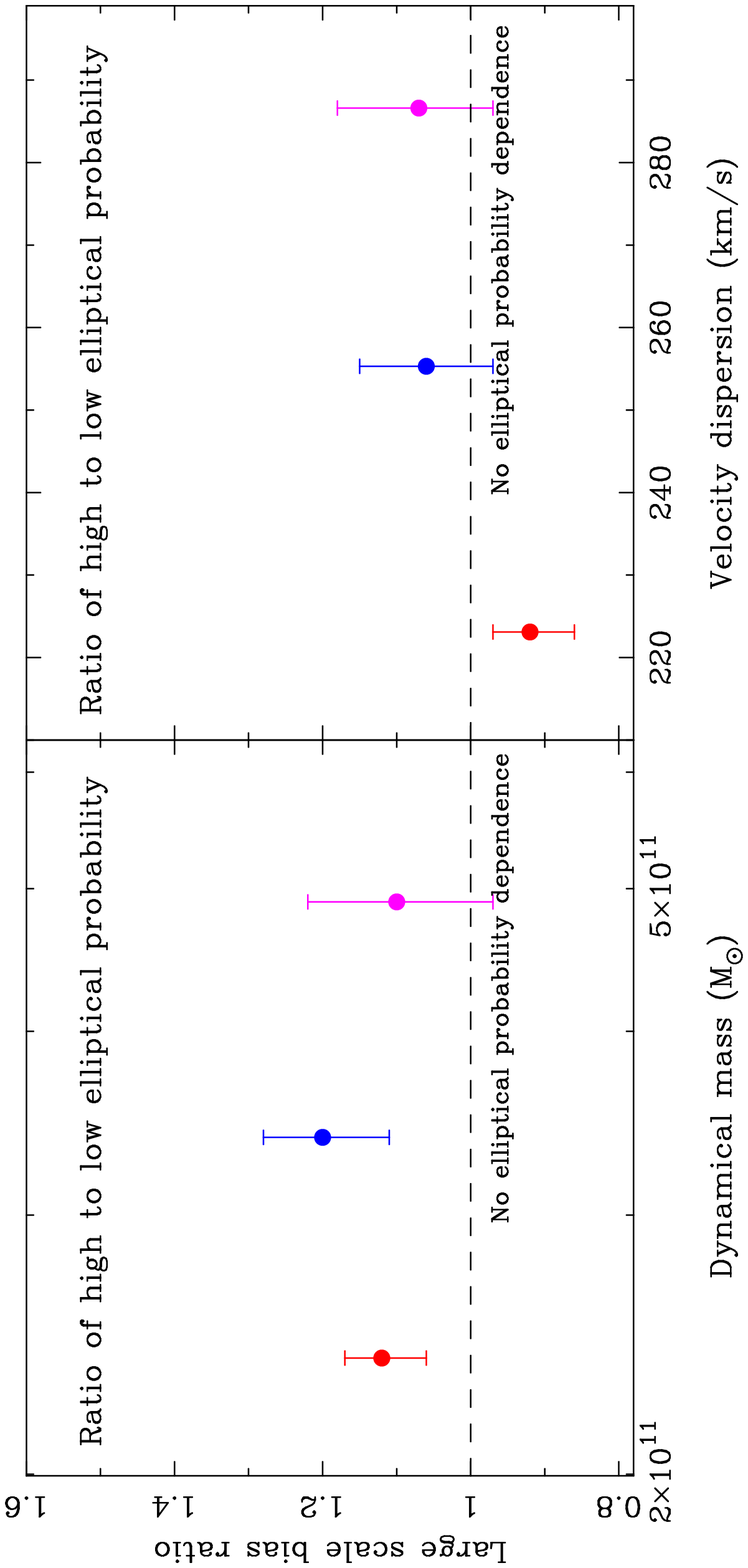}
\includegraphics{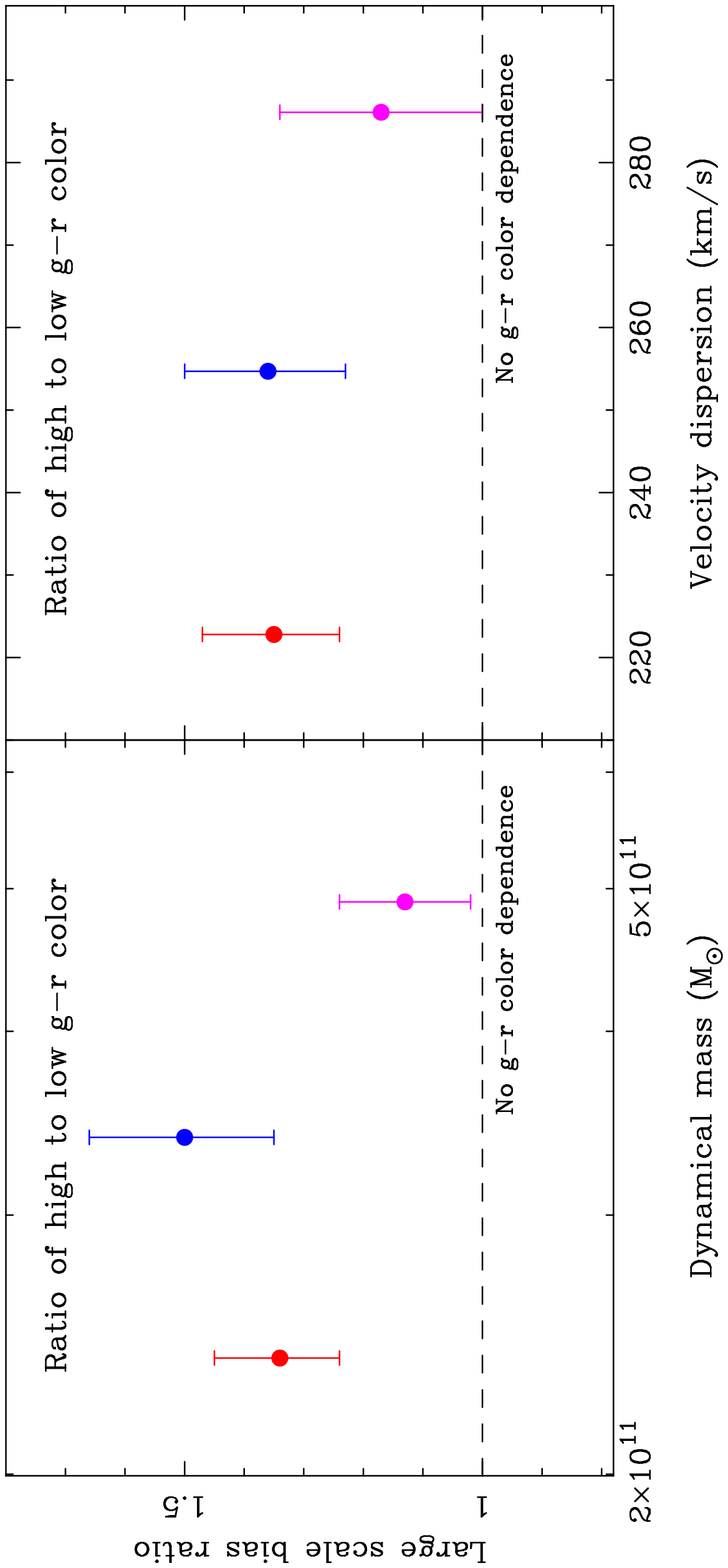}
\caption{\small The dependence of large scale bias on elliptical probability (top) and $g-r$ color (bottom) at fixed \Mdyn~(left)and \Vdisp~(right). The bias does depended on elliptical probability at fixed mass but not at fixed \Vdisp. However, there is a significant dependence of the bias on color at both fixed mass and \Vdisp.
\label{fig:biasrat2}}
\end{figure*}

\begin{figure*}

\vspace{12cm}
\includegraphics{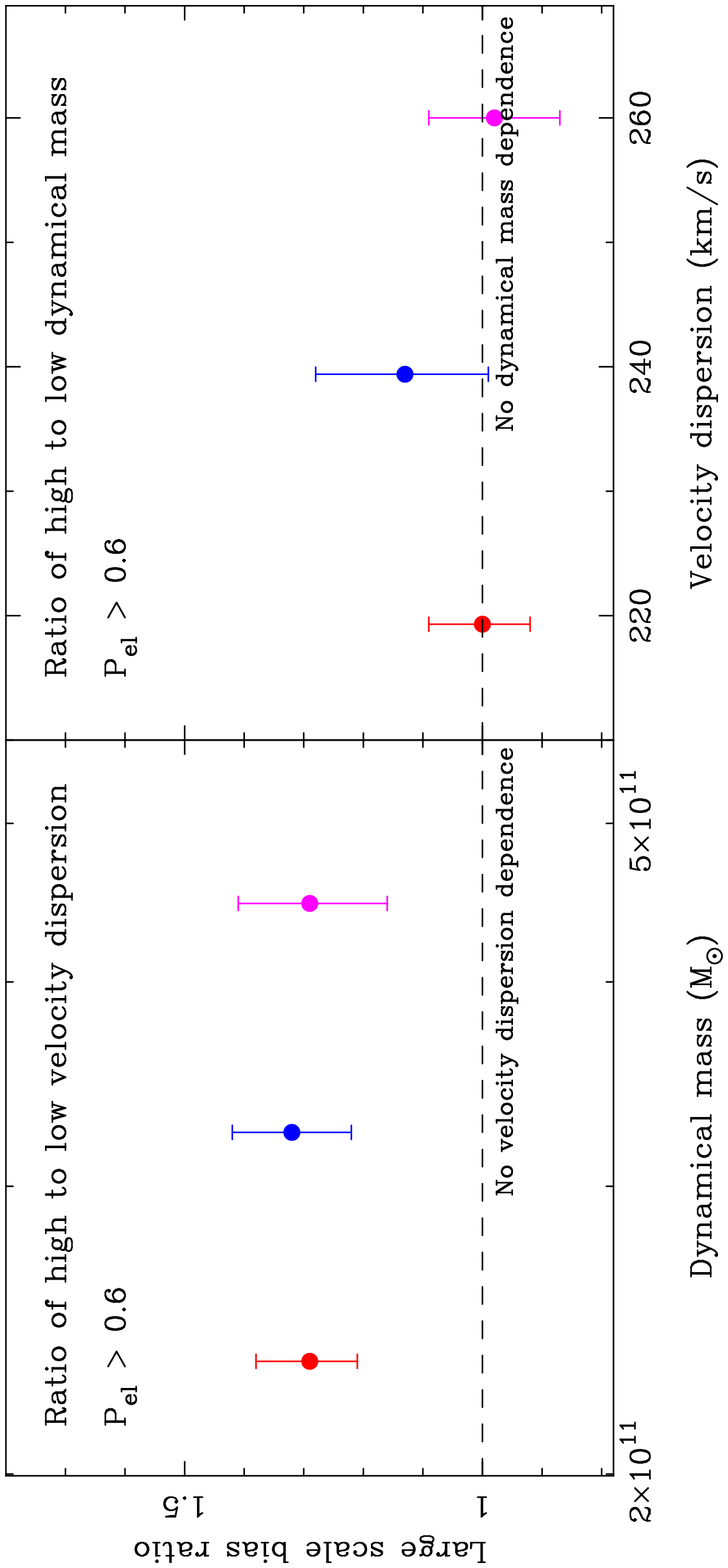}
\includegraphics{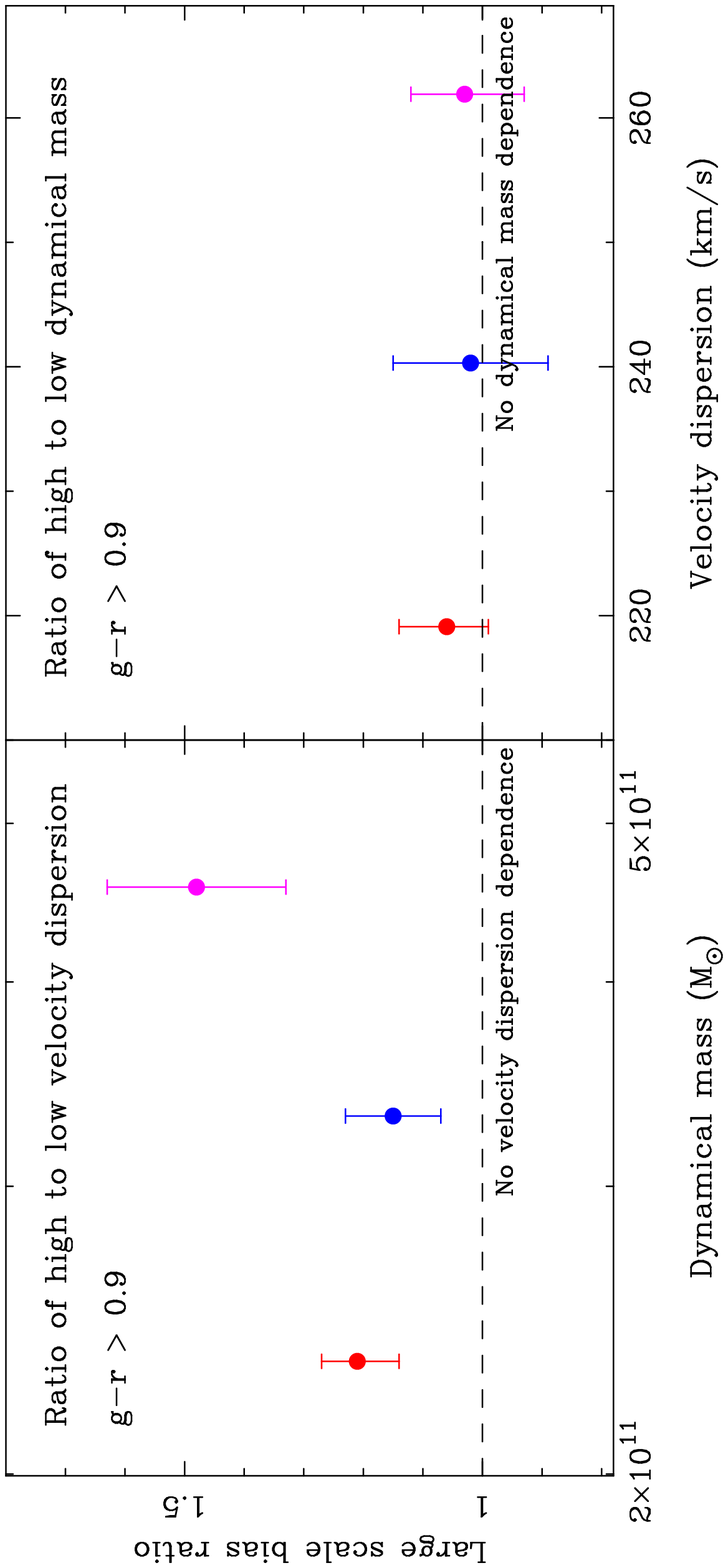}
\caption{\small The dependence of large scale bias on \Vdisp~at fixed \Mdyn~(left) and \Mdyn~at fixed \Vdisp~(right) for elliptical galaxies (P$_{el} >$ 0.6; top) and red galaxies ($g-r >$ 0.9; bottom). The same result as for the whole sample remains, there is significant dependence of the bias on \Vdisp~at fixed \Mdyn~but none on \Mdyn~at fixed \Vdisp.
\label{fig:biasrat3}}
\end{figure*}

We have included the distribution in elliptical probability for each of the samples in the plots in Appendix \ref{sec:appsamp} as it demonstrates one area that is both potentially a concern but also scientifically important for this analysis. \Vdisp~is much better correlated with morphology or color than stellar mass is. \Mden~is also a better discriminator than mass, but not as good as \Vdisp~(see Wake et al. 2012 for a detailed study of these trends). Because of this when we fix \Vdisp~we come pretty close to fixing morphology regardless of the mass and when we split into high and low \Vdisp~samples at fixed mass we are pretty close to splitting into disks and spheriods (we also see a splitting by color in the same way). This is of course the heart of our investigation: is a tighter correlation between \Vdisp~and halo properties (compared to the correlation between mass and halo properties) the explanation as to why \Vdisp~is a better indicator than stellar mass of a galaxy's stellar population? This does however raise a potential worry regarding systematics: are we measuring the same thing when we measure \Vdisp~for a disk galaxy as compared to a spheroidal galaxy? 

We tackle this question in several ways. Firstly we show in Figure \ref{fig:MstarMdyn} the observed relationship between \Mstar~and \Mdyn~for all the galaxies in SDSS with $0.04 < z < 0.113$ and \Vdisp~$>$ 75 km/s. We further split this sample into spheriods or disks based on the Galaxy Zoo classifications. The relationship is essentially identical regardless of morphology over practically the entire mass range. There is a small deviation at lower masses but this is below the mass range considered in this work. This gives us confidence that the velocity dispersion is measuring the same physical quantity for both disks and ellipticals within the SDSS fiber radius.

The second approach is a more direct one: we can see if morphology itself has any effect on the clustering amplitude of our galaxies at fixed mass or \Vdisp. We thus define samples with high and low elliptical probabilities in narrow ranges of \Mdyn~and \Vdisp. We note that we are unable to just take the quartiles of the distribution in elliptical probability as we have done for parameters previously as there is a redshift dependence to the probabilities assigned. This results from galaxies being harder to classify at higher redshift due to their smaller angular size and surface brightness dimming. Whilst there has been some correction for this effect applied to the Galaxy Zoo probabilities we find that it does not entirely remove this bias resulting in the samples split into upper and lower quartiles in \Pel~having different redshift distributions. We thus define our samples making use of the expectation that the morphological mix does not vary significantly over our narrow redshift range. We find the best fit linear relations between elliptical probability and redshift that produce the 25\% most and least likely ellipticals at each redshift and use these relations to define our samples. Details of these samples are given in Tables \ref{tab:PelatdM} and \ref{tab:PelatVd} and their distributions are plotted in Appendix \ref{sec:appsamp}. 

Figure \ref{fig:biasrat2} shows the large scale \wp~ratios of these high and low \Pel~samples at fixed mass and \Vdisp. As before we fit the \wp~ratio (shown in Figure \ref{fig:2ptPel} in Appendix \ref{sec:appsamp}) between $1.6 < r_p < 25.1$ h$^{-1}$Mpc and determine the probability that the ratio is consistent with one. Details of these fits are give in Table \ref{tab:ratfitcolel}. At fixed \Vdisp~there is no evidence for any dependence of the clustering amplitude on morphology at any scale. At fixed mass the ratios are all non-zero with galaxies with a higher \Pel~showing a higher clustering amplitude than those with a low \Pel. However, the significance of these positive \wp~ratios is low with all being consistent with no difference in the clustering amplitude. This implies that the correlation of morphology with dispersion is not the determining factor in the strong correlation of clustering amplitude with \Vdisp~at fixed mass. 

Whilst the strong correlation of color with \Vdisp~is unlikely to introduce any systematics in our measurements it is still informative to investigate whether there is any residual clustering dependence on color at either fixed mass or \Vdisp. Since color is better correlated with \Vdisp~than mass one might imagine that any residual clustering dependence on color would be lower at fixed \Vdisp~than at fixed mass if the halo properties are key to determining the color, much as is seen with morphology above. To test this we show in Figure \ref{fig:biasrat2} the best fit large scale correlation function ratios of high and low $g-r$ color samples at fixed mass and dispersion and list the fitting results in Table \ref{tab:ratfitcolel} (the full \wp~ratios are shown in Figure \ref{fig:2ptcol} in Appendix \ref{sec:appsamp}). Since the color evolves with redshift we have again had to split the samples in a similar way to the elliptical probability samples such that we have the reddest and bluest quartiles at any redshift.  
Now, unlike with the elliptical probability, we see a residual clustering dependence on color at both fixed mass and \Vdisp~such that red galaxies cluster more strongly than blue. 
This is most likely the result of the truncation of star formation in satellite galaxies in massive dark halos, resulting in satellite galaxies at fixed mass or \Vdisp~having redder colors than central galaxies with the same mass or \Vdisp~\citep[e.g.][]{Yang09}. The fact that we see a significant residual color dependence and not much of a residual morphology dependence suggests that the environmental mechanism for transforming the color of a galaxy from red to blue does not simultaneously change the morphology. A similar result was observed by \citet{Skibba09} using the marked correlation function applied to SDSS galaxies.

Since we do see some residual color dependence and perhaps a hint of morphology dependence in the clustering amplitude at fixed mass it is worth asking if the \Vdisp~dependence remains when morphology or color is fixed. We show in Figure \ref{fig:biasrat3} the large scale \wp~ratio of high and low \Vdisp~samples at fixed \Mdyn~and high and low \Mdyn~samples at fixed \Vdisp~where we have restricted the galaxies to be either red ($g-r > 0.9$) or have high elliptical probabilities ($P_{el} >$ 0.6). The fit details given in Table \ref{tab:ratfitcolel} and the full \wp~ratios shown in Figures \ref{fig:2ptVddMred} and \ref{fig:2ptVddMel} in Appendix \ref{sec:appsamp}. Despite the sample sizes being reduced and hence the errors increasing, the original trend of high \Vdisp~galaxies having a higher clustering amplitude at fixed mass remains significant and the large scale ratios stay essentially the same within the errors. There is a hint that the \wp~ratios are slightly reduced, but by removing the blue or spiral galaxies we have removed many lower \Vdisp~galaxies and reduced the difference between the high and low \Vdisp~samples.
This is a clear demonstration that galaxies with higher \Vdisp~cluster more strongly than galaxies with lower \Vdisp~ regardless of their mass, morphology or color.

\section{Discussion}
\label{sec:dis}

\begin{figure}

\vspace{11cm}
\includegraphics{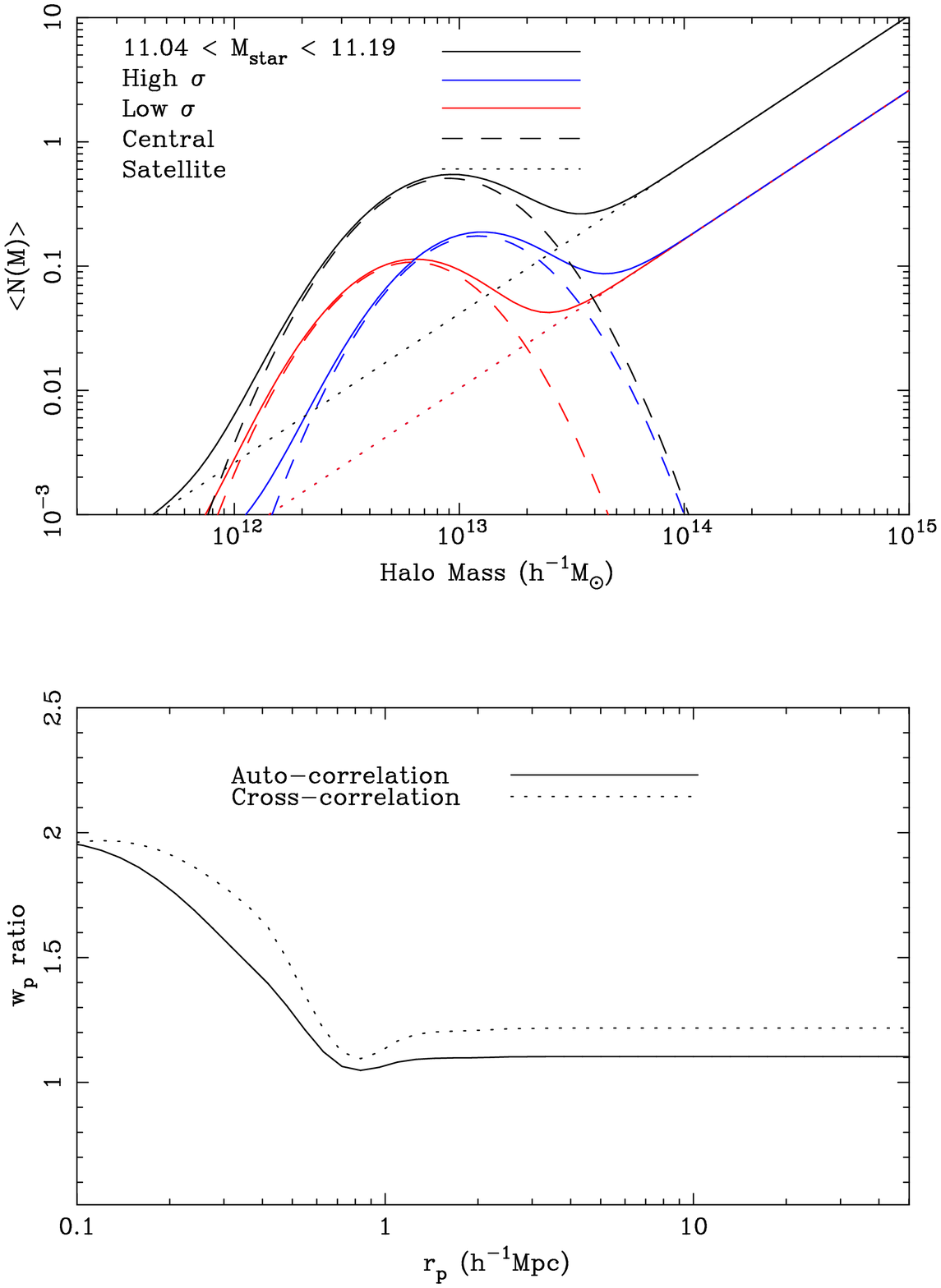}
\caption{\small Top: An example halo occupation distribution representing the case where correlation between \Vdisp~and dark matter halo mass is tighter than the correlation between \Mstar~and halo mass. The black line represents an HOD for a galaxy population with a narrow slice in \Mstar. The red and blue lines each contain 25\% of the \Mstar slice sample and have had their central halo mass thresholds adjusted to sample slightly higher and lower halo masses from within that sample, to represent high and low \Vdisp~samples. In each case the dashed lines show the central galaxies and the dotted the satellites. Bottom: The auto- (solid) and cross- (dotted) correlation function ratios between the example high and low \Vdisp HODs. They show the very similar characteristics to the measured correlation function ratios on both small and large scales.  
\label{fig:HOD}}
\end{figure}

We have shown that at fixed mass (stellar or dynamical) galaxies with high \Vdisp~(or \Mden) cluster more strongly than galaxies with low \Vdisp. Conversely, at fixed \Vdisp~(or \Mden) our measurements are consistent with high and low mass galaxies clustering the same. If the clustering of galaxies is dominated by the clustering of the DM halos in which they reside, as expected in the CDM paradigm, then this implies that even at fixed galaxy mass higher \Vdisp~galaxies live in more clustered halos than low \Vdisp~galaxies. The main determinant of a halo's clustering amplitude is its mass, but it has also been shown in simulations that halos with different formation histories also cluster differently, with older more centrally concentrated halos clustering more strongly \citep{Gao05,Wechsler06,Wetzel07,Gao07,Jing07,Li08}. This means that galaxies with higher \Vdisp~at fixed mass could be preferentially be living in either more massive or older more centrally concentrated halos. 

When considering how galaxies occupy halos it is also important to make the distinction between the galaxy that has been formed at the center of a given DM halo, the central galaxy, and those that initially formed at the centers of other halos and have later been accreted into this halo, satellite galaxies. The properties of a central galaxy will be largely determined by the formation history and properties of its host halo, whereas any effect the current host halo has on satellite galaxies is limited to the time after it has been accreted. 

In the following sections we address whether our result can be explained by a relationship between \Vdisp~and halo mass for central galaxies (Section \ref{sec:cenmass}) or for satellite galaxies (Section \ref{sec:satmass}), or by a relationship between \Vdisp~and halo formation history (Section \ref{sec:age})

\subsection{Central galaxy halo mass correlations}
\label{sec:cenmass}
Since DM halo mass has the largest effect on clustering amplitude, the simplest explanation for our results is that \Vdisp~is better correlated with its host halo mass than stellar or dynamical mass are. For galaxies living at the centers of halos, central galaxies, there is a strong correlation between stellar mass and halo mass \citep[e.g.][]{Yang09,Wake11}, but with a significant scatter \citep{Yang09,Yang11}.
If the scatter between \Vdisp~and halo mass for centrals is substantially lower than the scatter with \Mstar~this could result in clustering observations much as we have observed. 

It is striking in Figures \ref{fig:2ptVdSM} and \ref{fig:2ptVddM} that the clustering ratio between high and low \Vdisp~samples becomes even larger as the scale decreases. The clustering amplitude on such scales is determined by the separations of galaxies within individual halos, so between central and satellite galaxies. Thus these smaller scale clustering measurements can provide an additional constraint as to whether a tighter correlation between \Vdisp~and halo mass (\Mhalo) for central galaxies could produce the observed clustering results. 

We show in Figure \ref{fig:HOD} an illustrative attempt at modeling the effect of a smaller scatter between \Vdisp~and \Mhalo~than between \Mstar~and \Mhalo~using the halo model \citep[see][for details of our model]{Wake11}. We have fitted halo occupation distributions (HOD), the mean number of central and satellite galaxies as a function of halo mass, to the clustering measurements for a series of samples of SDSS galaxies limited in stellar mass (i.e. \Mstar $>$M$_{\rm low}$) \citep[see][for details of these measurements]{Wake12}. Subtracting two of the resulting HODs with slightly different M$_{\rm low}$ gives the HOD for a sample of galaxies with a narrow range in \Mstar. We show an HOD for such a sample as the black line in the top panel of Figure \ref{fig:HOD}, which has been chosen as it matches the lowest \Mstar~ sample we use ($11.04 <$ log(\Mstar) $< 11.19$). We can then attempt to mimic the effect of splitting into high and low \Vdisp~samples under the assumption that \Vdisp~for central galaxies is better correlated with halo mass than \Mstar~is, so that the high \Vdisp~sample will have higher average halo masses than the low \Vdisp~sample. We do this by slightly adjusting the central galaxy halo mass thresholds that define the high and low halo mass cut offs in our HODs. The HODs of these two samples, representing high and low \Vdisp~quartiles at fixed \Mstar, are shown as the red and blue lines in the top panel of Figure \ref{fig:HOD}. These mock high and low \Vdisp~samples each contain 25\% of the central and satellite galaxies of the full sample, with the size of the central HOD controlled by the mass threshold adjustments and the satellite just random sampled to 25\% of the number density, i.e. the form of the satellite HOD is left unchanged. The resulting \wp ratio is shown in the bottom panel of Figure \ref{fig:HOD} for both the auto- and cross-correlations. The cross-correlation is calculated with an HOD suitable for the parent sample we use throughout this work (i.e. log(\Mstar) $>$ 10.777).

The model \wp~ratio shown in Figure \ref{fig:HOD} is quite similar to the measured ratios shown in Figures \ref{fig:2ptVdSM}, \ref{fig:2ptVddM} and \ref{fig:2ptMdenSM}, showing a constant increase in the cross-correlation amplitude on intermediate and large scales and then an increasing amplitude at small scales. It is always tempting to interpret changes in small scale (1-halo) clustering as reflecting changes in the satellite population, but this is a clear example of how this is not necessarily the case. The satellite population has not changed between the mock low and high \Vdisp~samples, and it is the change in the central galaxies which is causing a change in the numbers of central-satellite pairs that causes the increased clustering amplitude on small scales. More explicitly, at a given stellar mass the higher \Vdisp~galaxies are in more massive halos, which typically contain more satellite galaxies. This results in an increase in the number of central-satellite pairs, thus increasing the small scale clustering amplitude. The converse is true for lower dispersion centrals i.e. they are in less massive halos, with fewer satellites and so fewer central-satellite pairs. The satellite-satellite pairs remain unchanged, but since for these massive galaxies the satellite fraction is low so there may only be at most one or two satellites in a given halo the central-satellite pairs make up a significant contribution to the small scale clustering amplitude.  

This exercise clearly demonstrates that if there is a much tighter correlation between \Vdisp~and halo mass than between \Mstar~and halo mass then the clustering ratios we measure would be expected. There is, of course, some theoretical expectation that this would be the case. The stellar mass of a central galaxy depends on a large number of complicated astrophysical baryonic processes such as gas cooling and feedback as well as mergers from satellite galaxies, which may give rise to a large scatter between halo mass and stellar mass. \Vdisp, on the other hand, is tracing the depth of the potential at the center of the halo and whilst this may be modified by the evolution of the baryons it is likely to be much more directly linked to the halo properties and in particular its mass. 

\subsection{Satellite galaxy halo mass correlations}
\label{sec:satmass}
Whilst our example halo model shows that our results can be produced just by central galaxies, it is of course possible to produce similar clustering trends by modifying the satellite distribution or both the central and satellite distributions. To increase the clustering amplitude of galaxies at fixed mass by modifying the satellites requires that higher \Vdisp~satellites are preferentially found in more massive halos. 

One possible process that could create such an effect is the stripping of the outer regions of a satellite galaxy by tidal interactions as it orbits it parent halo. This would likely have a minimal effect on \Vdisp~whilst reducing the mass and size of the galaxy. The likelihood of a galaxy getting disrupted in this way increases as the halo mass increases but decreases as the mass of the galaxy increases. Since, the galaxies in this study are all more massive than $6\times10^{10}\Msun$ it would only be effective for satellites passing very close to the center of the most massive halos an occurrence that would be quite rare. 

However, assuming that this process was an effective one we would expect to see its affects on the clustering both when the mass is held fixed and \Vdisp~varied and visa-versa. Satellite galaxies at fixed mass would have higher \Vdisp~in higher mass halos, since they would have been stripped more.  This would mean that the satellite galaxies in the higher \Vdisp~samples would be preferentially found in higher mass halos and hence would cluster more strongly, just as we have observed. The opposite would be true of satellite galaxies with fixed \Vdisp; one would expect them to have lower masses in more massive halos since again they would have experienced more tidal stripping. 

One could then imagine a scenario that could explain our observations. If we assume that the more massive a central galaxy or the higher its \Vdisp~then the more massive its parent halo, then for centrals one would observe higher clustering amplitudes both for higher mass galaxies at fixed \Vdisp~and higher \Vdisp~galaxies at fixed mass. At fixed \Vdisp~the higher mass satellites would typically be in lower mass halos (where they haven't been stripped) and visa-versa, resulting in an overall reduction in the clustering ratio between the high and low mass samples. The opposite would be true at fixed mass, where the higher \Vdisp~satellite galaxies would preferentially be in higher mass halos (where the stripping is effective), increasing the clustering ratio between high and low \Vdisp~samples. A similar effect could also occur when considering \Mdyn~rather than \Mstar~since the size of a galaxy would be reduced as it is stripped, thus reducing \Mdyn. This does seem to be a viable explanation of our results, although we note that the fraction of satellite galaxies should be relatively low in our samples, $<$ 20\%.

\subsection{Central galaxy halo age correlations}
\label{sec:age}
Finally, as already mentioned, there is an alternative to the halo mass dependence outlined above, whereby instead of a tight correlation between central galaxy \Vdisp~and halo mass there is a residual correlation between \Vdisp~and a halo's formation history or age. N-body simulations show that at a given mass halos that formed earlier (and are more concentrated) are more clustered than younger halos \citep{Gao05,Wechsler06,Wetzel07,Gao07,Jing07,Li08}. This is an appealing explanation since it would link together both the correlation between \Vdisp~and clustering amplitude as well as that between \Vdisp~and the star formation history of galaxies. At a given mass galaxies with higher \Vdisp~exhibit older stellar populations which would be naturally explained by the earlier formation of their parent halos and thus their higher clustering amplitude. 

\subsection{Consequences of these relationships}
\begin{figure}
\vspace{11cm}
\includegraphics{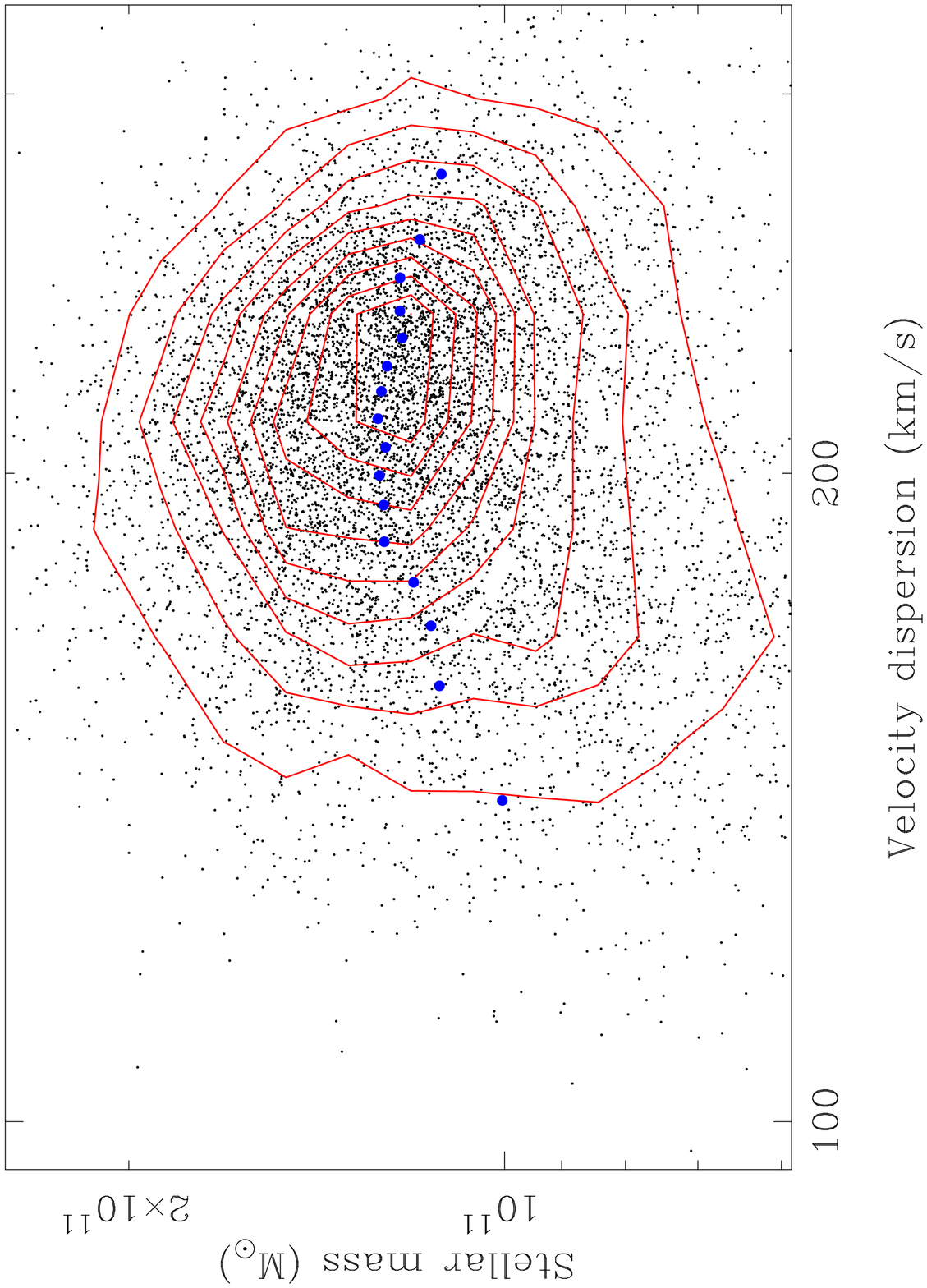}
\includegraphics{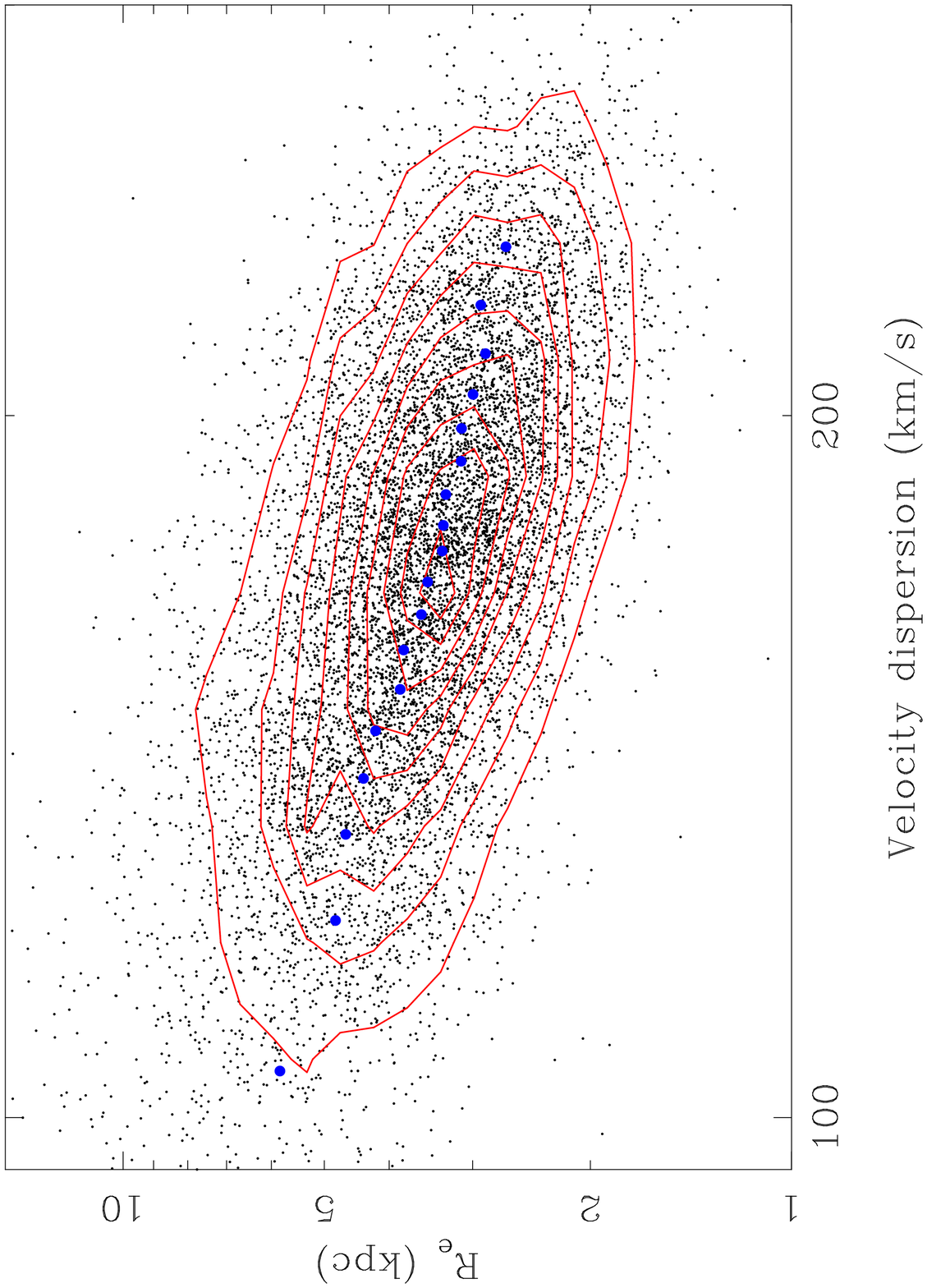}
\caption{\small Top: The relationship between \Re~and \Vdisp~at fixed \Mstar~($10.85 <$ log(\Mstar) $< 10.9$). The blue points show the running median. At a given \Mstar~galaxies with higher \Vdisp~are smaller. Bottom: The relationship between \Mstar~and \Vdisp~at fixed \Mdyn~($11.35 <$ log(\Mdyn) $< 11.45$), showing very little trend.
\label{fig:ReatVdisp}}
\end{figure}

Assuming that a tighter relationship exists between \Vdisp~and halo properties than between galaxy mass and halo properties for central galaxies, as outlined in Sections \ref{sec:cenmass} and \ref{sec:age}, we can use our measurements to infer how \Vdisp, \Mstar, \Mdyn~and \Re~interrelate with one another, and with halo mass and concentration. We expect that \Vdisp, \Mstar~and \Mdyn~all depend on halo mass and each other as follows: 
\begin{equation}
\sigma^2 \sim \frac{GM_{dyn}}{R_e} \propto  \frac{M_{star}}{R_e} + \frac{M_{halo}}{R_{vir}/c}
\end{equation}
where \Mhalo~is the halo mass, $R_{vir}$ is the halo virial radius and $c$ is the halo concentration \citep{Hopkins09}. 
%%\begin{equation}
%%M_{dyn}\propto \sigma^2 R_e.
%%\end{equation}
 Therefore, when \Mstar~is fixed and \Vdisp~varied either \Re, \Mhalo~or $c$ must vary. Since the clustering amplitude does change then either \Mhalo~or $c$ must also change, such that higher dispersion galaxies must live in more massive or more concentrated (older) halos. Whilst it is the case that galaxies with higher \Vdisp~at a fixed \Mstar~typically have a smaller \Re~(see Figure \ref{fig:ReatVdisp}) they must also occupy more massive or more concentrated halos. Conversely as \Mstar~is increased at fixed \Vdisp~\Re~must be changing as well such that more massive galaxies have larger \Re, since there is no change in the clustering amplitude so \Mhalo~and $c$ aren't varying. This means that at fixed \Vdisp~increasing \Mstar~increases the ratio between \Mstar~and \Mhalo.  

We see a similar effect with \Mdyn. When \Mdyn~is held fixed and \Vdisp~varied \Re~must also vary such that higher \Vdisp~leads to lower \Re. We also know that either \Mhalo~or $c$ increases with \Vdisp~at fixed \Mdyn~since the clustering amplitude increases. \Mstar~could potentially vary in either direction with the size of the change depending on the ratio of the contributions of dark and stellar mass to the potential and the magnitude of the change in \Mhalo~or $c$ with \Vdisp. In fact Figure \ref{fig:ReatVdisp} shows that at fixed \Mdyn~there is very little average change in \Mstar~as \Vdisp~increases, indicating that \Mhalo~or $c$ have to increase with \Vdisp~at fixed \Mdyn~just as we observe.
When \Mdyn~is varied at fixed \Vdisp~\Re~must also change, increasing as \Mdyn~increases. Since the clustering amplitude does not change, \Mhalo~and $c$ do not change and so \Mstar~must be increasing. This implies that at fixed \Vdisp~galaxies with higher \Mdyn~will have both larger \Mstar~and \Re.

\section{Summary and Conclusions}

We have investigated how the clustering of galaxies depends on stellar mass, dynamical mass, velocity dispersion, surface mass density, color and morphology using large samples of massive galaxies from the SDSS. We split the samples into narrow ranges in mass, \Vdisp~or \Mden~and measure any residual clustering dependence on the other parameters. We find the following:

\begin{enumerate}
\item When \Mstar~or \Mdyn~are fixed there is a significant dependence of the clustering amplitude on \Vdisp~on all scales, such that galaxies with higher \Vdisp~are more strongly clustered. Conversely when \Vdisp~is fixed there is no dependence of the clustering amplitude on either \Mstar~or \Mdyn. These trends remain when we limit the samples to galaxies that are red or morphologically elliptical. We see a similar trend using \Mden~instead of \Vdisp.

\item When \Mstar~or \Mdyn~are fixed there is a weak dependence of the clustering amplitude on morphology. When \Vdisp~ is fixed there is no significant morphological dependent clustering. 

\item There always remains a strong dependence of the clustering amplitude on a galaxy's $g-r$ color at fixed mass or \Vdisp. This is most likely caused by satellite galaxies in massive halos always being preferentially redder than central galaxies of the same mass or \Vdisp.

\end{enumerate}

We suggest that for our main finding, point 1. above, that there are three possible explanations; the relationship between \Vdisp~or \Mden~and halo mass for central galaxies is tighter than that between \Mstar~or \Mdyn~and halo mass; the relationship between \Vdisp~or \Mden~and halo age (or concentration) for central galaxies is tighter than that between \Mstar~or \Mdyn~and  halo age (or concentration); tidal stripping of satellite galaxies reduces the size and mass of the galaxy whilst having a minimal effect on \Vdisp~and is more effective in more massive halos. Or it could be a combination of all three effects. If the host halo mass or age are indeed better correlated with \Vdisp~and \Mden~than with \Mstar, these halo properties may also drive the star formation history of galaxies, explaining why \citet{Kauffmann03b} found that star formation rate is better correlated with \Mden~than \Mstar, and why \citet{Franx08} found that color and star formation rate are better correlated with \Vdisp~than \Mstar. 
       
\acknowledgments{

DAW would like to thank Nikhil Padmanabhan, Carlton Baugh, Violeta Gonzalez, Rachel Bezanson, Britt Lundgren and Ravi Sheth for helpful discussions on this work. We would like to thank the NYU and MPA-JHU groups for making their value added data products freely available. 

This work makes use of data from the Sloan Digital Sky Survey (SDSS). Funding for the Sloan Digital Sky Survey (SDSS) has been provided by the Alfred P. Sloan Foundation, the Participating Institutions, the National Aeronautics and Space Administration, the National Science Foundation, the U.S. Department of Energy, the Japanese Monbukagakusho, and the Max Planck Society. The SDSS Web site is http://www.sdss.org/.

The SDSS is managed by the Astrophysical Research Consortium (ARC) for the Participating Institutions. The Participating Institutions are The University of Chicago, Fermilab, the Institute for Advanced Study, the Japan Participation Group, The Johns Hopkins University, Los Alamos National Laboratory, the Max-Planck-Institute for Astronomy (MPIA), the Max-Planck-Institute for Astrophysics (MPA), New Mexico State University, University of Pittsburgh, Princeton University, the United States Naval Observatory, and the University of Washington. 
}

\clearpage
\appendix
\section{A. Sample properties}
\label{sec:appsamp}

\begin{table}[h]
  \begin{center}
    \caption{\label{tab:SMatVd} Details of the high and low stellar mass samples at fixed velocity dispersion}
	\begin{tabular}{ccccccccc}
	\tableline\tableline
      	\multicolumn{4}{c}{Velocity Dispersion (km/s)} &
      	\multicolumn{4}{c}{log(Stellar Mass) ($\Msun$)} &
      	\multicolumn{1}{c}{} \\
      	\multicolumn{1}{c}{Min} &
      	\multicolumn{1}{c}{Max} &
      	\multicolumn{1}{c}{Mean} &
      	\multicolumn{1}{c}{Median} &
      	\multicolumn{1}{c}{Min} &
      	\multicolumn{1}{c}{Max} &
      	\multicolumn{1}{c}{Mean} &
      	\multicolumn{1}{c}{Median} &
      	\multicolumn{1}{c}{Ngal}\\
	\tableline
210.0 & 242.3 & 223.5 &  222.5 & 10.78 & 10.91 & 10.85 & 10.85 &  3242\\
210.0 & 242.3 & 226.4 &  226.3 & 11.16 & 11.93 & 11.30 &  11.26 &  3242\\
242.3 & 274.6 & 255.2 &  253.8 & 10.78 & 10.98 & 10.89 &  10.89 &  1619\\
242.3 & 274.6 & 257.8 &  257.5 & 11.27 & 11.88 & 11.40 &  11.37 &  1619\\
274.5 & 306.4 & 286.3 &  284.5 & 10.78 & 11.07 & 10.95 &  10.96 &   591\\
274.5 & 306.8 & 288.6 &  287.2 & 11.37 & 11.91 & 11.50 &  11.46 &   590\\

	\tableline
    \end{tabular}
%%\tablecomments{}
\end{center}
\end{table}

\begin{table}[h]
  \begin{center}
    \caption{\label{tab:VdatSM} Details of the high and low velocity dispersion samples at fixed stellar mass}
	\begin{tabular}{ccccccccc}
	\tableline\tableline
      	\multicolumn{4}{c}{log(Stellar Mass) ($\Msun$)} &
      	\multicolumn{4}{c}{Velocity Dispersion (km/s)} &
      	\multicolumn{1}{c}{} \\
      	\multicolumn{1}{c}{Min} &
      	\multicolumn{1}{c}{Max} &
      	\multicolumn{1}{c}{Mean} &
      	\multicolumn{1}{c}{Median} &
      	\multicolumn{1}{c}{Min} &
      	\multicolumn{1}{c}{Max} &
      	\multicolumn{1}{c}{Mean} &
      	\multicolumn{1}{c}{Median} &
      	\multicolumn{1}{c}{Ngal}\\
	\tableline

11.04 & 11.19 & 11.10 &  11.09 & 79.0 & 177.2 & 155.5 &  159.3 &  3242\\
11.04 & 11.19 & 11.12 &  11.12 & 232.5 & 921.7 & 259.7 &  252.3 &  3242\\
11.19 & 11.33 & 11.24 &  11.24 & 92.6 & 198.4 & 173.6 &  178.0 &  1601\\
11.19 & 11.33 & 11.26 &  11.26 & 253.6 & 565.1 & 278.5 &  273.0 &  1600\\
11.33 & 11.48 & 11.39 &  11.38 & 108.3 & 219.6 & 192.3 &  198.3 &   610\\
11.33 & 11.48 & 11.40 &  11.40 & 273.7 & 562.7 & 297.3 &  291.8 &   610\\

	\tableline
    \end{tabular}
%%\tablecomments{}
\end{center}
\end{table}

\begin{table}[h]
  \begin{center}
    \caption{\label{tab:dMatVd} Details of the high and low dynamical mass samples at fixed velocity dispersion}
	\begin{tabular}{ccccccccc}
	\tableline\tableline
      	\multicolumn{4}{c}{Velocity Dispersion (km/s)} &
      	\multicolumn{4}{c}{log(Dynamical Mass) ($\Msun$)} &
      	\multicolumn{1}{c}{} \\
      	\multicolumn{1}{c}{Min} &
      	\multicolumn{1}{c}{Max} &
      	\multicolumn{1}{c}{Mean} &
      	\multicolumn{1}{c}{Median} &
      	\multicolumn{1}{c}{Min} &
      	\multicolumn{1}{c}{Max} &
      	\multicolumn{1}{c}{Mean} &
      	\multicolumn{1}{c}{Median} &
      	\multicolumn{1}{c}{Ngal}\\
	\tableline
210.0 & 242.3 & 222.4 & 221.0 & 10.67 & 11.21 & 11.12 & 11.13 & 3242\\
210.0 & 242.3 & 227.0 & 227.2 & 11.45 & 12.22 & 11.59 & 11.55 & 3242\\
242.3 & 274.6 & 254.7 & 253.0 & 10.51 & 11.33 & 11.23 & 11.24 & 1619\\
242.3 & 274.6 & 258.3 & 257.7 & 11.60 & 12.20 & 11.73 & 11.70 & 1619\\
274.6 & 306.4 & 286.1 & 284.0 & 10.66 & 11.44 & 11.33 & 11.34 & 591\\
274.6 & 306.8 & 289.0 & 287.9 & 11.72 & 12.28 & 11.86 & 11.82 & 590\\

	\tableline
    \end{tabular}
%%\tablecomments{}
\end{center}
\end{table}

\begin{table}[h]
  \begin{center}
    \caption{\label{tab:VdatdM} Details of the high and low velocity dispersion samples at fixed dynamical mass}
	\begin{tabular}{ccccccccc}
	\tableline\tableline
      	\multicolumn{4}{c}{log(Dynamical Mass) ($\Msun$)} &
      	\multicolumn{4}{c}{Velocity Dispersion (km/s)} &
      	\multicolumn{1}{c}{} \\
      	\multicolumn{1}{c}{Min} &
      	\multicolumn{1}{c}{Max} &
      	\multicolumn{1}{c}{Mean} &
      	\multicolumn{1}{c}{Median} &
      	\multicolumn{1}{c}{Min} &
      	\multicolumn{1}{c}{Max} &
      	\multicolumn{1}{c}{Mean} &
      	\multicolumn{1}{c}{Median} &
      	\multicolumn{1}{c}{Ngal}\\
	\tableline
11.31 & 11.47 & 11.37 & 11.36 & 81.4 & 183.2 & 162.2 & 166.3 & 3242\\
11.31 & 11.47 & 11.39 & 11.39 & 235.0 & 454.5 & 258.5 & 252.7 & 3242\\
11.47 & 11.62 & 11.53 & 11.52 & 102.9 & 205.9 & 183.0 & 187.6 & 1647\\
11.47 & 11.62 & 11.55 & 11.54 & 257.4 & 410.7 & 280.7 & 275.4 & 1646\\
11.62 & 11.78 & 11.68 & 11.68 & 115.9 & 231.9 & 207.0 & 213.1 & 695\\
11.62 & 11.78 & 11.70 & 11.70 & 276.7 & 500.7 & 298.2 & 292.7 & 695\\

	\tableline
    \end{tabular}
%%\tablecomments{}
\end{center}
\end{table}

\begin{table}[h]
  \begin{center}
    \caption{\label{tab:SMatMden} Details of the high and low stellar mass samples at fixed surface stellar mass density}
	\begin{tabular}{ccccccccc}
	\tableline\tableline
      	\multicolumn{4}{c}{log(Surface mass density) ($\Msun Kpc^{-2}$)} &
      	\multicolumn{4}{c}{log(Stellar Mass) ($\Msun$)} &
      	\multicolumn{1}{c}{} \\
      	\multicolumn{1}{c}{Min} &
      	\multicolumn{1}{c}{Max} &
      	\multicolumn{1}{c}{Mean} &
      	\multicolumn{1}{c}{Median} &
      	\multicolumn{1}{c}{Min} &
      	\multicolumn{1}{c}{Max} &
      	\multicolumn{1}{c}{Mean} &
      	\multicolumn{1}{c}{Median} &
      	\multicolumn{1}{c}{Ngal}\\
	\tableline
 9.00  & 9.16 &  9.08 &   9.08 & 10.78 & 10.87  &10.82 &  10.82  & 3541\\
 9.00 &  9.16 &  9.08  &  9.07 & 11.10 & 11.94  &11.24  & 11.21 &  3541\\
 9.16 &  9.32 &  9.24 &   9.24 & 10.78 & 10.86 & 10.82  & 10.82 &  2272\\
 9.16 &  9.32 &  9.24 &   9.23 & 11.08 & 12.07 & 11.22 &  11.18 &  2272\\
 9.32 &  9.48 &  9.39 &   9.39 & 10.78 & 10.84 & 10.81 &  10.81 &  1075\\
 9.32 &  9.48 &  9.39 &   9.38 & 11.05 & 11.75 & 11.18 &  11.15 &  1075\\
 9.48 &  9.64 &  9.55 &   9.54 & 10.78 & 10.84 & 10.80 &  10.81 &   383\\
 9.48 &  9.64 &  9.55  &  9.54 & 11.04 & 11.78 & 11.17 &  11.14 &   383\\

	\tableline
    \end{tabular}
%%\tablecomments{}
\end{center}
\end{table}

\begin{table}[h]
  \begin{center}
    \caption{\label{tab:MdenatSM} Details of the high and low surface stellar mass density samples at fixed stellar mass}
	\begin{tabular}{ccccccccc}
	\tableline\tableline
      	\multicolumn{4}{c}{log(Stellar Mass) ($\Msun$)} &
      	\multicolumn{4}{c}{log(Surface mass density) ($\Msun Kpc^{-2}$)} &
      	\multicolumn{1}{c}{} \\
      	\multicolumn{1}{c}{Min} &
      	\multicolumn{1}{c}{Max} &
      	\multicolumn{1}{c}{Mean} &
      	\multicolumn{1}{c}{Median} &
      	\multicolumn{1}{c}{Min} &
      	\multicolumn{1}{c}{Max} &
      	\multicolumn{1}{c}{Mean} &
      	\multicolumn{1}{c}{Median} &
      	\multicolumn{1}{c}{Ngal}\\
	\tableline
10.80 & 10.87 & 10.83 &  10.83 &  6.58 &  8.71 &  8.46 &   8.53 &  3541 \\
10.80 & 10.87 & 10.83 &  10.83 &  9.15 & 10.51 &  9.33 &   9.29 &  3541 \\
10.87 & 10.93 & 10.90 &  10.90 &  6.58 &  8.73 &  8.48 &   8.56 &  2761 \\
10.87 & 10.93 & 10.90 &  10.90 &  9.15 & 10.24 &  9.32 &   9.29 &  2761 \\
10.93 & 10.99 & 10.96 &  10.96 &  6.77 &  8.74 &  8.49 &   8.57 &  2377 \\
10.93 & 10.99 & 10.96 &  10.96 &  9.14 & 10.41 &  9.31 &   9.28 &  2377 \\
10.99 & 11.05 & 11.02 &  11.02 &  7.05 &  8.74 &  8.51 &   8.58 &  1963 \\
10.99 & 11.05 & 11.02 &  11.02 &  9.14 & 10.37 &  9.31 &   9.27 &  1963 \\

	\tableline
    \end{tabular}
%%\tablecomments{}
\end{center}
\end{table}

\begin{table}[h]
  \begin{center}
    \caption{\label{tab:PelatdM} Details of the high and low elliptical probability samples at fixed dynamical mass}
	\begin{tabular}{ccccccccc}
	\tableline\tableline
      	\multicolumn{4}{c}{log(Dynamical mass) ($\Msun$)} &
      	\multicolumn{4}{c}{Elliptical probability} &
      	\multicolumn{1}{c}{} \\
      	\multicolumn{1}{c}{Min} &
      	\multicolumn{1}{c}{Max} &
      	\multicolumn{1}{c}{Mean} &
      	\multicolumn{1}{c}{Median} &
      	\multicolumn{1}{c}{Min} &
      	\multicolumn{1}{c}{Max} &
      	\multicolumn{1}{c}{Mean} &
      	\multicolumn{1}{c}{Median} &
      	\multicolumn{1}{c}{Ngal}\\
	\tableline
11.31 & 11.47 & 11.38 &  11.37 &  0.00 &  0.29 &  0.09 &   0.07 &  3911\\
11.31 & 11.47 & 11.39 &  11.39 &  0.84 &  1.00 &  0.94 &   0.94 &  2011\\
11.47 & 11.62 & 11.53 &  11.52 &  0.00 &  0.29 &  0.09 &   0.06 &  1405\\
11.47 & 11.62 & 11.54 &  11.54 &  0.84 &  1.00 &  0.94 &   0.94 &  1869\\
11.62 & 11.78 & 11.69 &  11.68 &  0.00 &  0.29 &  0.09 &   0.07 &   364\\
11.62 & 11.78 & 11.70 &  11.69 &  0.84 &  1.00 &  0.95 &   0.95 &  1272\\

	\tableline
    \end{tabular}
%%\tablecomments{}
\end{center}
\end{table}

\begin{table}[h]
  \begin{center}
    \caption{\label{tab:PelatVd} Details of the high and low elliptical probability samples at fixed velocity dispersion}
	\begin{tabular}{ccccccccc}
	\tableline\tableline
      	\multicolumn{4}{c}{Velocity dispersion (km/s)} &
      	\multicolumn{4}{c}{Elliptical probability} &
      	\multicolumn{1}{c}{} \\
      	\multicolumn{1}{c}{Min} &
      	\multicolumn{1}{c}{Max} &
      	\multicolumn{1}{c}{Mean} &
      	\multicolumn{1}{c}{Median} &
      	\multicolumn{1}{c}{Min} &
      	\multicolumn{1}{c}{Max} &
      	\multicolumn{1}{c}{Mean} &
      	\multicolumn{1}{c}{Median} &
      	\multicolumn{1}{c}{Ngal}\\
	\tableline
210.0 & 242.3 & 223.1 &  221.7 &  0.00 &  0.76 &  0.35 &   0.37 &  4026\\
210.1 & 242.3 & 226.1 &  225.9 &  0.86 &  1.00 &  0.95 &   0.94 &  2471\\
242.3 & 274.6 & 255.3 &  254.1 &  0.00 &  0.77 &  0.40 &   0.43 &  1169\\
242.3 & 274.7 & 256.4 &  255.4 &  0.86 &  1.00 &  0.95 &   0.95 &  1925\\
274.7 & 306.4 & 286.6 &  284.9 &  0.02 &  0.74 &  0.44 &   0.47 &   307\\
274.7 & 306.8 & 287.6 &  286.0 &  0.86 &  1.00 &  0.95 &   0.95 &   871\\

	\tableline
    \end{tabular}
%%\tablecomments{}
\end{center}
\end{table}

\begin{table}[h]
  \begin{center}
    \caption{\label{tab:colatdM} Details of the high and low g-r color samples at fixed dynamical mass}
	\begin{tabular}{ccccccccc}
	\tableline\tableline
      	\multicolumn{4}{c}{log(Dynamical mass) ($\Msun$)} &
      	\multicolumn{4}{c}{g-r color} &
      	\multicolumn{1}{c}{} \\
      	\multicolumn{1}{c}{Min} &
      	\multicolumn{1}{c}{Max} &
      	\multicolumn{1}{c}{Mean} &
      	\multicolumn{1}{c}{Median} &
      	\multicolumn{1}{c}{Min} &
      	\multicolumn{1}{c}{Max} &
      	\multicolumn{1}{c}{Mean} &
      	\multicolumn{1}{c}{Median} &
      	\multicolumn{1}{c}{Ngal}\\
	\tableline
11.31 & 11.47 & 11.38 &  11.37 &  0.00 &  0.91 &  0.75 &   0.76 &  3260\\
11.31 & 11.47 & 11.39 &  11.38 &  0.96 &  1.87 &  0.99 &   0.98 &  3203\\
11.47 & 11.62 & 11.53 &  11.52 &  0.00 &  0.95 &  0.79 &   0.81 &  1628\\
11.47 & 11.62 & 11.54 &  11.53 &  0.97 &  1.79 &  1.00 &   0.99 &  1660\\
11.62 & 11.78 & 11.69 &  11.68 &  0.53 &  0.96 &  0.85 &   0.87 &   713\\
11.62 & 11.78 & 11.70 &  11.69 &  0.98 &  1.95 &  1.01 &   1.00 &   680\\

	\tableline
    \end{tabular}
%%\tablecomments{}
\end{center}
\end{table}

\begin{table}[h]
  \begin{center}
    \caption{\label{tab:colatVd} Details of the high and low g-r color samples at fixed velocity dispersion}
	\begin{tabular}{ccccccccc}
	\tableline\tableline
      	\multicolumn{4}{c}{Velocity dispersion (km/s)} &
      	\multicolumn{4}{c}{g-r color} &
      	\multicolumn{1}{c}{} \\
      	\multicolumn{1}{c}{Min} &
      	\multicolumn{1}{c}{Max} &
      	\multicolumn{1}{c}{Mean} &
      	\multicolumn{1}{c}{Median} &
      	\multicolumn{1}{c}{Min} &
      	\multicolumn{1}{c}{Max} &
      	\multicolumn{1}{c}{Mean} &
      	\multicolumn{1}{c}{Median} &
      	\multicolumn{1}{c}{Ngal}\\
	\tableline
210.0 & 242.3 & 222.8 &  221.3 &  0.00 &  0.94 &  0.87 &   0.89 &  3280\\
210.0 & 242.3 & 225.9 &  225.9 &  0.97 &  1.95 &  1.00 &   0.99 &  3228\\
242.4 & 274.6 & 254.7 &  252.9 &  0.00 &  0.95 &  0.91 &   0.92 &  1643\\
242.3 & 274.6 & 257.1 &  256.6 &  0.98 &  1.49 &  1.01 &   1.00 &  1606\\
274.6 & 306.6 & 286.1 &  284.5 &  0.44 &  0.96 &  0.93 &   0.94 &   583\\
274.7 & 306.9 & 288.0 &  286.6 &  0.99 &  1.50 &  1.02 &   1.01 &   593\\

	\tableline
    \end{tabular}
%%\tablecomments{}
\end{center}
\end{table}

\begin{table}[h]
  \begin{center}
    \caption{\label{tab:dMatVdel} Details of the high and low dynamical mass samples at fixed velocity dispersion and elliptical probability $>$ 0.6}
	\begin{tabular}{ccccccccc}
	\tableline\tableline
      	\multicolumn{4}{c}{Velocity Dispersion (km/s)} &
      	\multicolumn{4}{c}{log(Dynamical Mass) ($\Msun$)} &
      	\multicolumn{1}{c}{} \\
      	\multicolumn{1}{c}{Min} &
      	\multicolumn{1}{c}{Max} &
      	\multicolumn{1}{c}{Mean} &
      	\multicolumn{1}{c}{Median} &
      	\multicolumn{1}{c}{Min} &
      	\multicolumn{1}{c}{Max} &
      	\multicolumn{1}{c}{Mean} &
      	\multicolumn{1}{c}{Median} &
      	\multicolumn{1}{c}{Ngal}\\
	\tableline
210.3 & 230.8 & 219.3 &  218.8 & 10.68 & 11.20 & 11.11 &  11.12 &  1483\\
210.3 & 230.8 & 221.4 &  221.7 & 11.43 & 12.16 & 11.56 &  11.52 &  1483\\
230.8 & 251.3 & 239.4 &  238.7 & 10.51 & 11.28 & 11.19 &  11.20 &  1160\\
230.9 & 251.3 & 241.9 &  242.1 & 11.53 & 12.19 & 11.67 &  11.63 &  1160\\
251.4 & 271.8 & 260.0 &  259.3 & 10.88 & 11.35 & 11.26 &  11.26 &   765\\
251.4 & 271.8 & 261.3 &  260.9 & 11.62 & 12.18 & 11.75 &  11.71 &   765\\

	\tableline
    \end{tabular}
%%\tablecomments{}
\end{center}
\end{table}

\begin{table}[h]
  \begin{center}
    \caption{\label{tab:VdatdMel} Details of the high and low velocity dispersion samples at fixed dynamical mass and elliptical probability $>$ 0.6.}
	\begin{tabular}{ccccccccc}
	\tableline\tableline
      	\multicolumn{4}{c}{log(Dynamical Mass) ($\Msun$)} &
      	\multicolumn{4}{c}{Velocity Dispersion (km/s)} &
      	\multicolumn{1}{c}{} \\
      	\multicolumn{1}{c}{Min} &
      	\multicolumn{1}{c}{Max} &
      	\multicolumn{1}{c}{Mean} &
      	\multicolumn{1}{c}{Median} &
      	\multicolumn{1}{c}{Min} &
      	\multicolumn{1}{c}{Max} &
      	\multicolumn{1}{c}{Mean} &
      	\multicolumn{1}{c}{Median} &
      	\multicolumn{1}{c}{Ngal}\\
	\tableline
11.31 & 11.45 & 11.37 &  11.36 & 132.0 & 206.8 & 189.7 &  193.1 &  1483\\
11.31 & 11.45 & 11.39 &  11.39 & 246.7 & 454.5 & 268.3 &  261.9 &  1483\\
11.45 & 11.59 & 11.51 &  11.50 & 136.0 & 221.0 & 204.2 &  207.8 &  1030\\
11.45 & 11.59 & 11.52 &  11.52 & 261.4 & 386.6 & 283.9 &  278.5 &  1030\\
11.59 & 11.73 & 11.65 &  11.64 & 156.9 & 237.6 & 220.6 &  223.7 &   574\\
11.59 & 11.73 & 11.66 &  11.66 & 277.2 & 500.7 & 298.0 &  292.2 &   574\\

	\tableline
    \end{tabular}
%%\tablecomments{}
\end{center}
\end{table}

\begin{table}[h]
  \begin{center}
    \caption{\label{tab:dMatVdred} Details of the high and low dynamical mass samples at fixed velocity dispersion and $g-r >$ 0.9}
	\begin{tabular}{ccccccccc}
	\tableline\tableline
      	\multicolumn{4}{c}{Velocity Dispersion (km/s)} &
      	\multicolumn{4}{c}{log(Dynamical Mass) ($\Msun$)} &
      	\multicolumn{1}{c}{} \\
      	\multicolumn{1}{c}{Min} &
      	\multicolumn{1}{c}{Max} &
      	\multicolumn{1}{c}{Mean} &
      	\multicolumn{1}{c}{Median} &
      	\multicolumn{1}{c}{Min} &
      	\multicolumn{1}{c}{Max} &
      	\multicolumn{1}{c}{Mean} &
      	\multicolumn{1}{c}{Median} &
      	\multicolumn{1}{c}{Ngal}\\
	\tableline
210.0 & 231.5 & 219.1 &  218.4 & 10.73 & 11.19 & 11.11 &  11.12  & 1973\\
210.0 & 231.5 & 221.6 &  221.9 & 11.42 & 12.16 & 11.55 &  11.51 &  1972\\
231.5 & 253.0 & 240.3 &  239.5 & 10.71 & 11.27 & 11.18 &  11.19 &  1442\\
231.5 & 252.9 & 243.0 &  243.3 & 11.53 & 12.19 & 11.66 &  11.63 &  1442\\
253.0 & 274.4 & 261.9 &  261.3 & 10.96 & 11.34 & 11.25 &  11.25 &   872\\
253.0 & 274.4 & 263.6 &  263.7 & 11.62 & 12.20 & 11.75 &  11.71 &   872\\

	\tableline
    \end{tabular}
%%\tablecomments{}
\end{center}
\end{table}

\begin{table}[h]
  \begin{center}
    \caption{\label{tab:VdatdMred} Details of the high and low velocity dispersion samples at fixed dynamical mass and $g-r >$ 0.9.}
	\begin{tabular}{ccccccccc}
	\tableline\tableline
      	\multicolumn{4}{c}{log(Dynamical Mass) ($\Msun$)} &
      	\multicolumn{4}{c}{Velocity Dispersion (km/s)} &
      	\multicolumn{1}{c}{} \\
      	\multicolumn{1}{c}{Min} &
      	\multicolumn{1}{c}{Max} &
      	\multicolumn{1}{c}{Mean} &
      	\multicolumn{1}{c}{Median} &
      	\multicolumn{1}{c}{Min} &
      	\multicolumn{1}{c}{Max} &
      	\multicolumn{1}{c}{Mean} &
      	\multicolumn{1}{c}{Median} &
      	\multicolumn{1}{c}{Ngal}\\
	\tableline
11.31 & 11.46 & 11.37 &  11.36 & 125.5 & 203.8 & 187.0 &  190.3 &  1973\\
11.31 & 11.46 & 11.39 &  11.39 & 245.4 & 454.5 & 267.0 &  260.7 &  1972\\
11.46 & 11.60 & 11.52 &  11.51 & 129.3 & 220.0 & 202.4 &  206.1 &  1213\\
11.46 & 11.60 & 11.53 &  11.53 & 262.3 & 410.7 & 284.9 &  279.0 &  1212\\
11.60 & 11.75 & 11.66 &  11.65 & 159.5 & 237.7 & 219.9 &  223.5  &  624\\
11.60 & 11.75 & 11.67 &  11.67 & 278.4 & 500.7 & 299.5 &  293.8  &  624\\

	\tableline
    \end{tabular}
%%\tablecomments{}
\end{center}
\end{table}

\clearpage

\begin{figure}[h]

\vspace{22.2cm}
\includegraphics{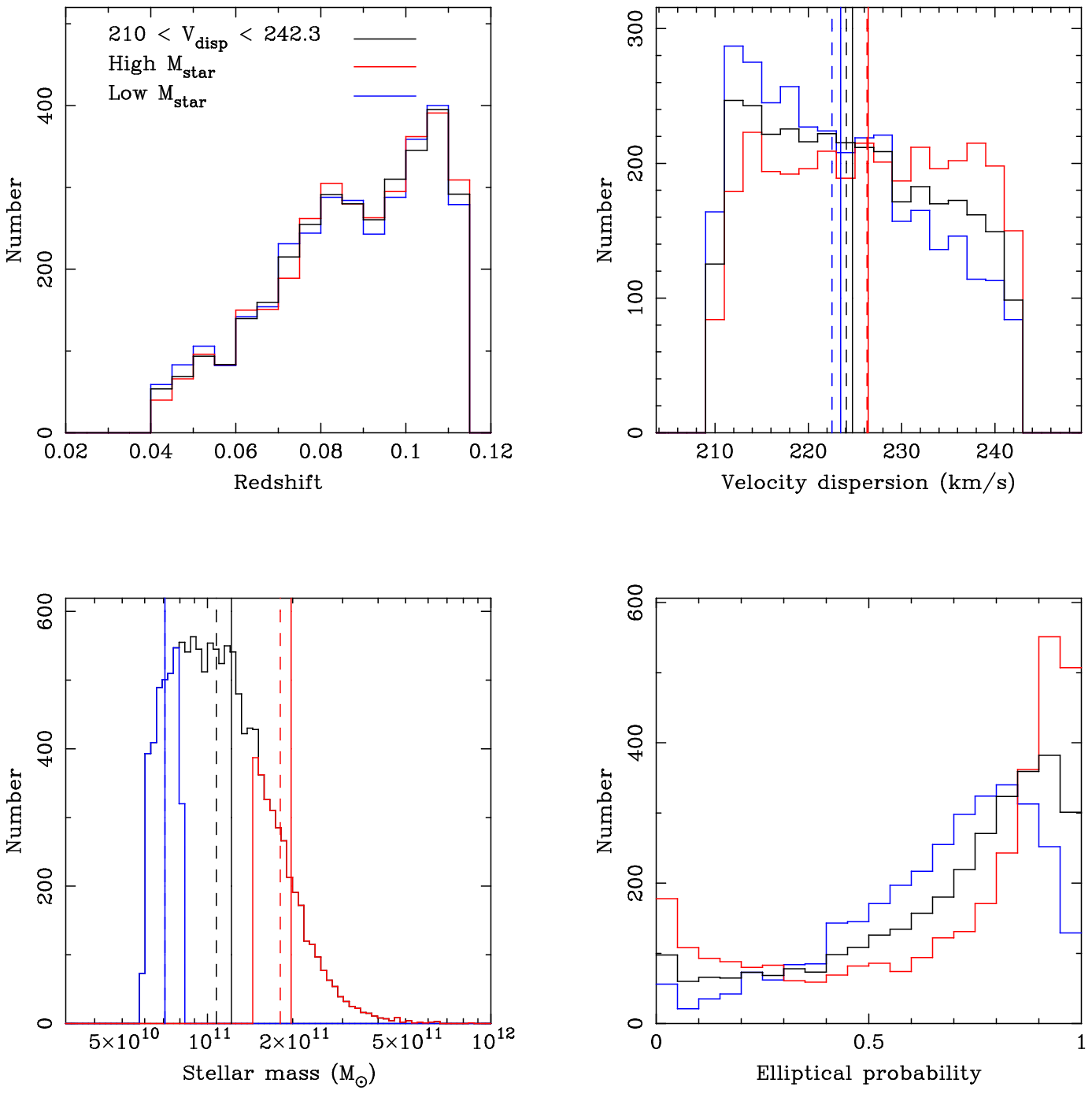}
\includegraphics{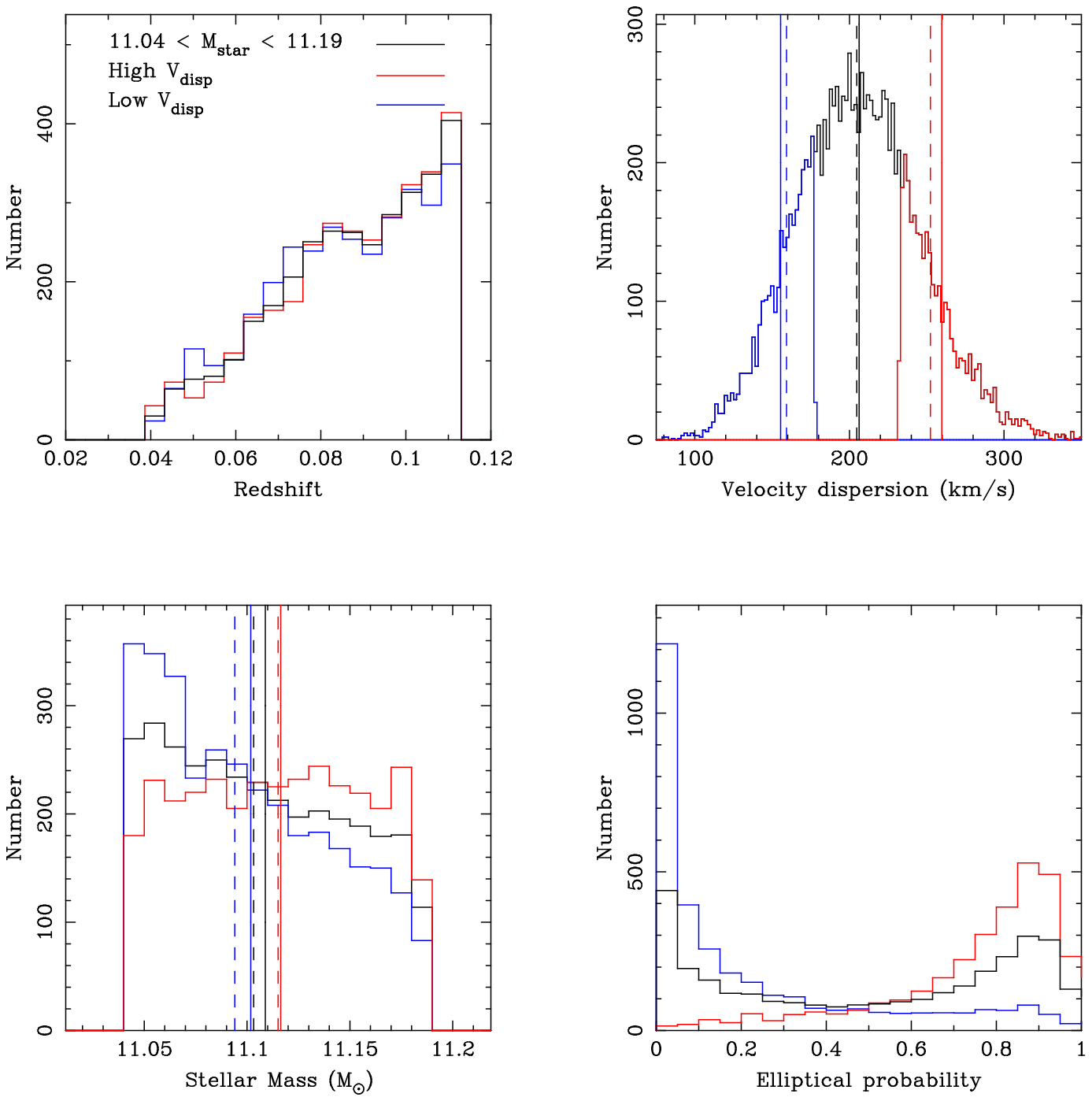}
\includegraphics{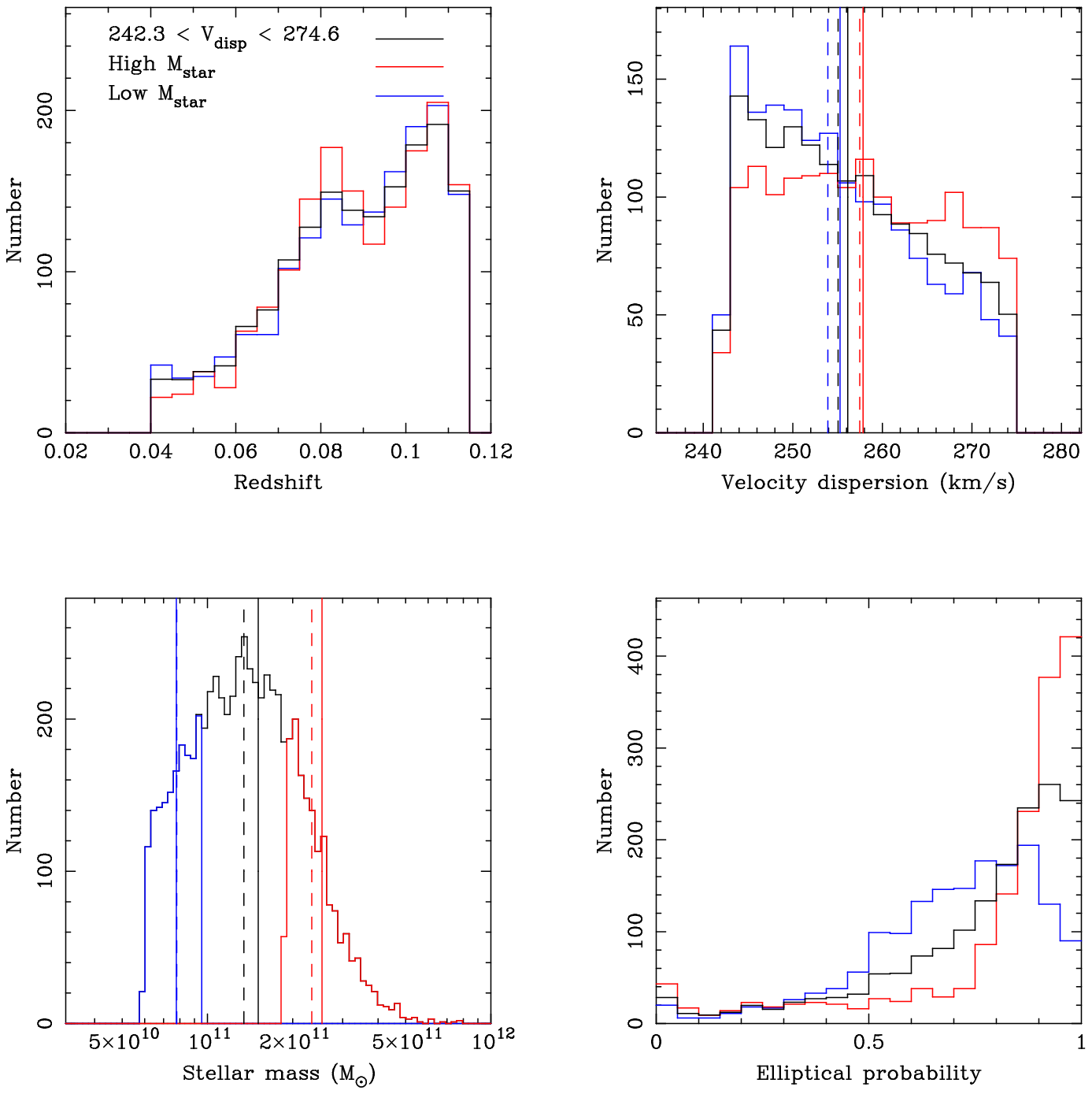}
\includegraphics{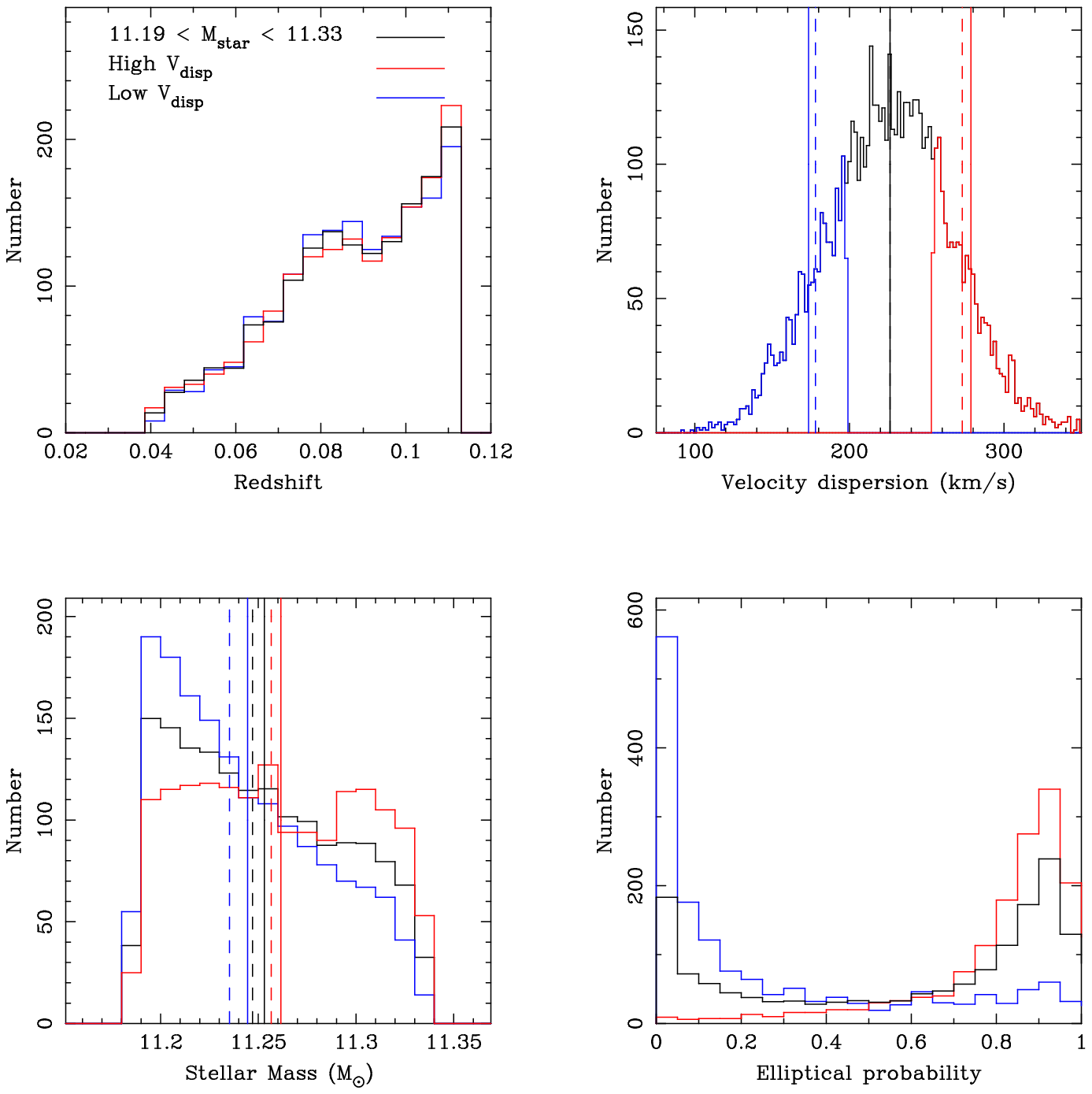}
\includegraphics{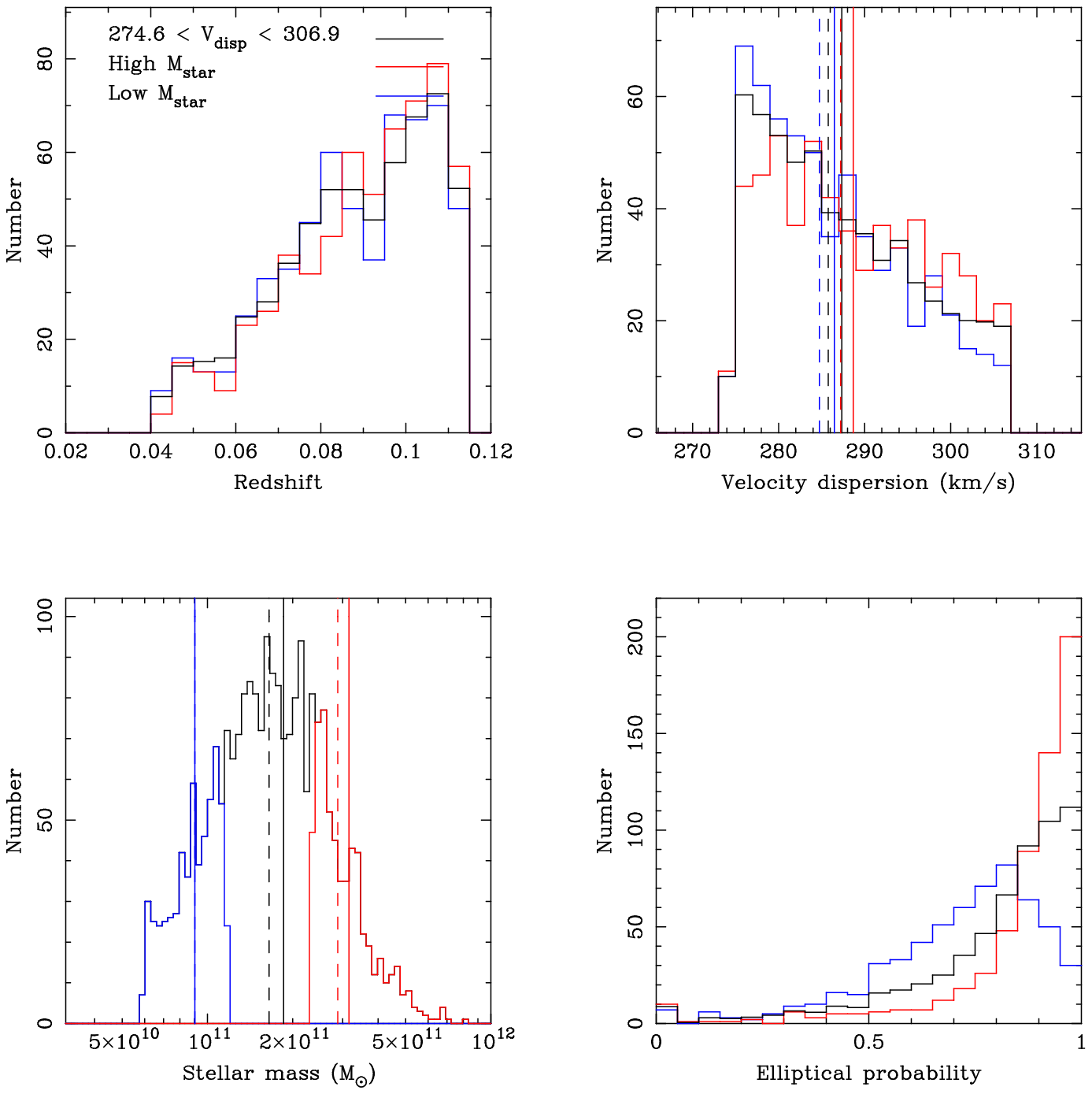}
\includegraphics{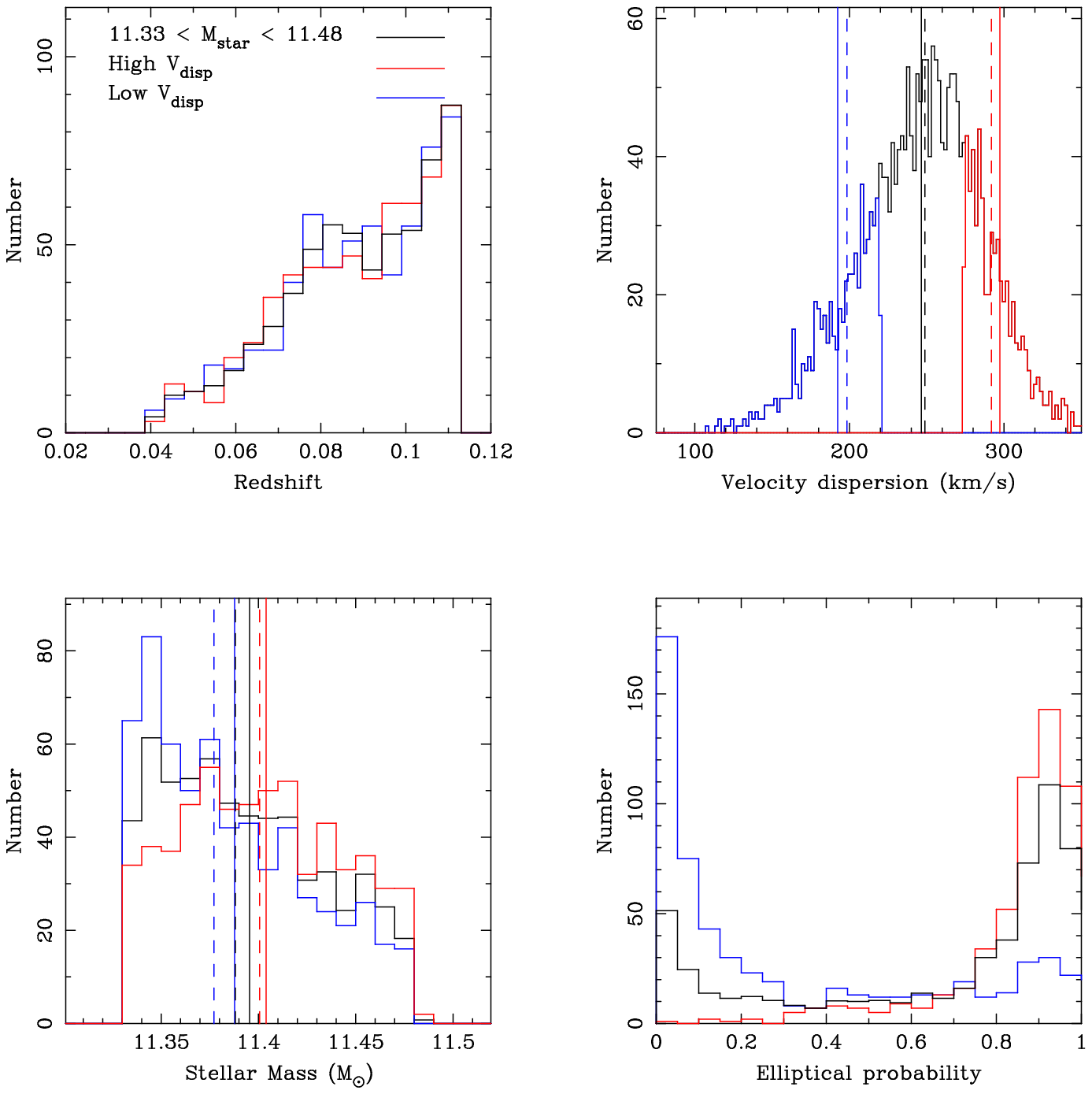}
\caption{\small The redshift, velocity dispersion, stellar mass and elliptical probability distributions for the high (red) and low (blue) stellar mass at fixed velocity dispersion samples (left) and the high and low  velocity dispersion at fixed stellar mass samples (right) The solid and dashed vertical lines in the velocity dispersion and stellar mass plots show the mean and median of the distributions respectively.
\label{fig:dSMVddist}}
\end{figure}

\begin{figure}[h]

\vspace{22.2cm}
\includegraphics{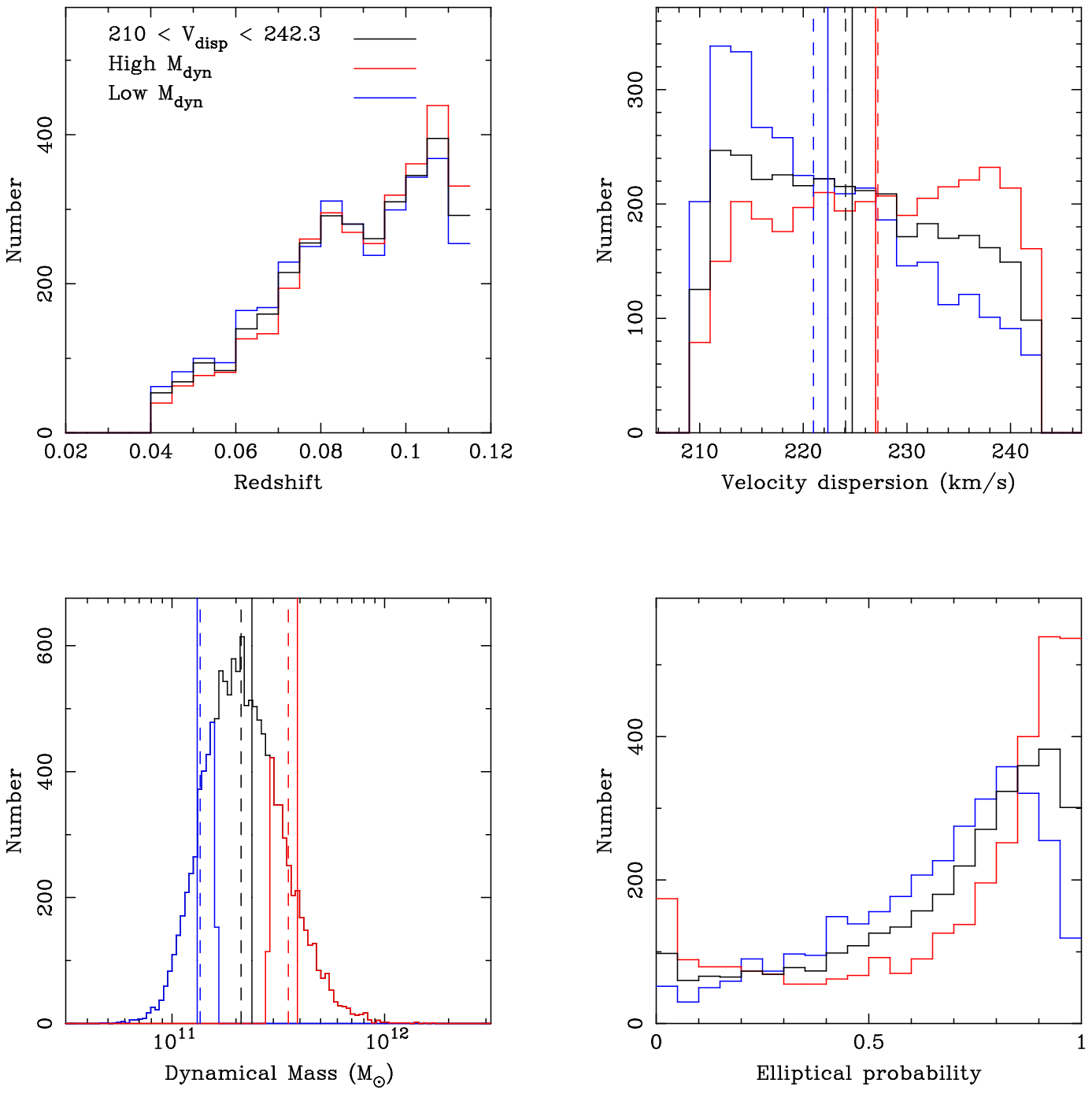}
\includegraphics{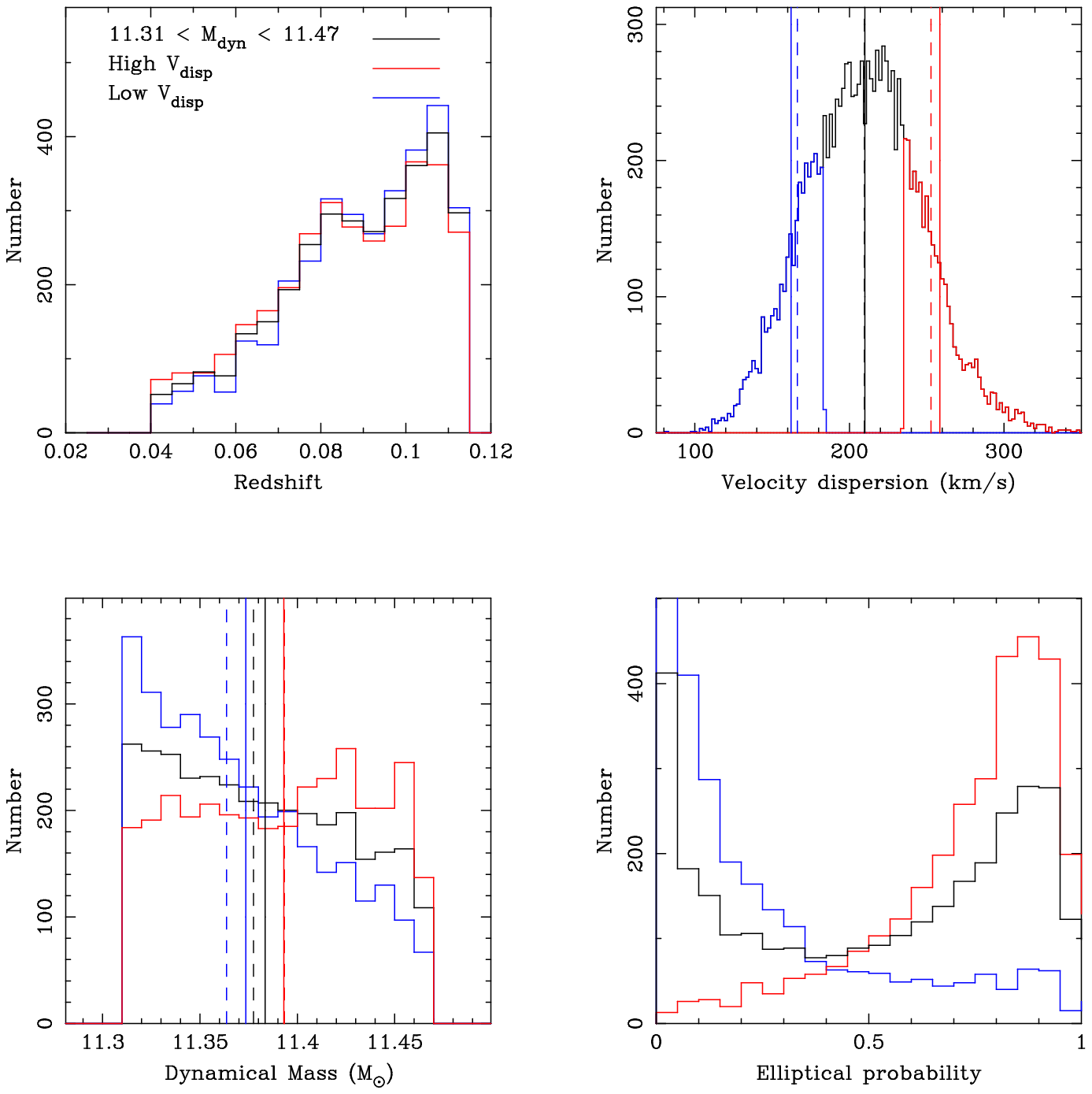}
\includegraphics{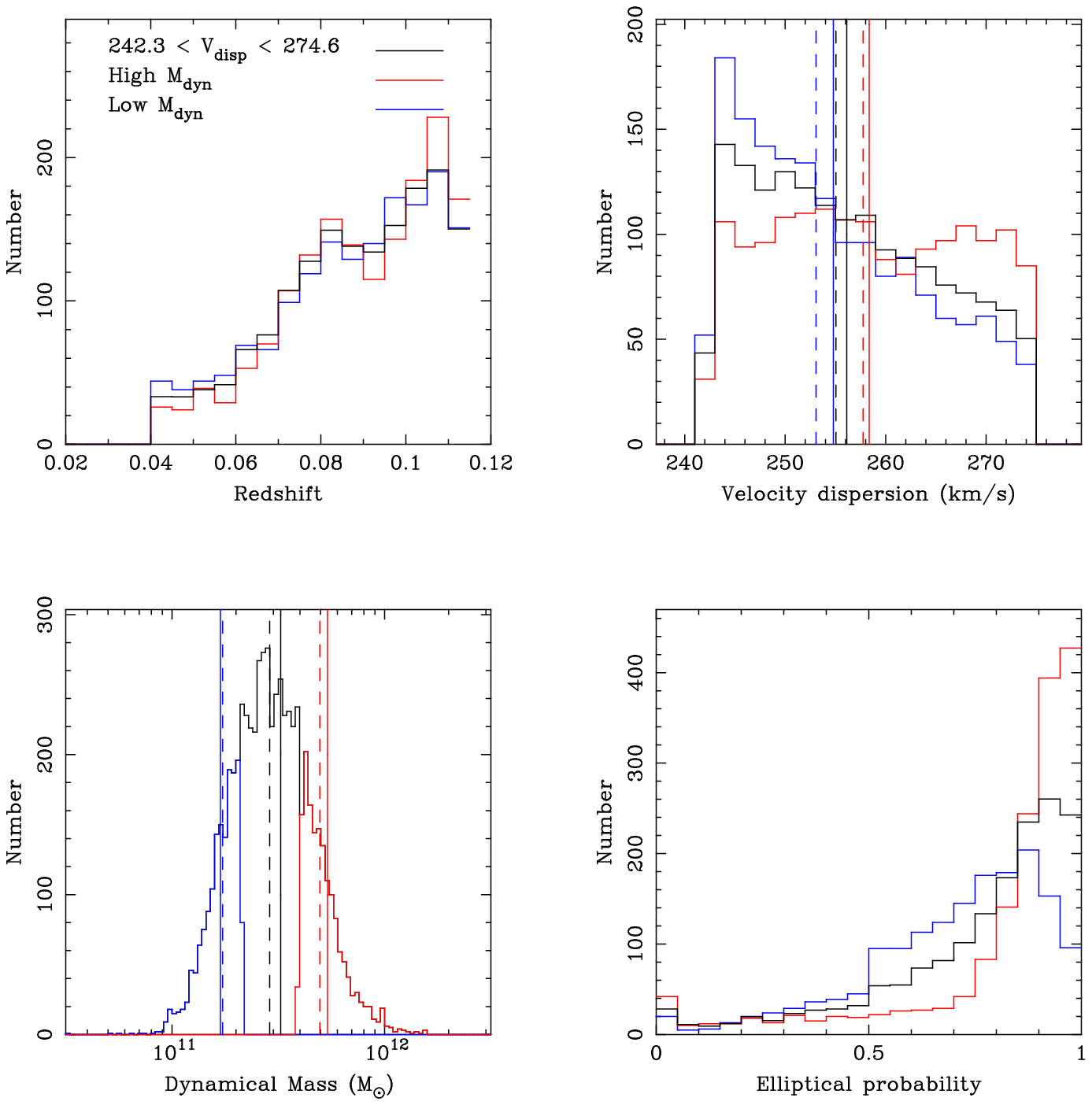}
\includegraphics{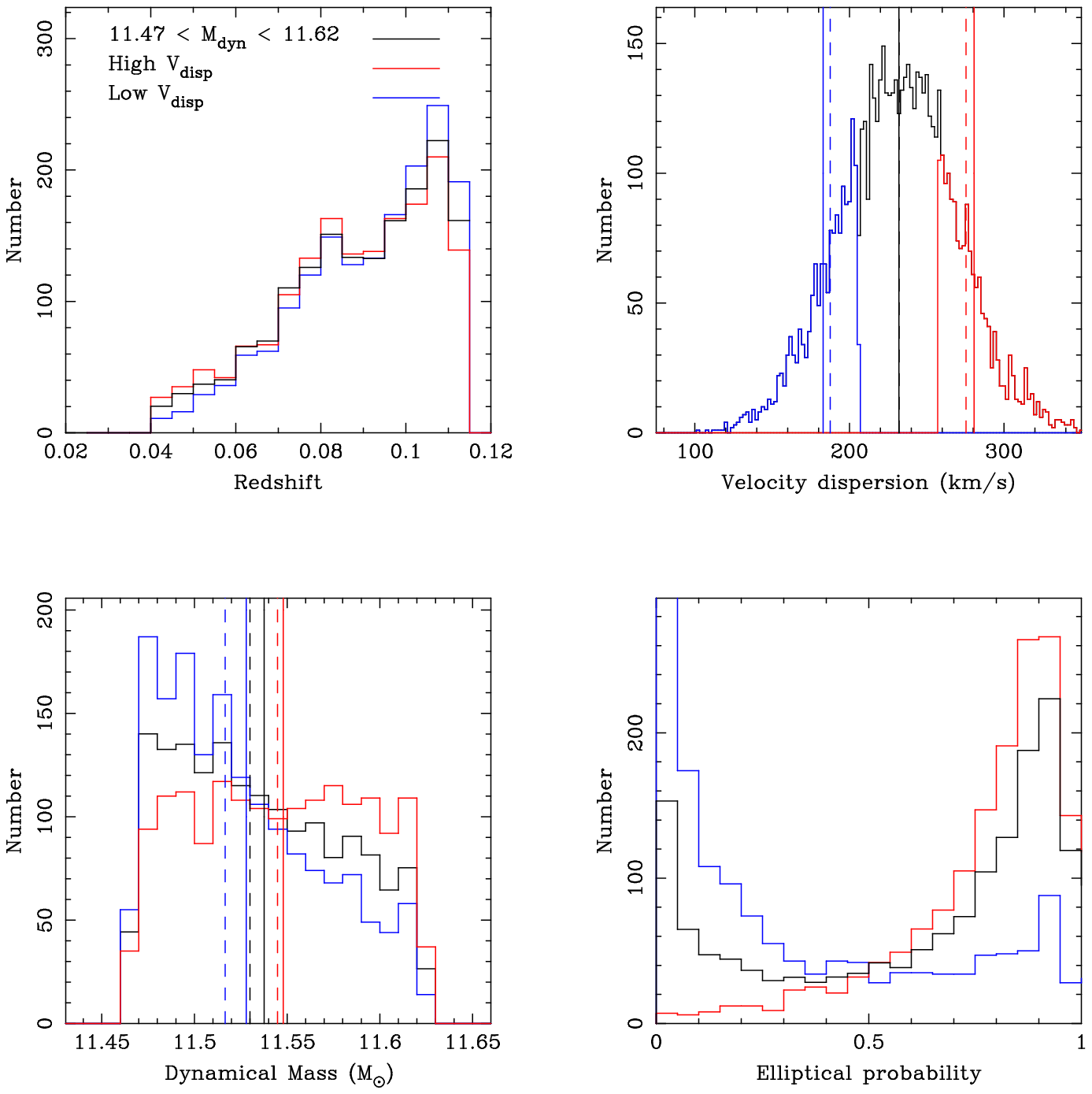}
\includegraphics{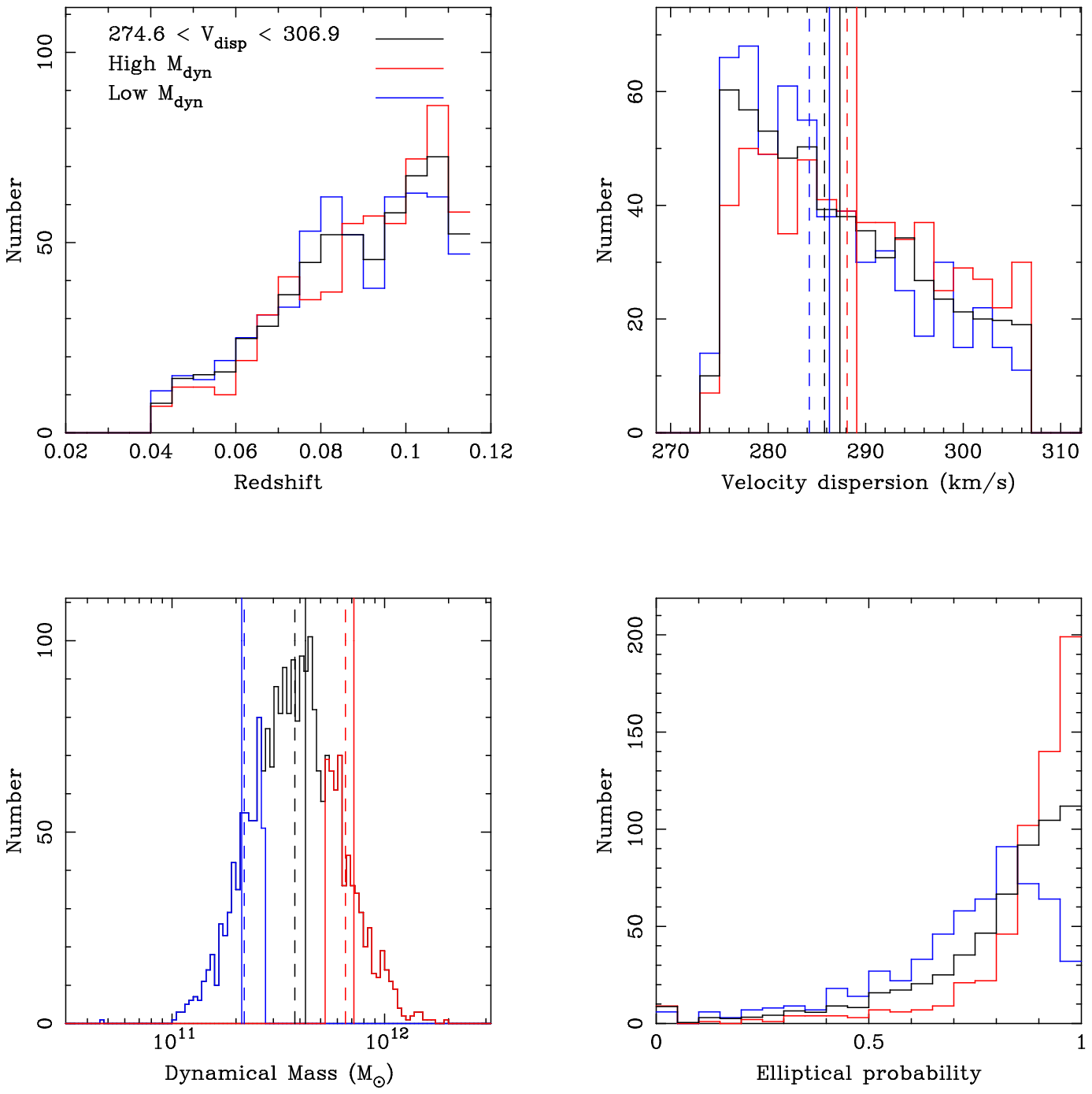}
\includegraphics{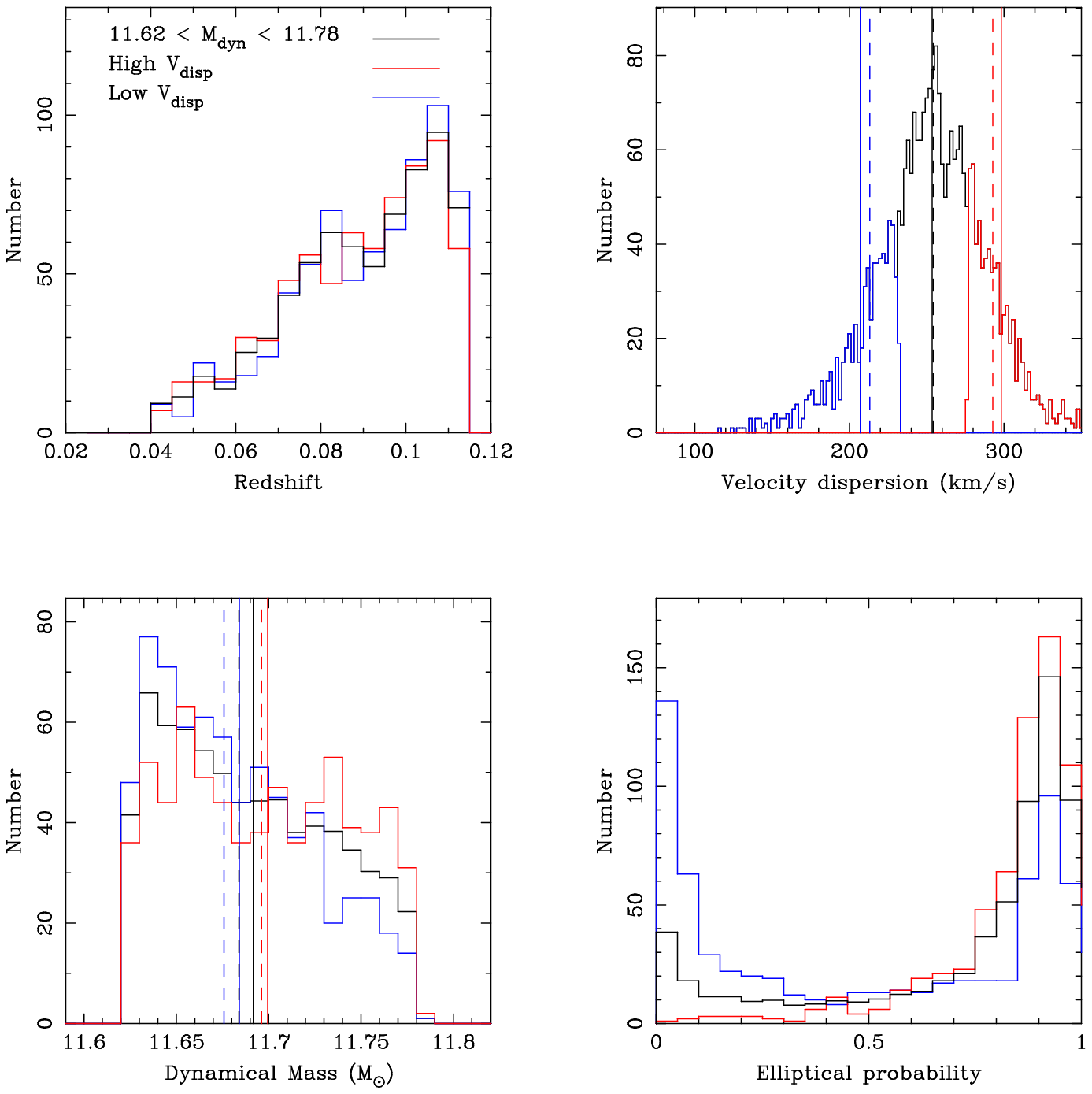}
\caption{\small The redshift, velocity dispersion, dynamical mass and elliptical probability distributions for the high (red) and low (blue) dynamical mass at fixed velocity dispersion samples (left) and the high and low  velocity dispersion at fixed dynamical mass samples (right). The solid and dashed vertical lines in the velocity dispersion and stellar mass plots show the mean and median of the distributions respectively.
\label{fig:ddMVddist}}
\end{figure}

\begin{figure}[h]

\vspace{22.2cm}
\includegraphics{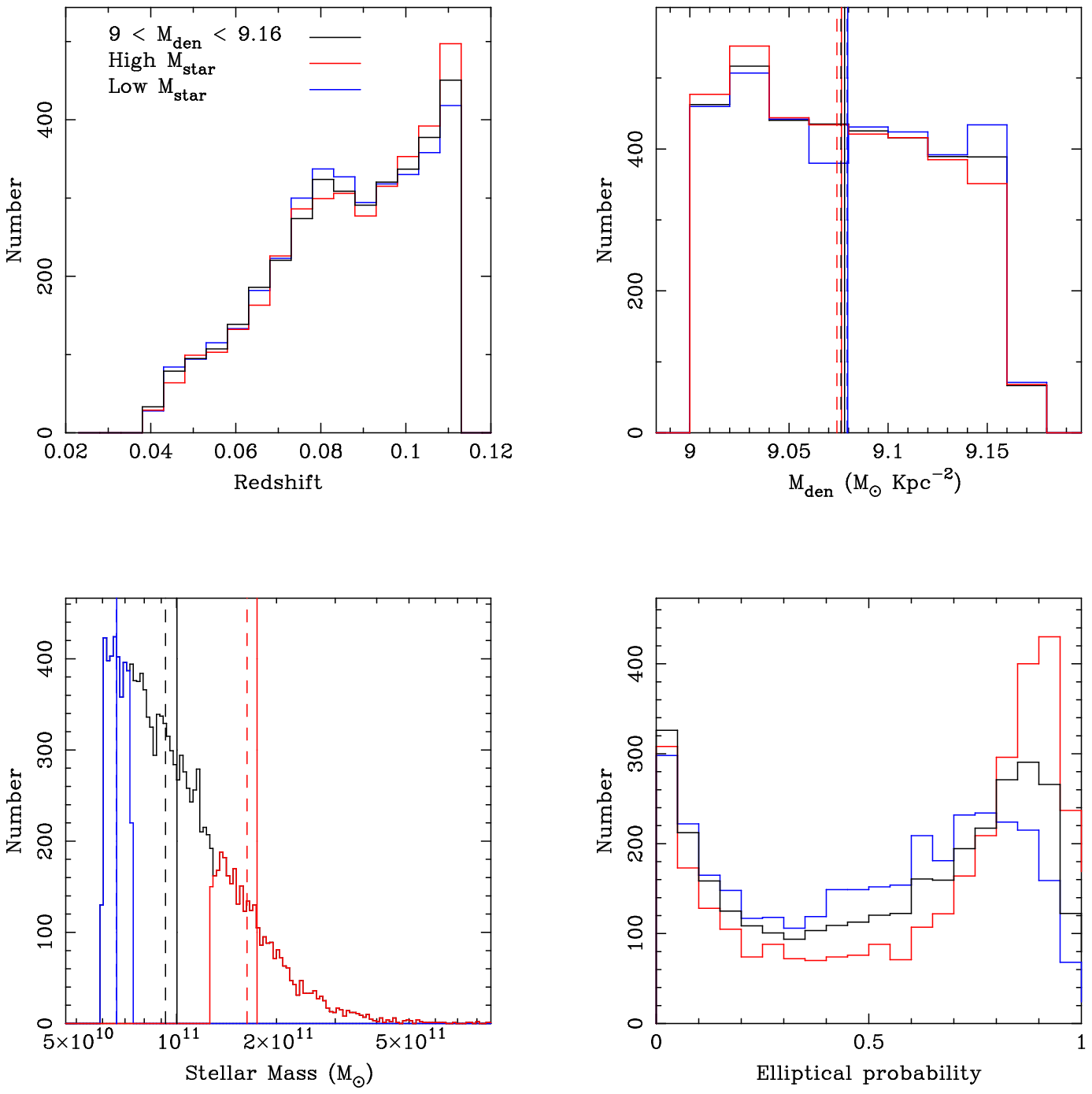}
\includegraphics{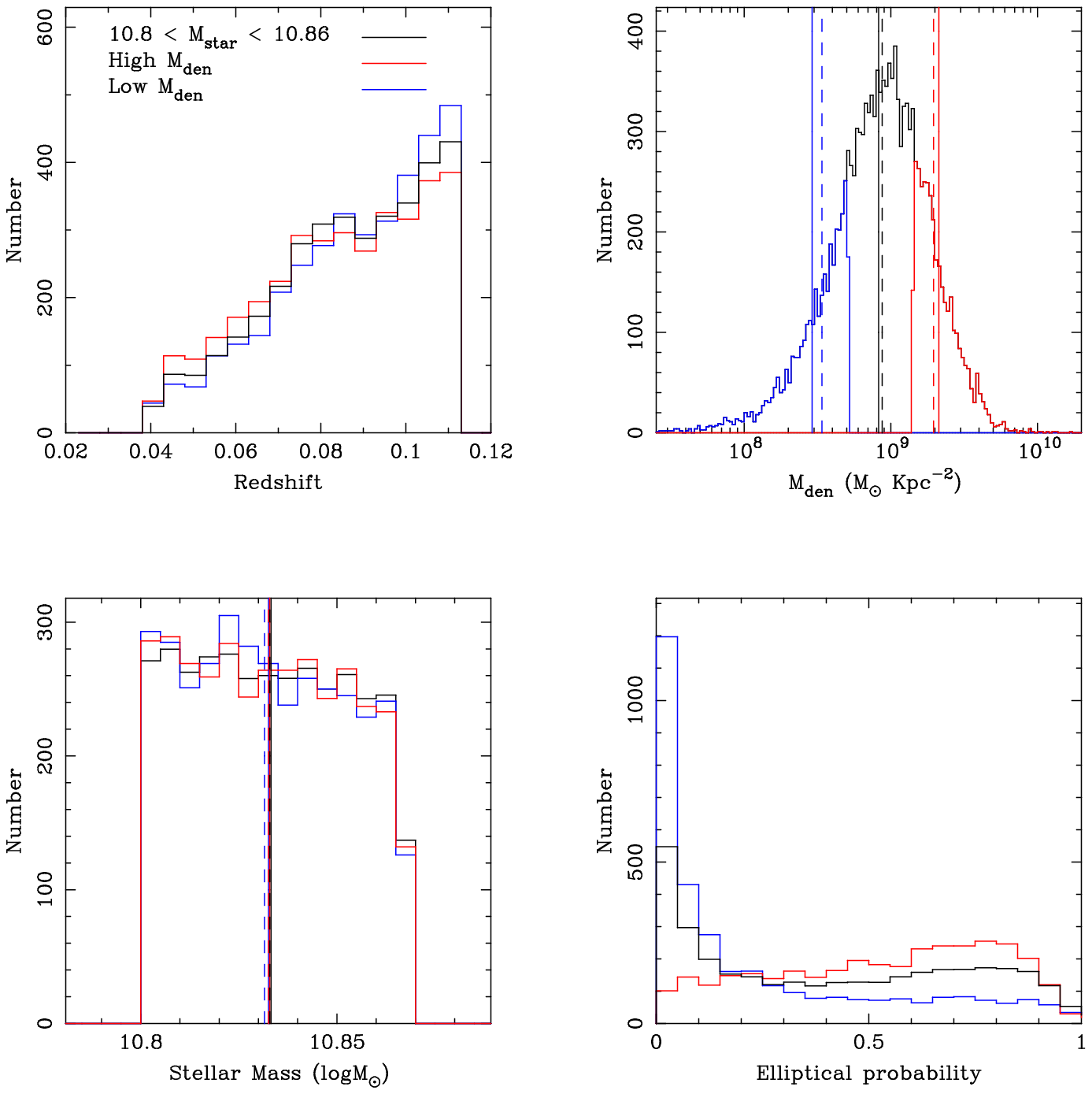}
\includegraphics{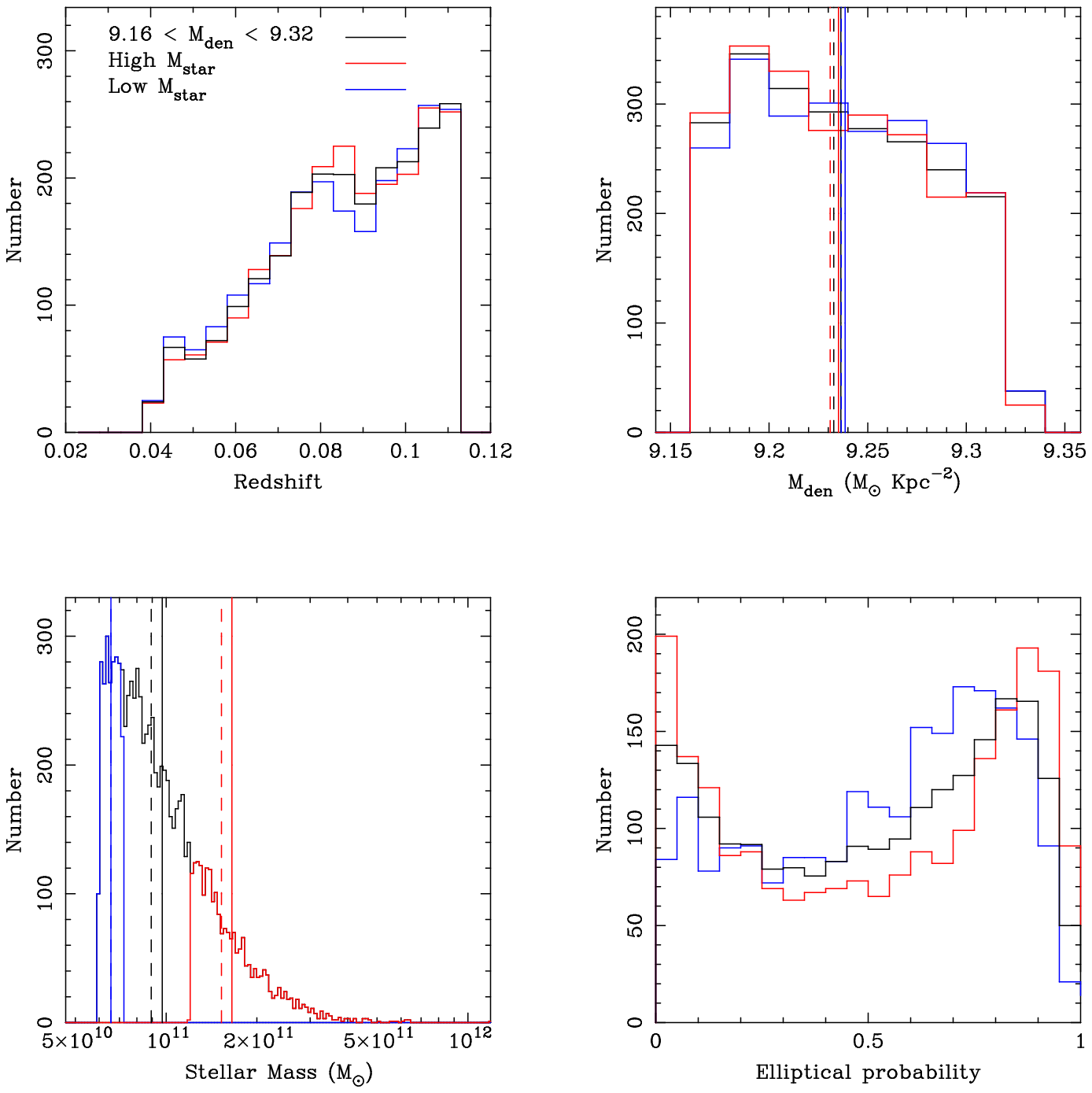}
\includegraphics{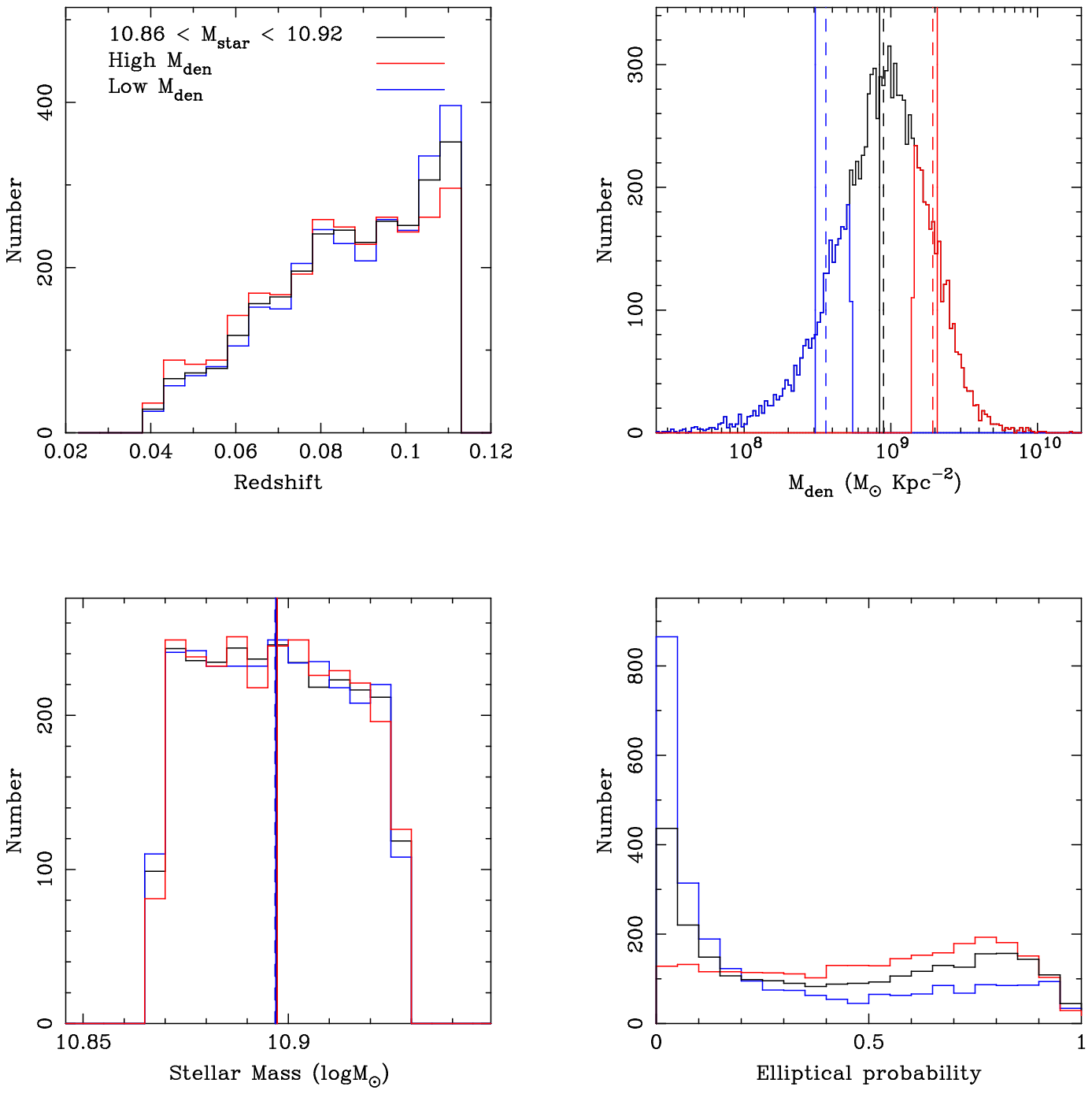}
\includegraphics{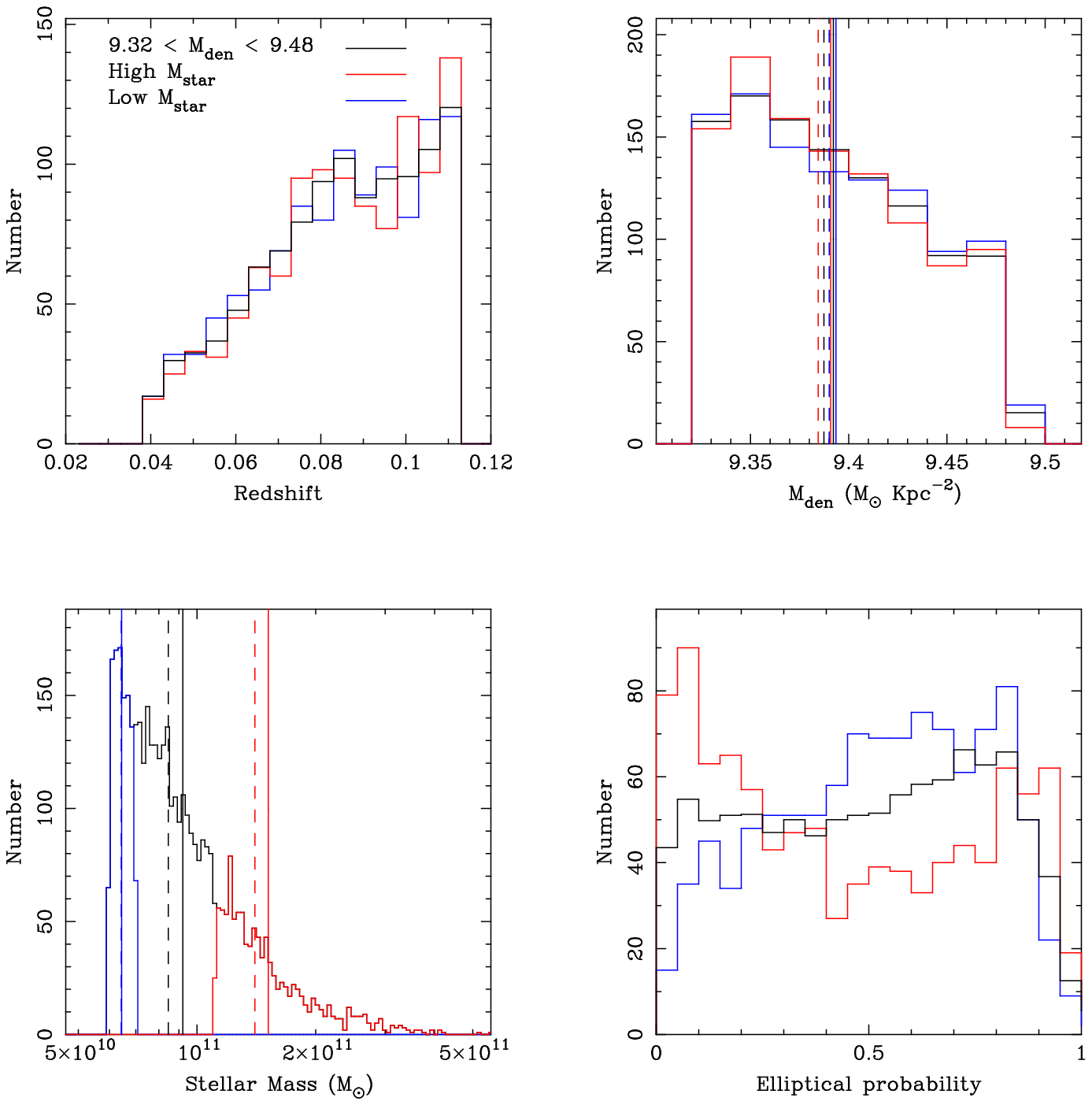}
\includegraphics{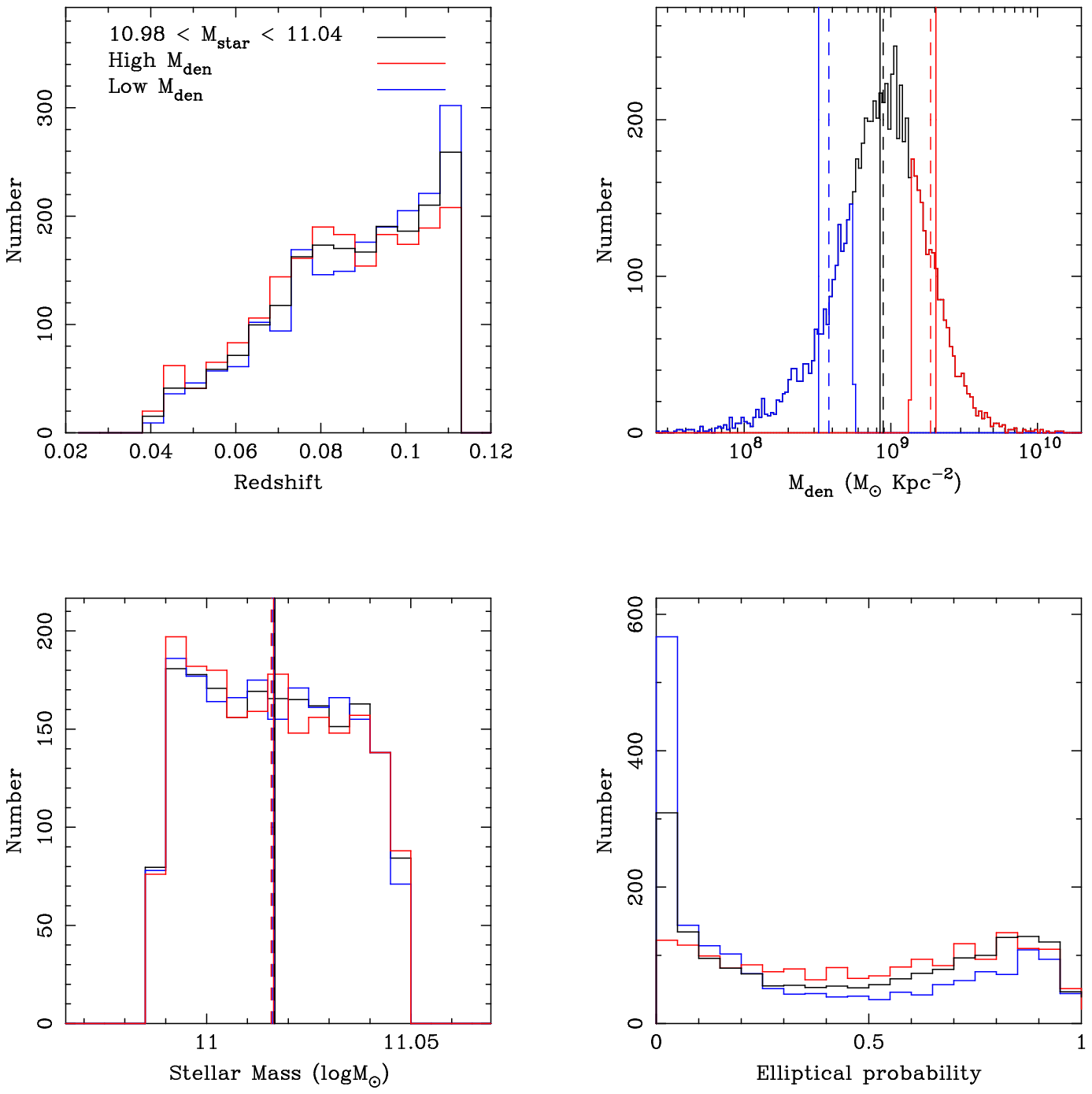}
\caption{\small The redshift, surface mass density, stellar mass and elliptical probability distributions for the high (red) and low (blue) stellar mass at fixed surface mass density samples (left) and the high and low surface mass density at fixed stellar mass samples (right). The solid and dashed vertical lines in the surface mass density and stellar mass plots show the mean and median of the distributions respectively.
\label{fig:dMdendist}}
\end{figure}

\begin{figure}[h]

\vspace{22.2cm}
\includegraphics{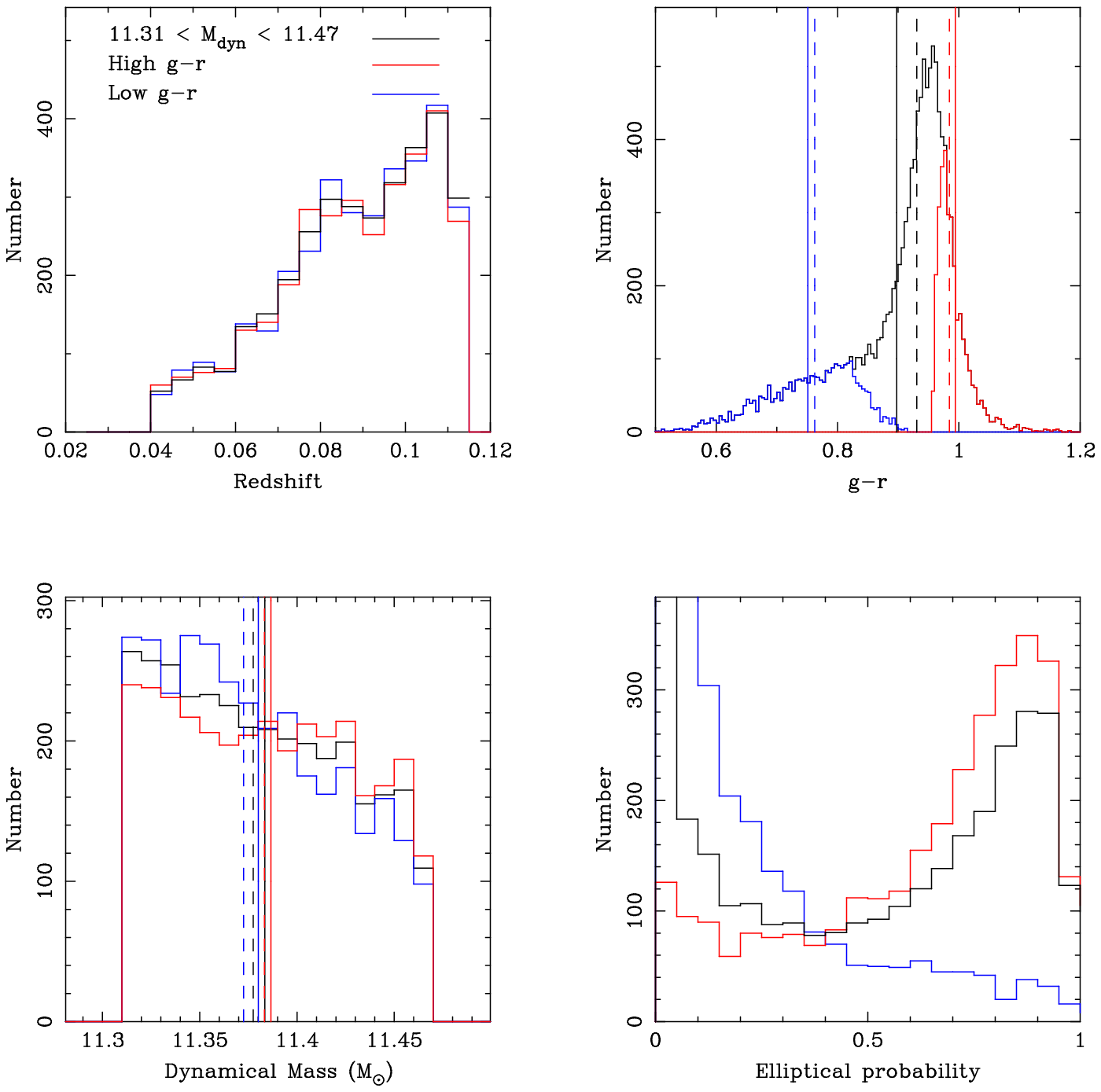}
\includegraphics{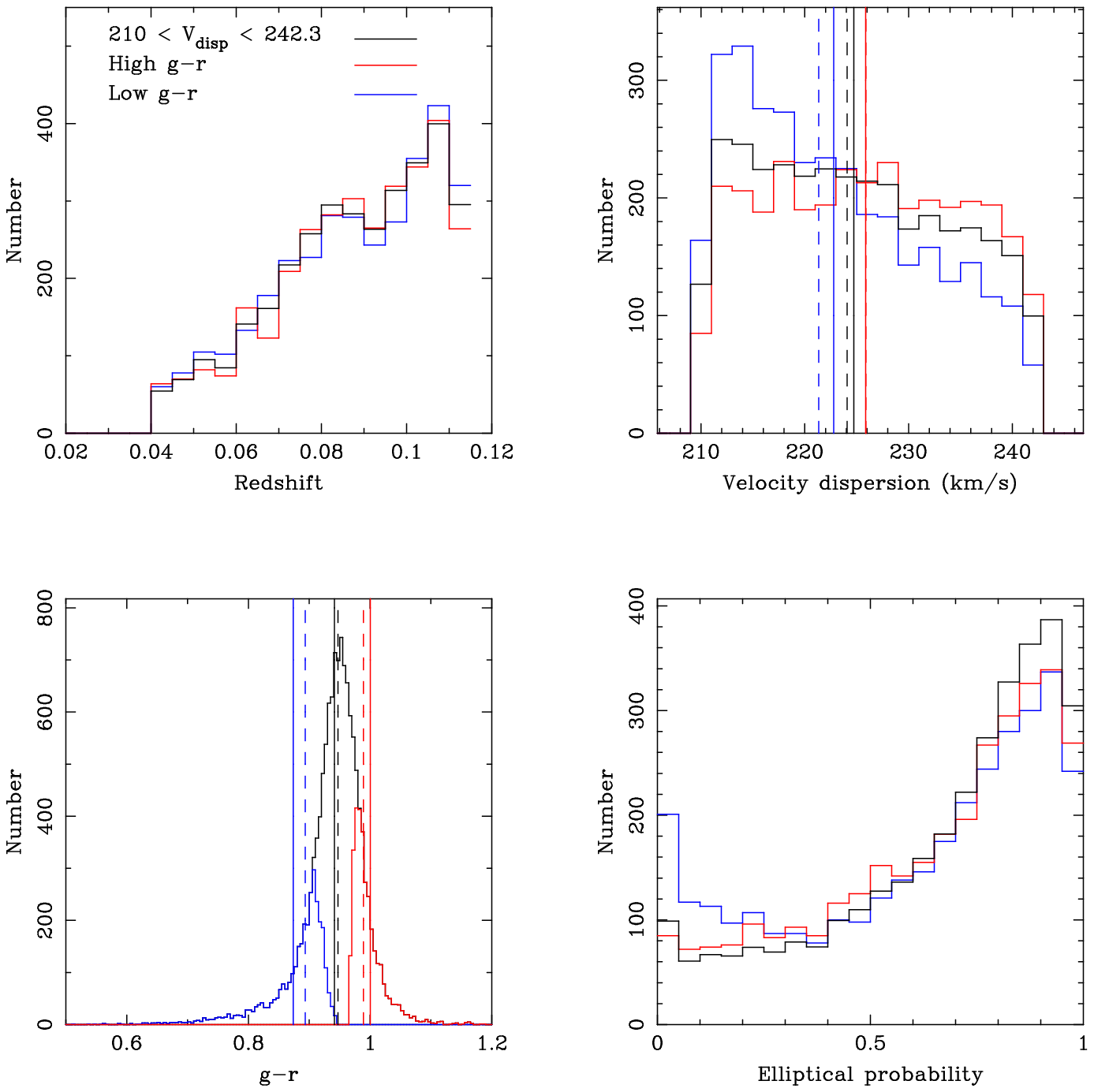}
\includegraphics{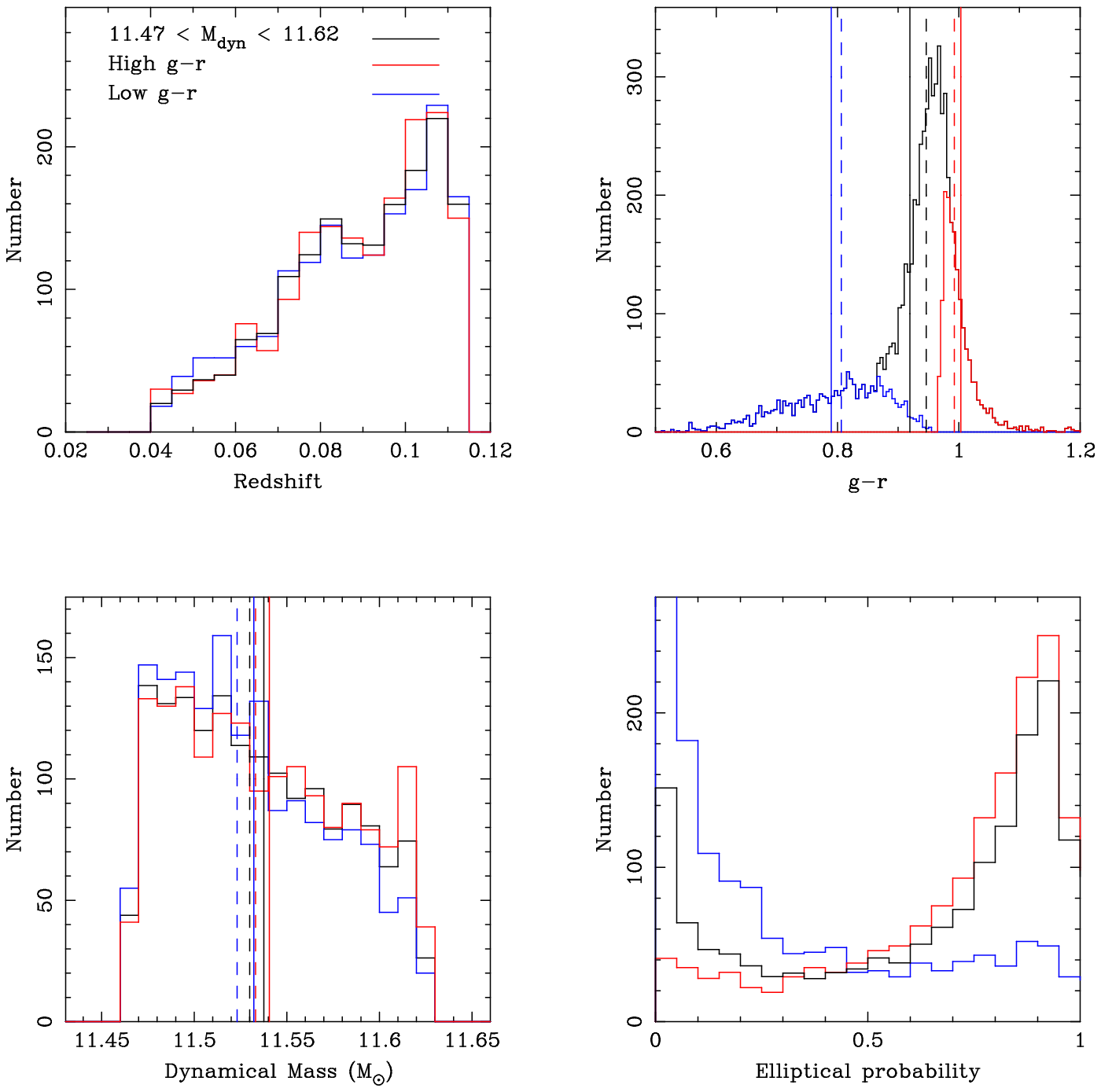}
\includegraphics{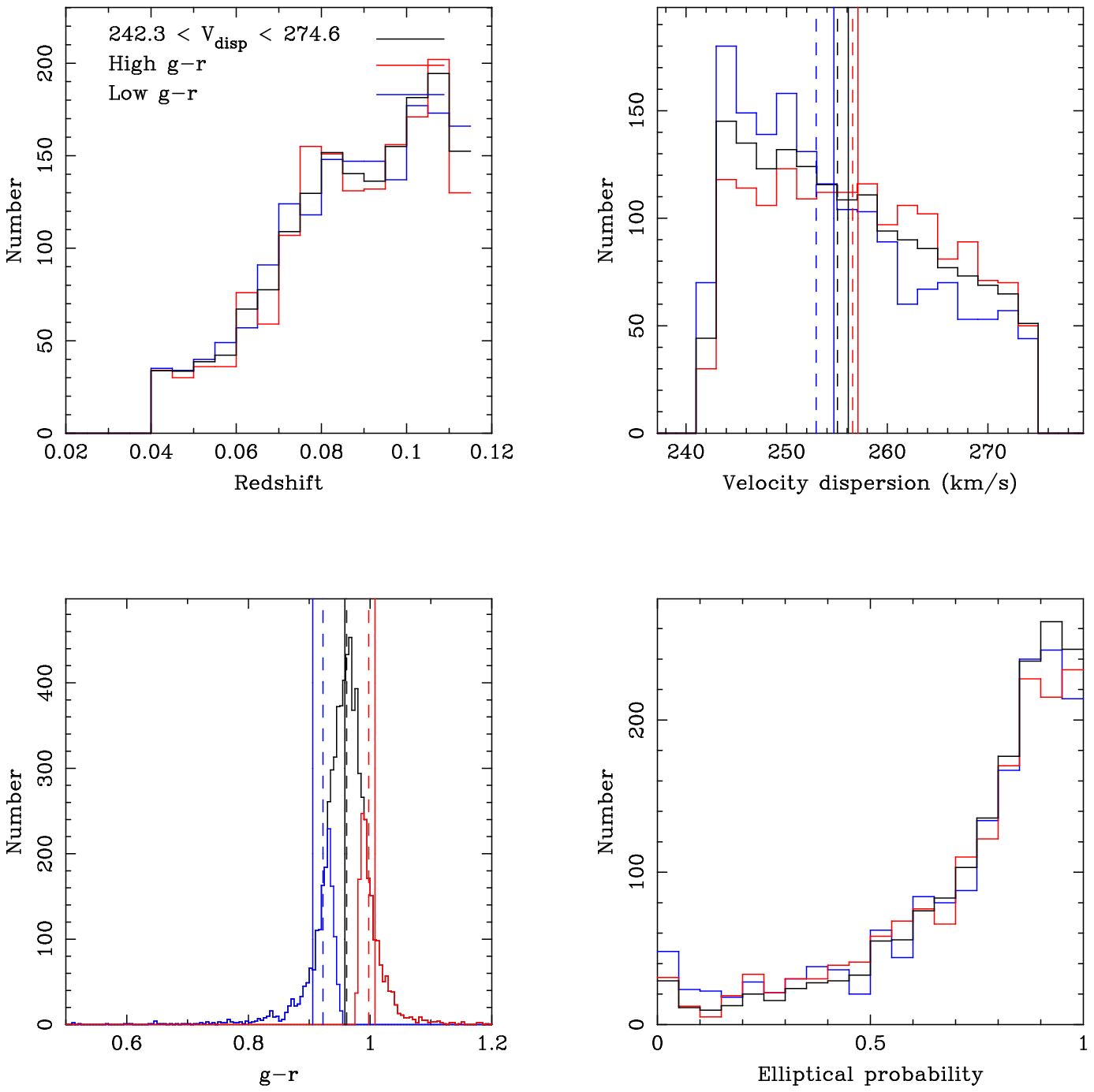}
\includegraphics{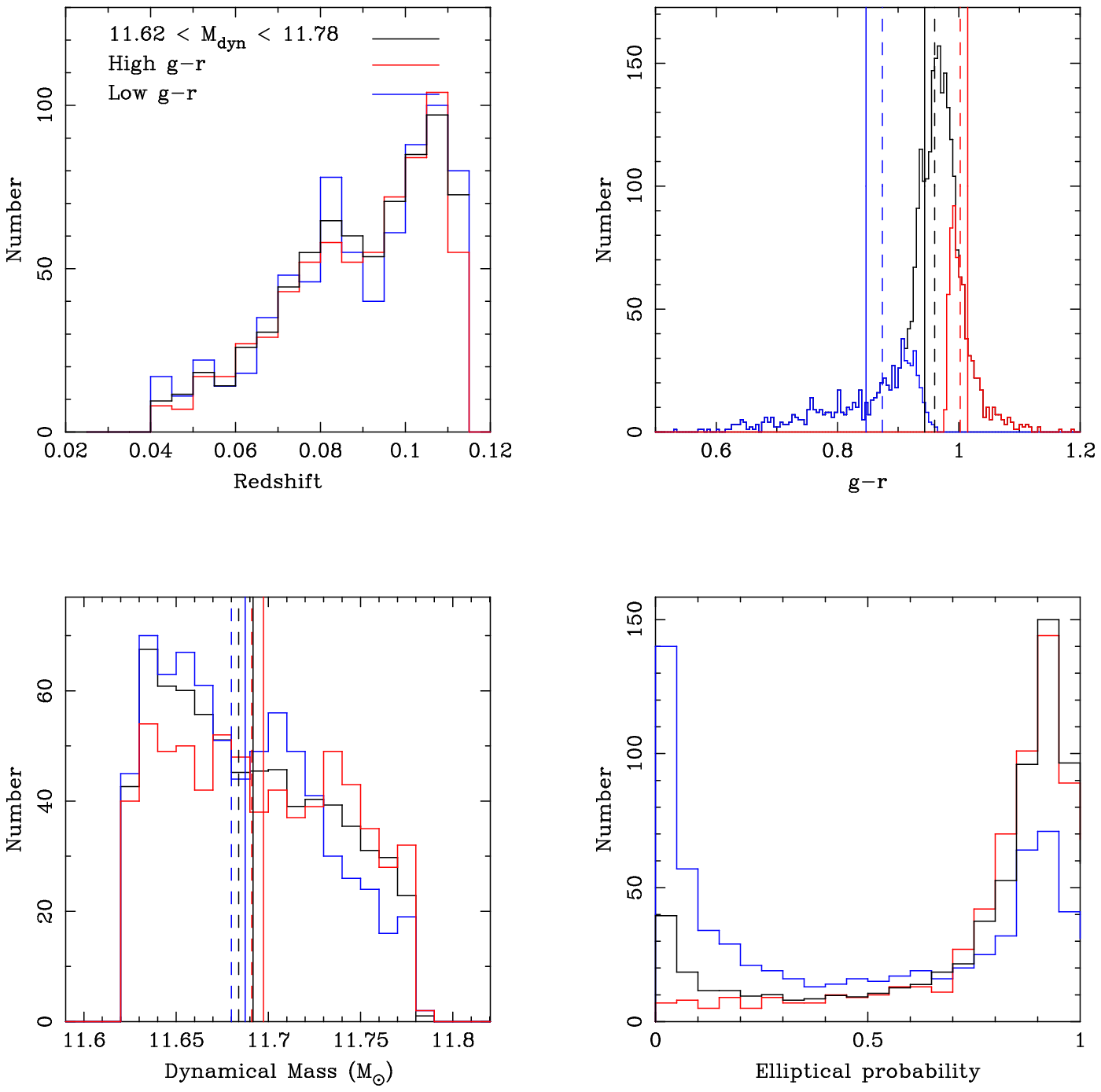}
\includegraphics{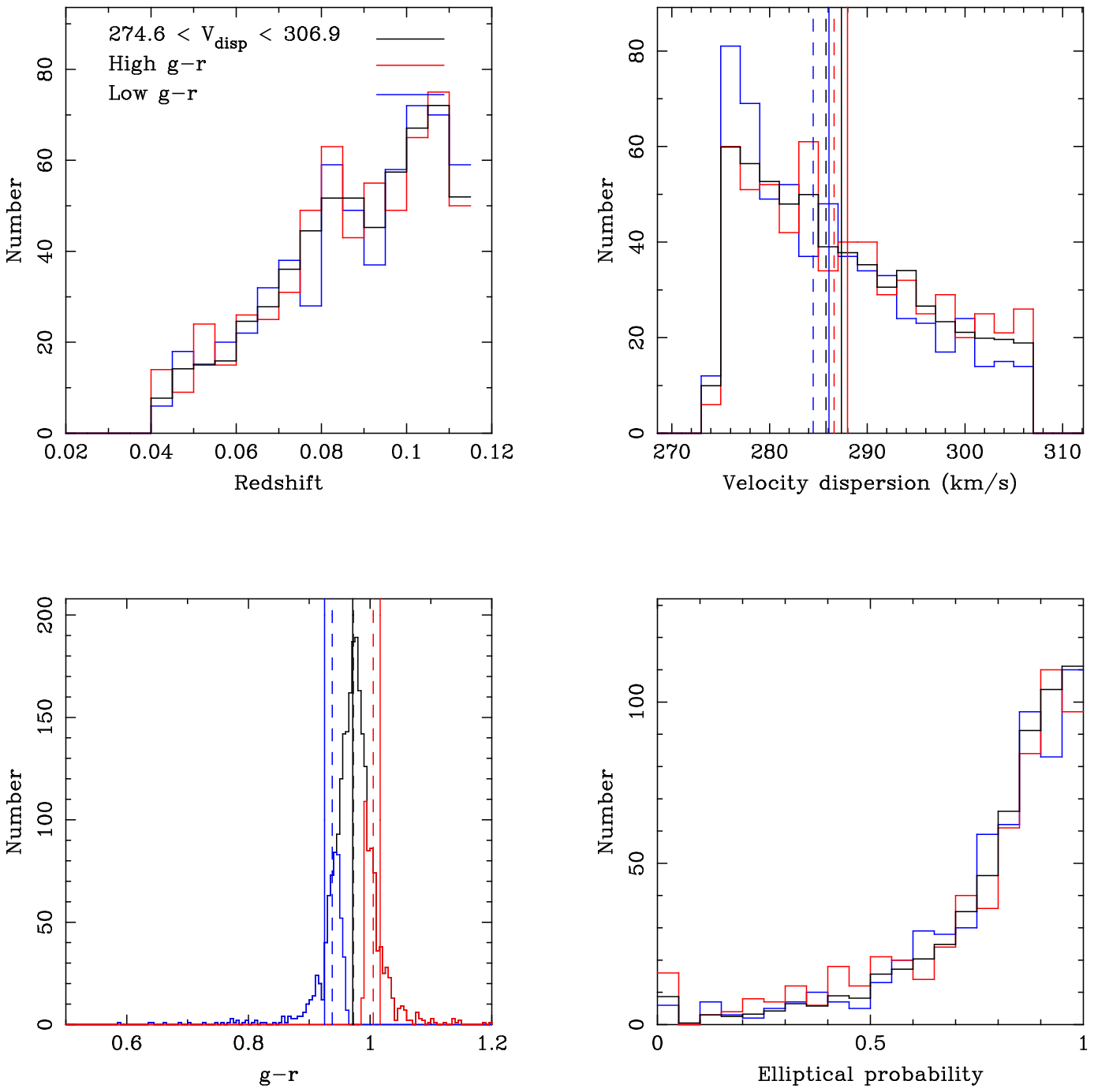}
\caption{\small The redshift, $g-r$ color, dynamical mass and elliptical probability distributions for the high (red) and low (blue) $g-r$ color at fixed dynamical mass samples (left). The redshift, velocity dispersion, $g-r$ color, and elliptical probability distributions for the high (red) and low (blue) $g-r$ color at fixed velocity dispersion samples (right). The solid and dashed vertical lines show the mean and median of the distributions respectively.
\label{fig:dcoldist}}
\end{figure}

\begin{figure}[h]

\vspace{22.2cm}
\includegraphics{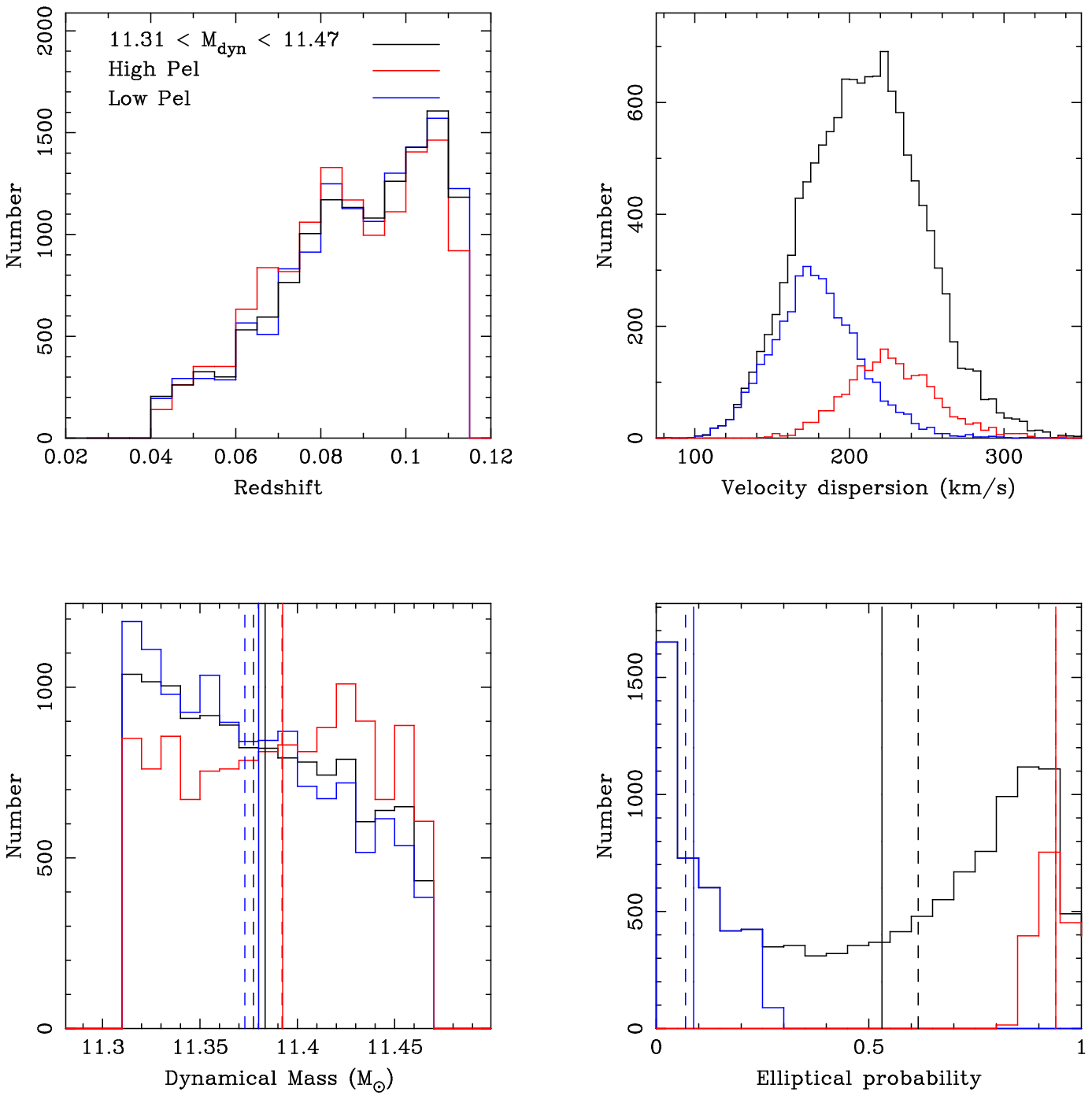}
\includegraphics{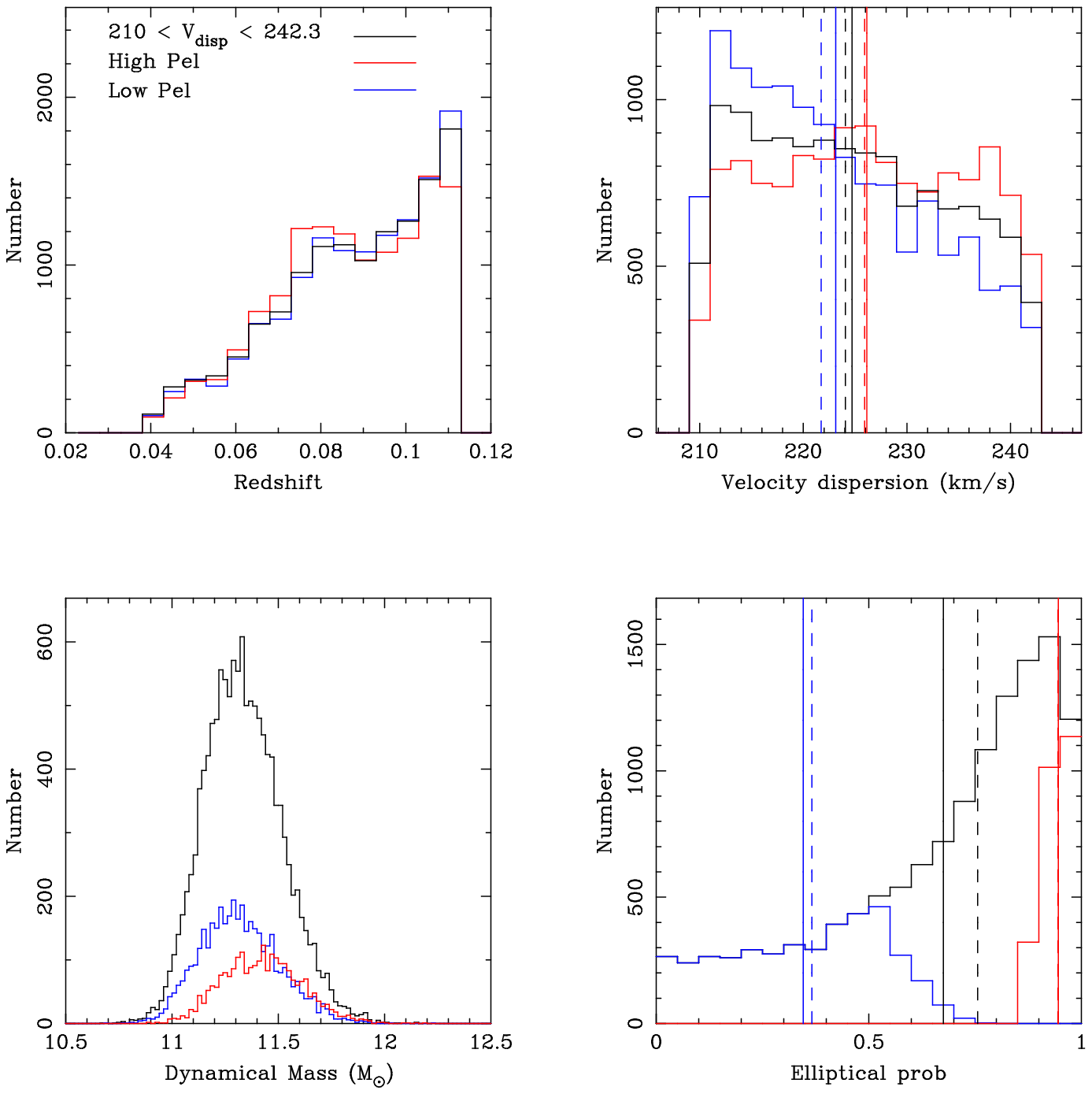}
\includegraphics{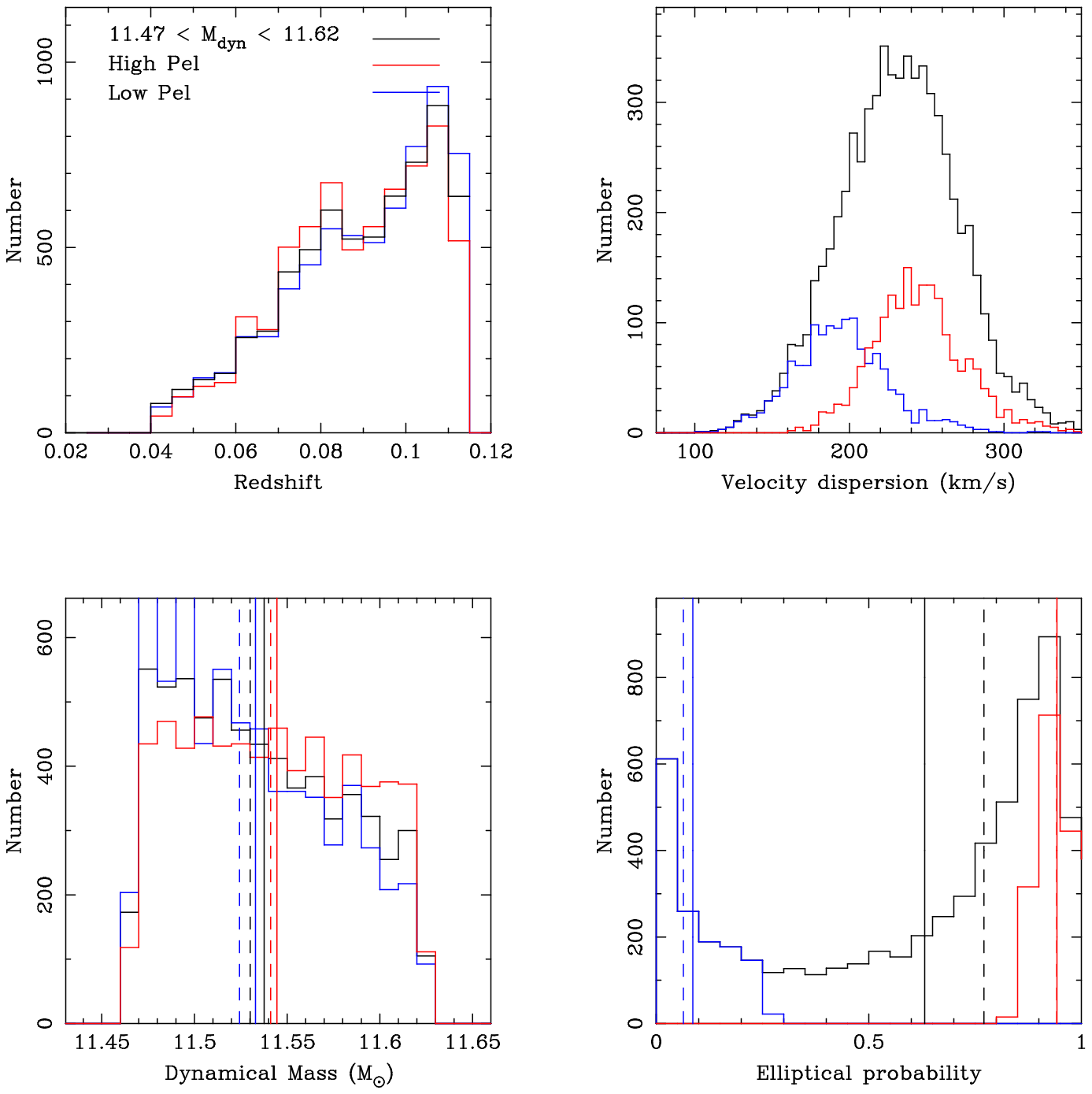}
\includegraphics{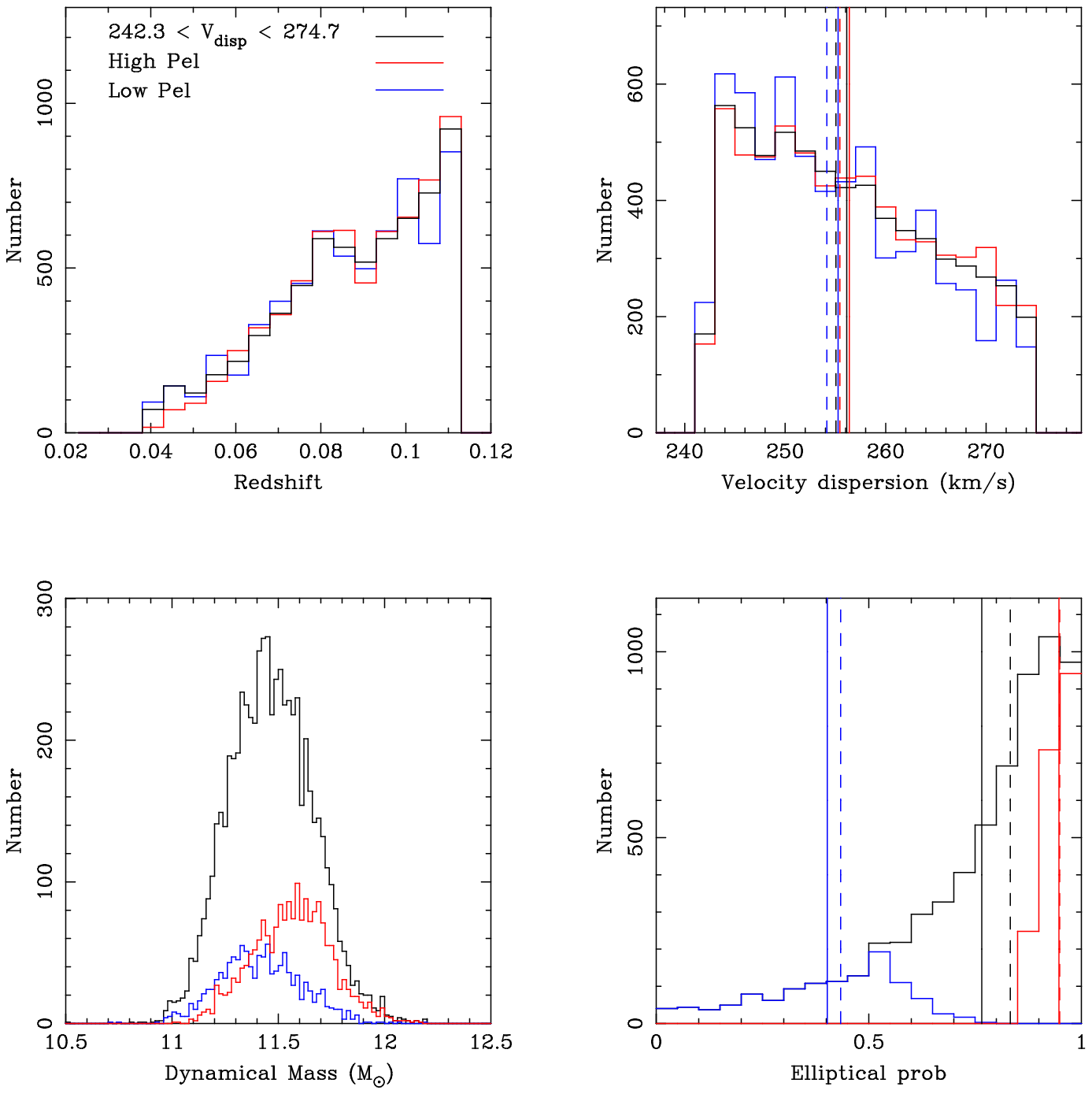}
\includegraphics{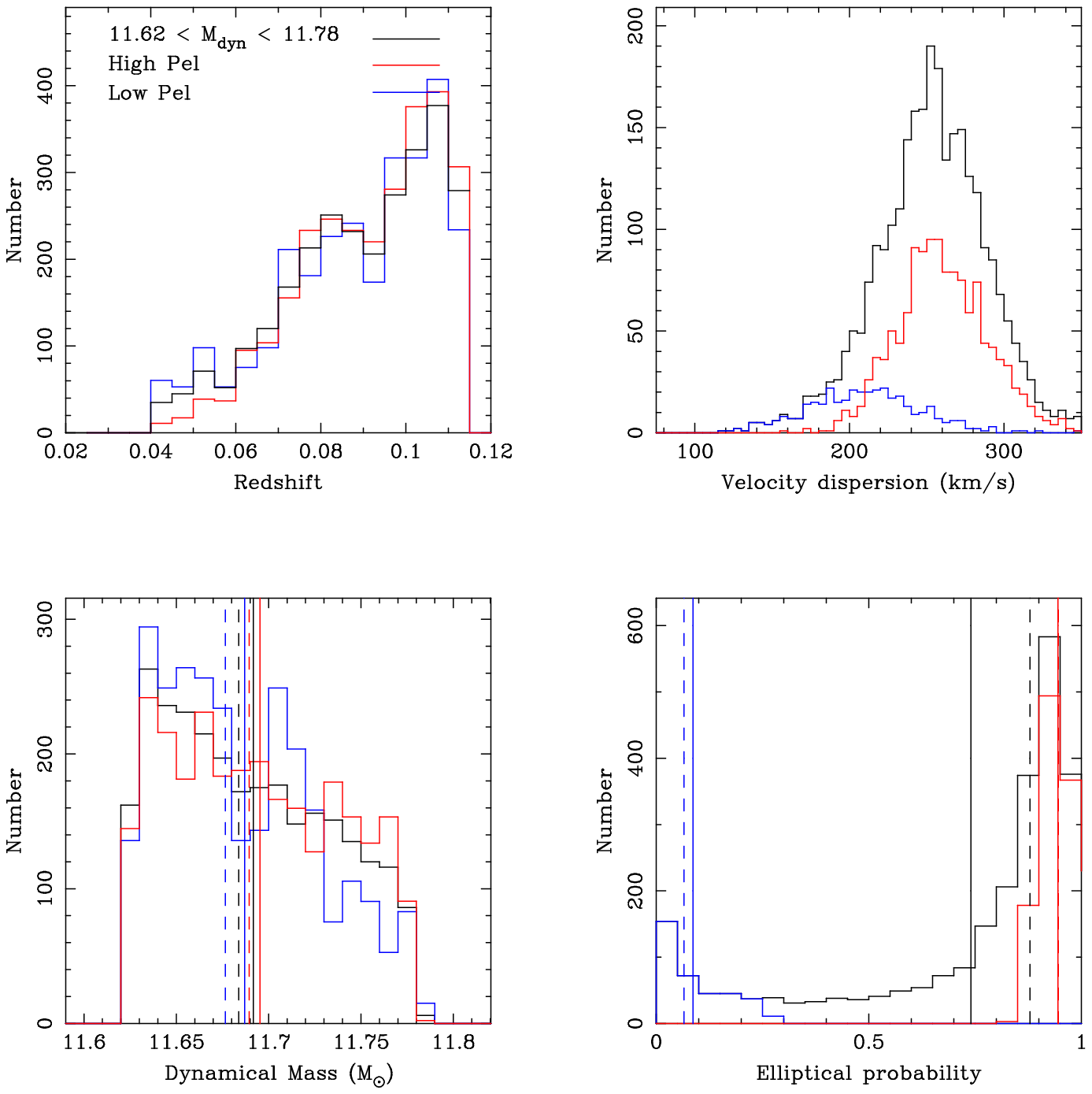}
\includegraphics{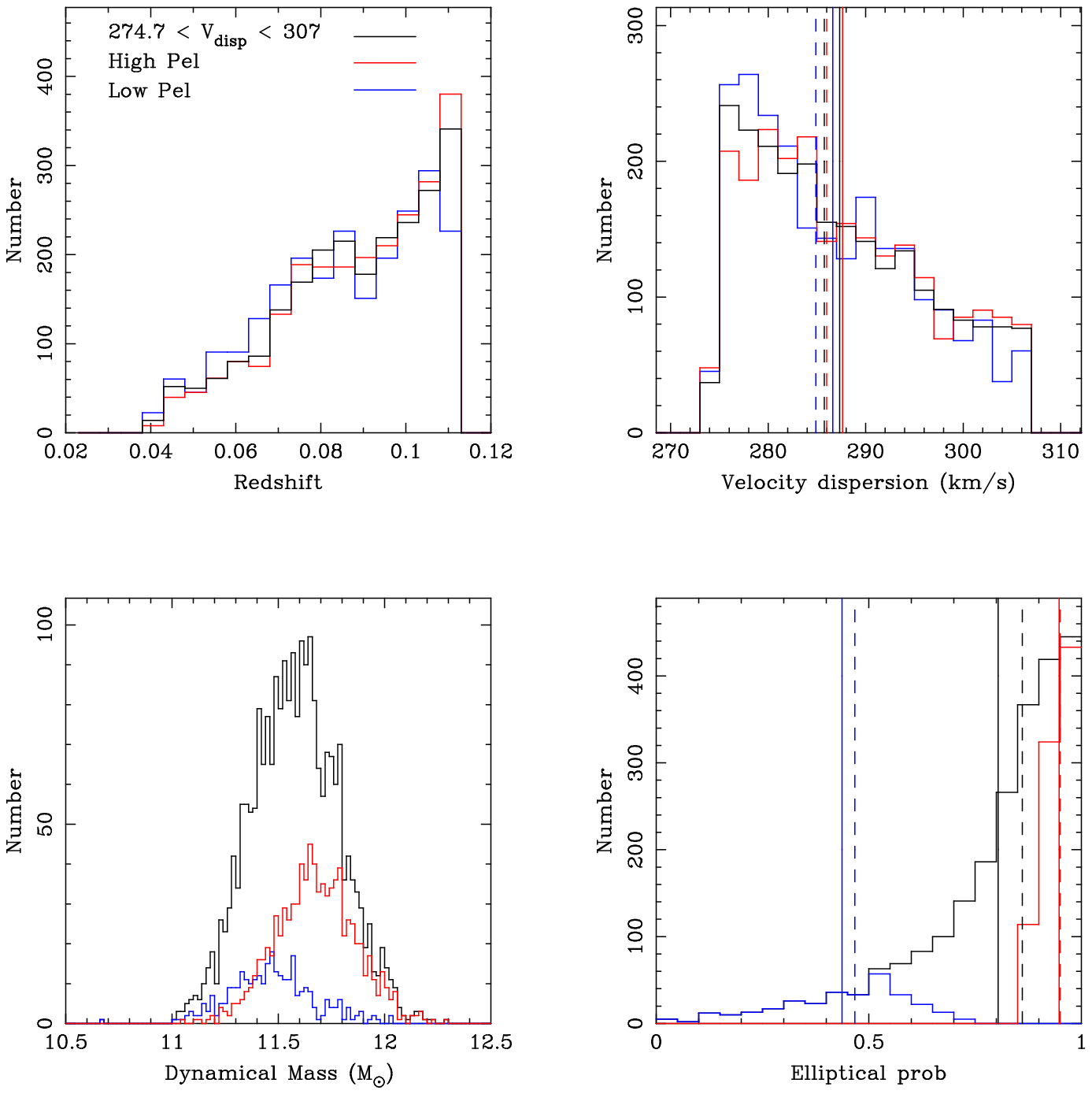}
\caption{\small The redshift, velocity dispersion, dynamical mass and elliptical probability distributions for the high (red) and low (blue) elliptical probability at fixed dynamical mass density samples (left) and at fixed velocity dispersion samples (right). The solid and dashed vertical lines show the mean and median of the distributions respectively. The distributions in the redshift and dynamical mass plots have been scaled to the same number of objects.
\label{fig:dmorphdist}}
\end{figure}

\clearpage
\section{B. \wp~ratios for samples cut by elliptical probability or color}

\begin{figure}[h]

\vspace{10cm}
\includegraphics{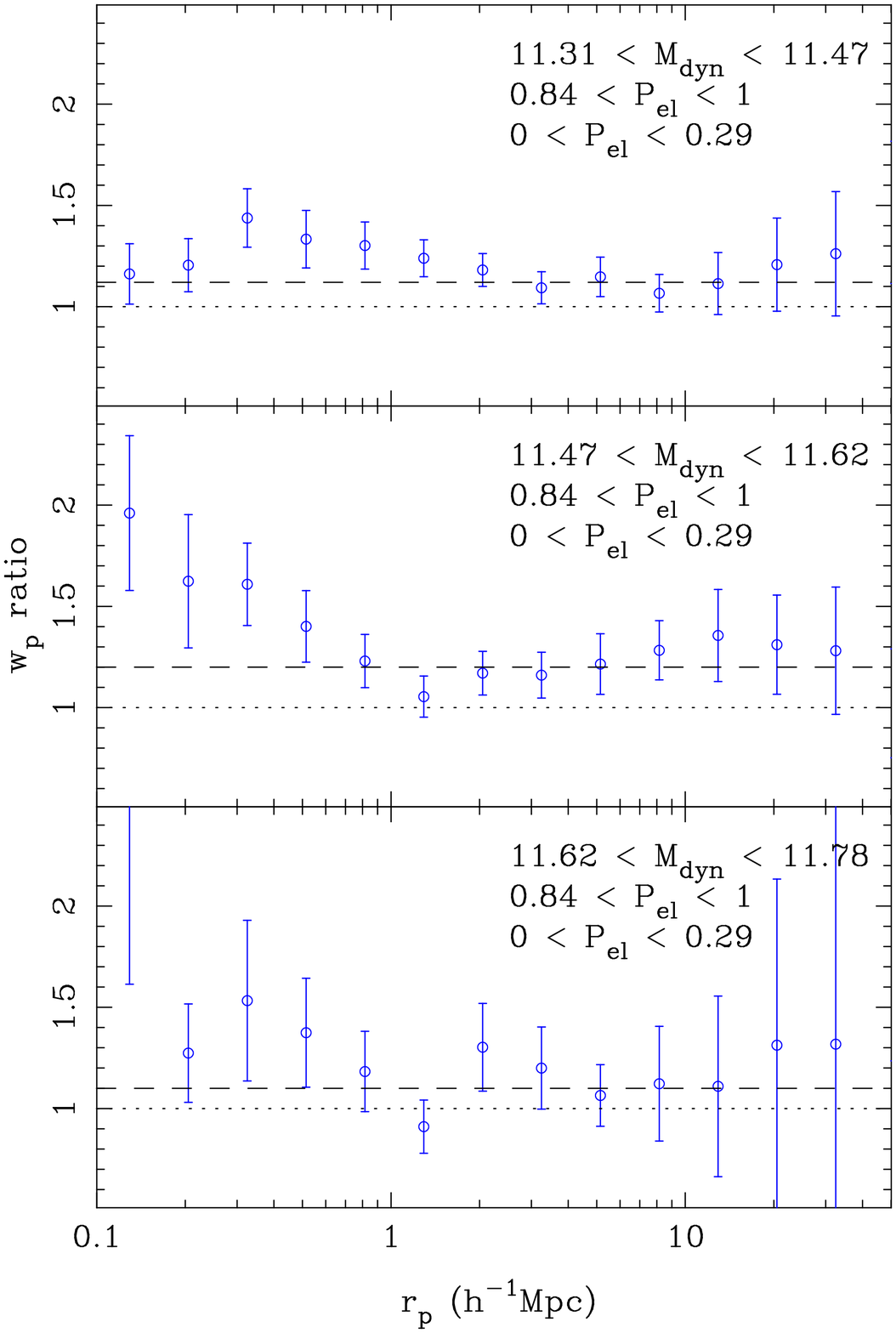}
\includegraphics{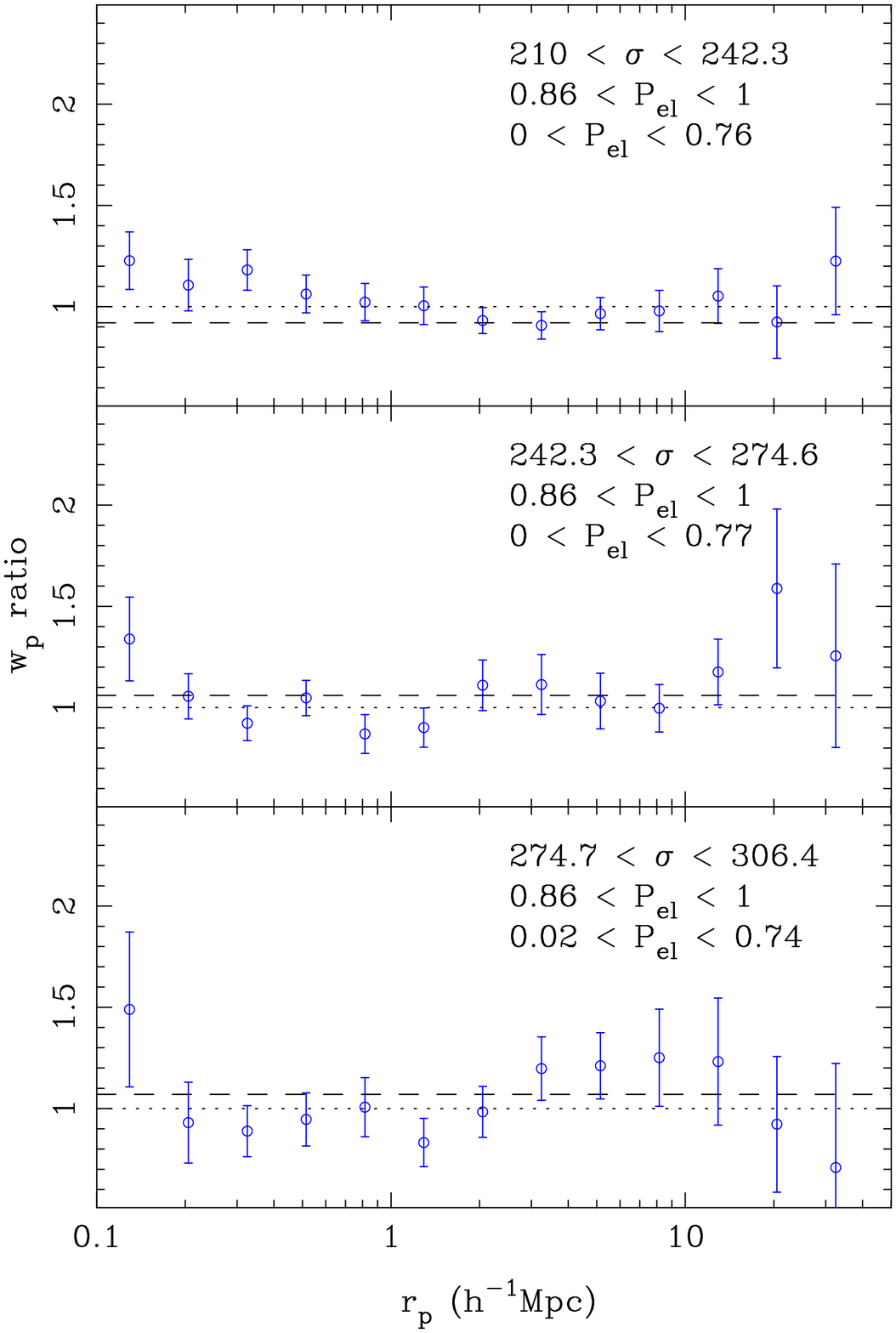}
\caption{\small The ratio of the projected correlation functions between high and elliptical probability samples at fixed dynamical mass (left) and velocity dispersion (right). The best fit ratio on scales $1.6 < r_p < 25.1$ h$^{-1}$Mpc is shown as the dashed line.
\label{fig:2ptPel}}
\end{figure}

\begin{figure}[h]

\vspace{10cm}
\includegraphics{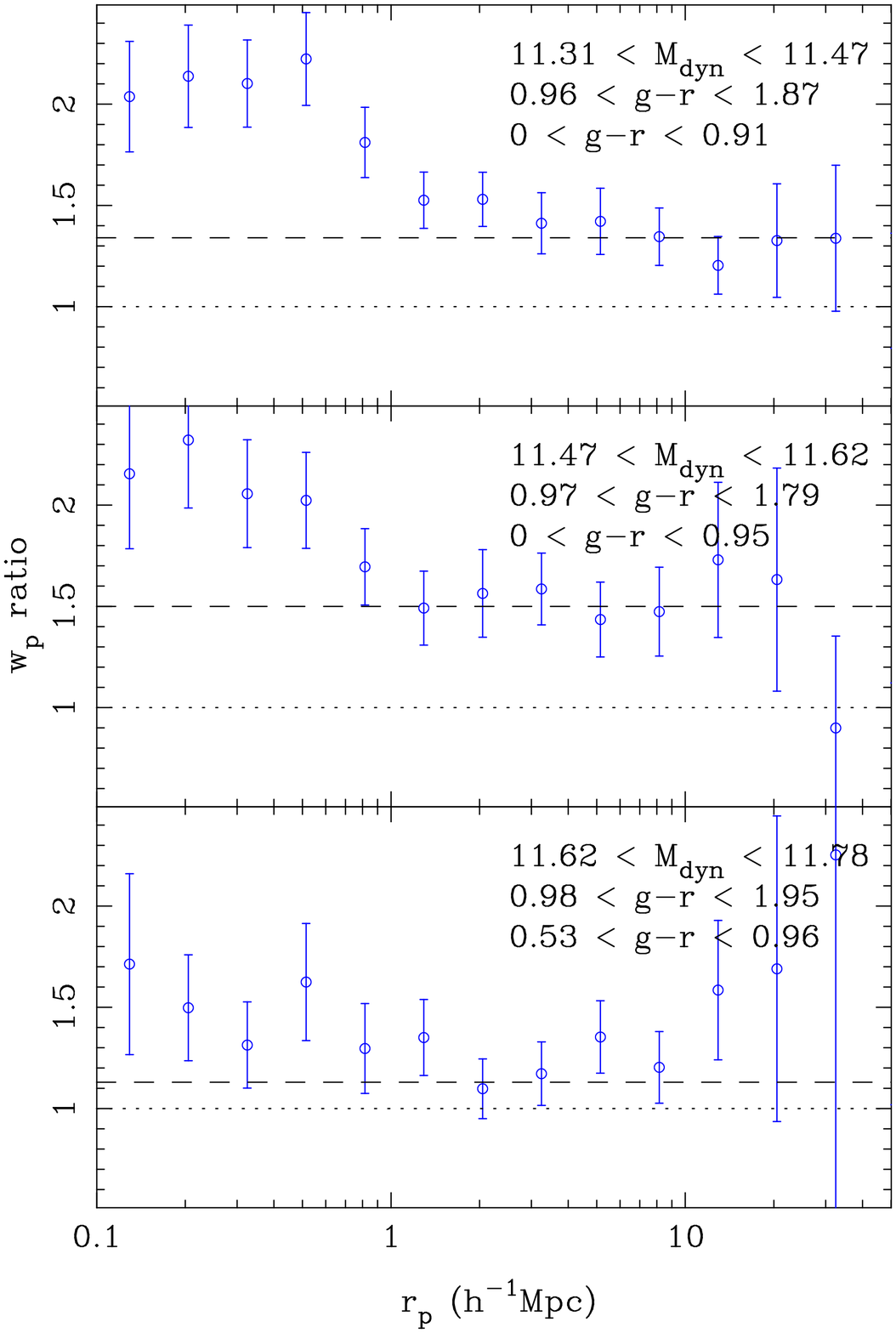}
\includegraphics{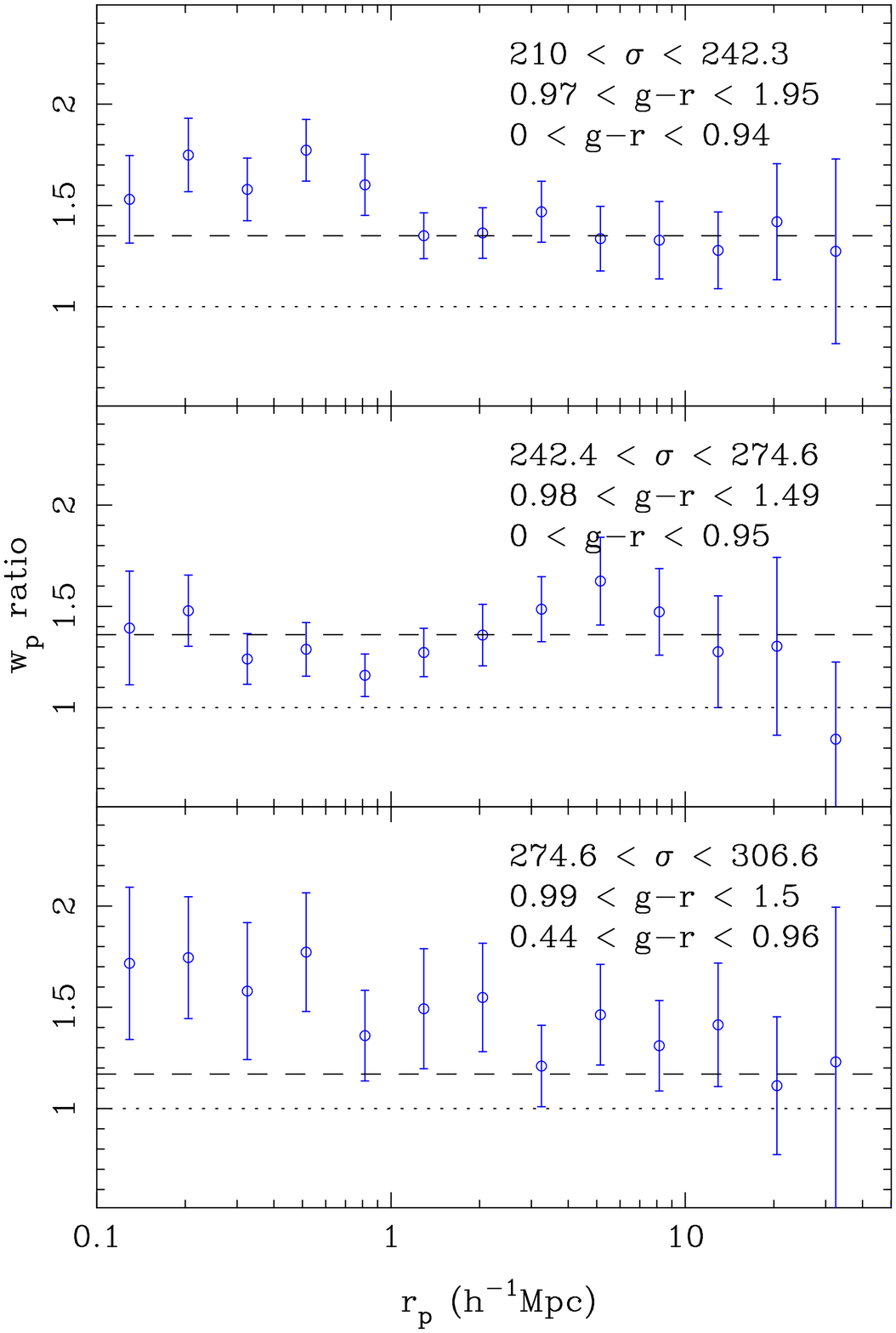}
\caption{\small The ratio of the projected correlation functions between red and blue g-r color samples at fixed dynamical mass (left) and velocity dispersion (right). The best fit ratio on scales $1.6 < r_p < 25.1$ h$^{-1}$Mpc is shown as the dashed line.
\label{fig:2ptcol}}
\end{figure}

\clearpage

\begin{figure}

\vspace{10cm}
\includegraphics{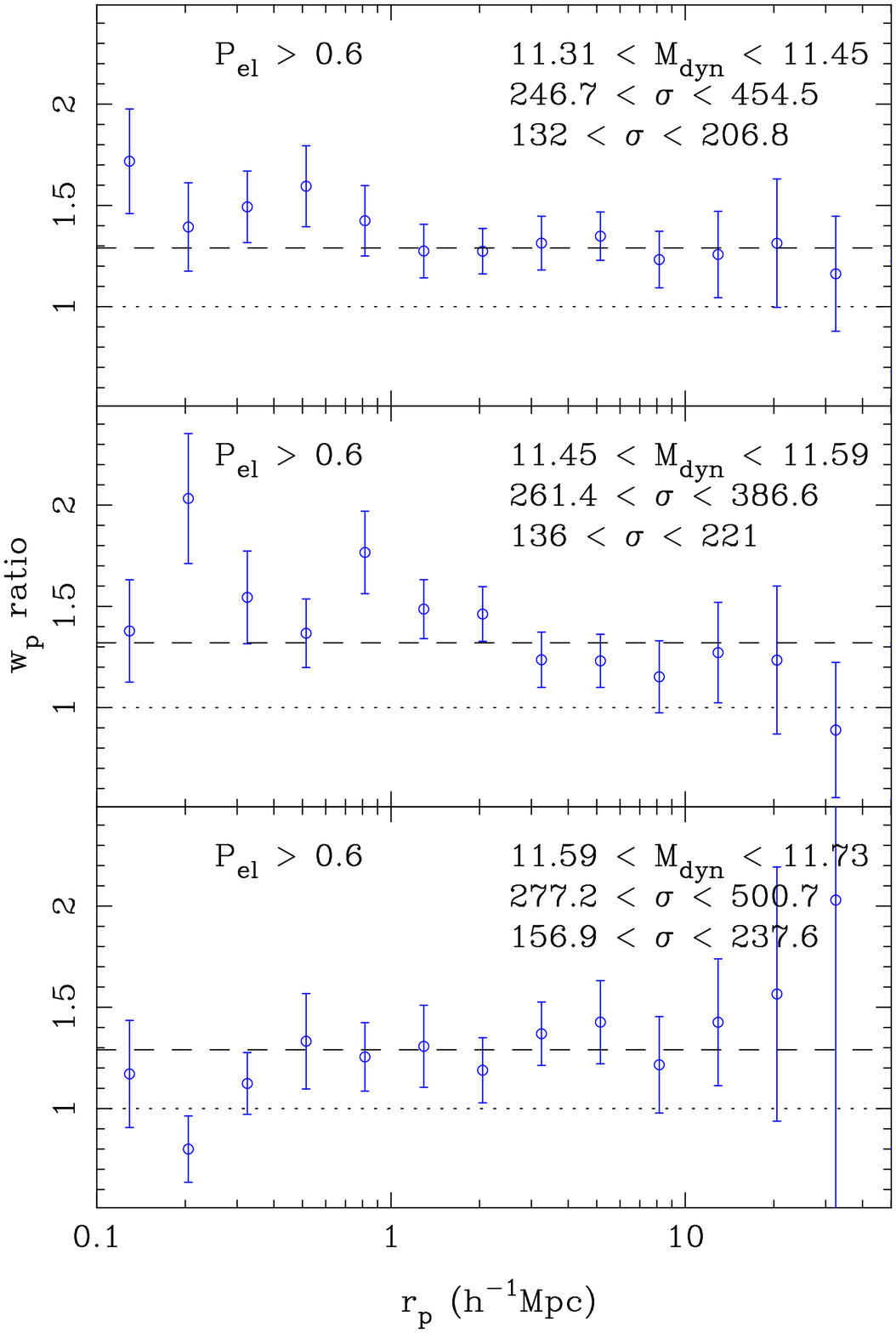}
\includegraphics{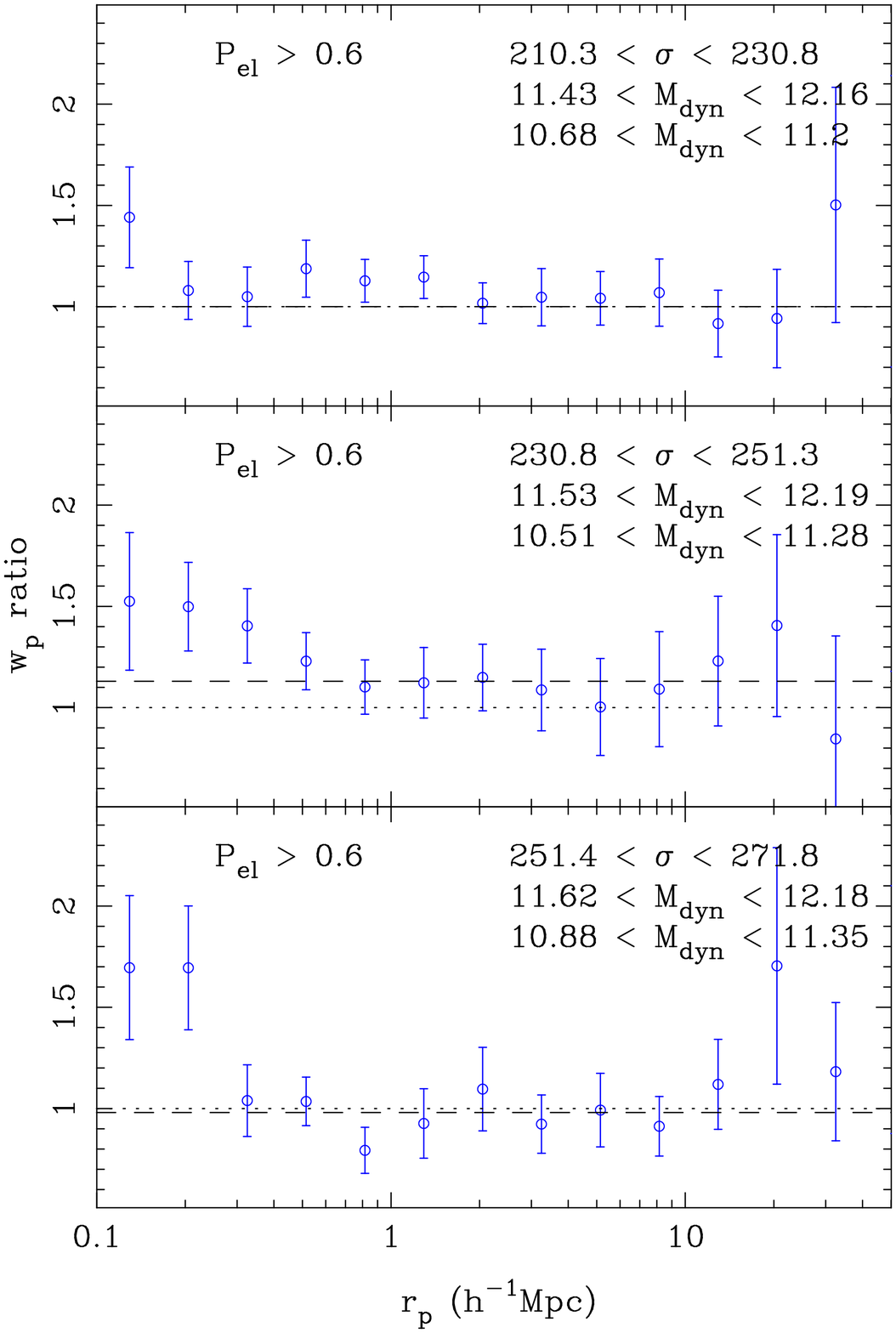}
\caption{\small The ratio of the projected correlation functions between high and low \Vdisp~samples at fixed \Mdyn~for Elliptical galaxies (P$_{el} >$ 0.6 left) and between high and low \Mdyn~samples at fixed \Vdisp~for Elliptical galaxies (P$_{el} >$ 0.6 left). The best fit ratio on scales $1.6 < r_p < 25.1$ h$^{-1}$Mpc is shown as the dashed line.
\label{fig:2ptVddMel}}
\end{figure}

\begin{figure}

\vspace{10cm}
\includegraphics{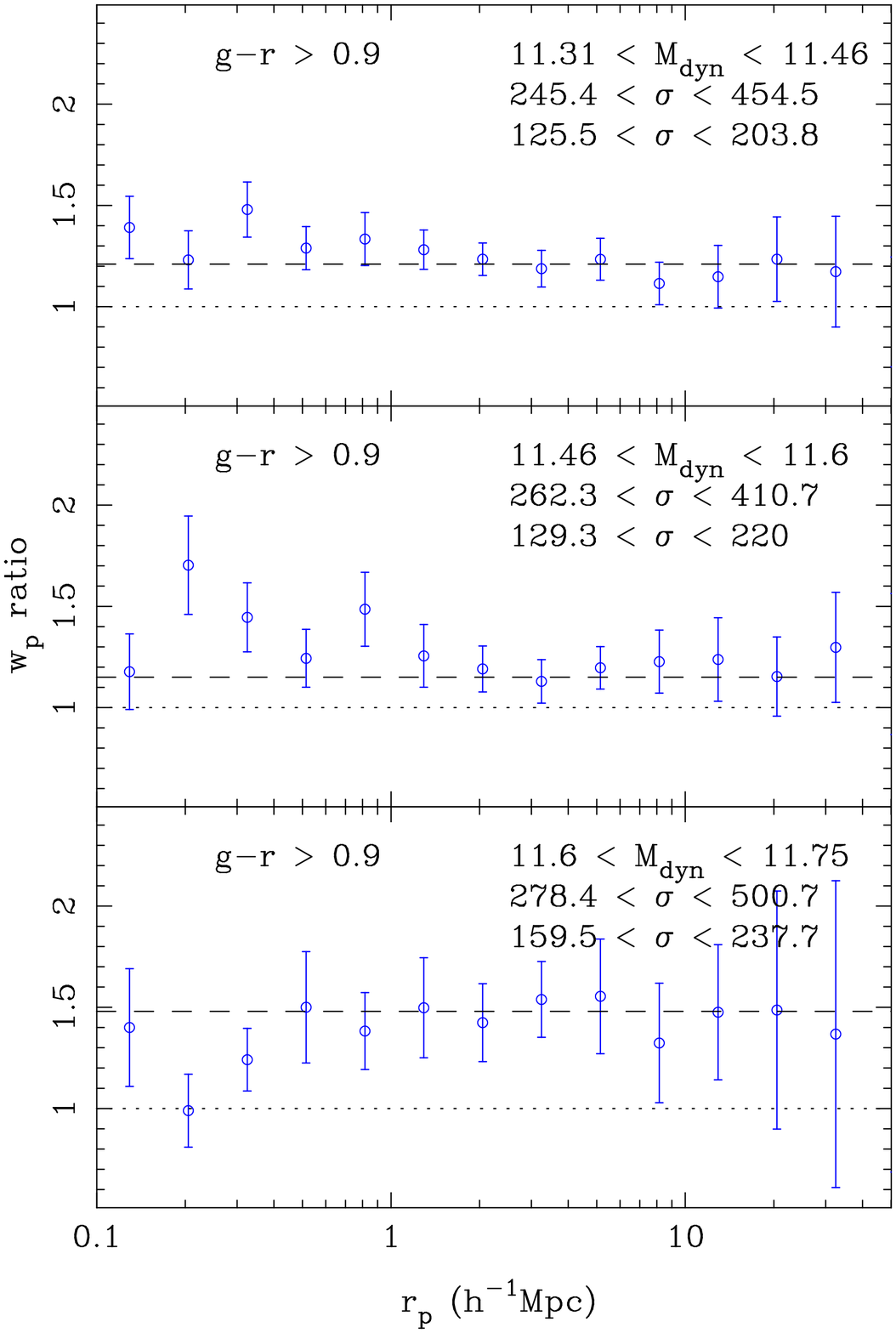}
\includegraphics{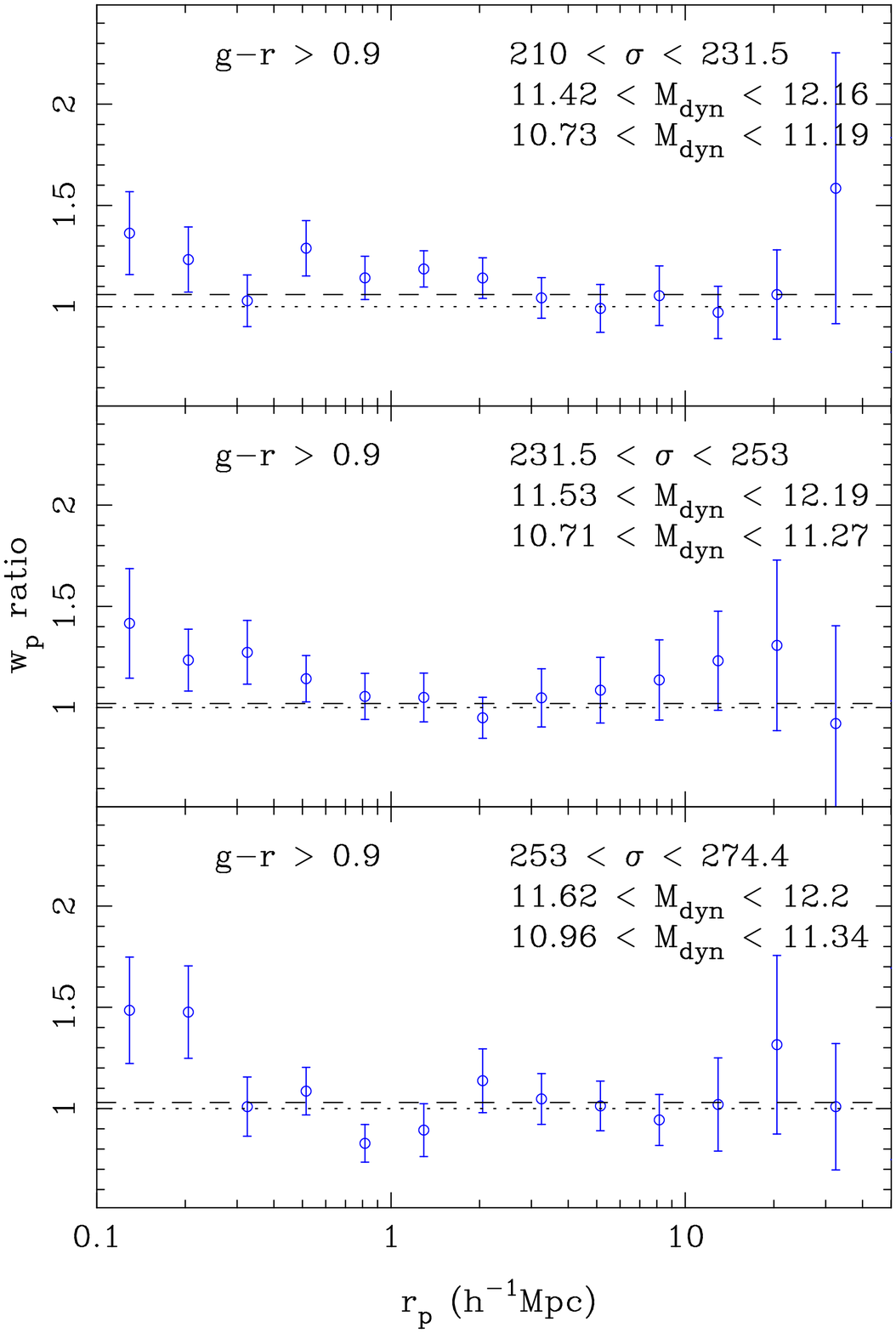}
\caption{\small The ratio of the projected correlation functions between high and low \Vdisp~samples at fixed \Mdyn~for red galaxies ($g-r >$ 0.9 left) and between high and low \Mdyn~samples at fixed \Vdisp~for red galaxies ($g-r >$ 0.9 left). The best fit ratio on scales $1.6 < r_p < 25.1$ h$^{-1}$Mpc is shown as the dashed line.
\label{fig:2ptVddMred}}
\end{figure}

\clearpage

\end{document}